\documentclass[fleqn,usenatbib]{mnras}
\usepackage{newtxtext,newtxmath}
\usepackage{soul}
\usepackage[T1]{fontenc}
\usepackage{ae,aecompl}
\usepackage{graphicx}	
\usepackage{amsmath}	
\usepackage{amssymb}	
\usepackage{xcolor}
\usepackage{pdflscape}	
\newcommand{\rsun}{\mathrm{R}_{\odot}}
\newcommand{\msun}{\mathrm{M}_{\odot}}
\title[Planetary nebulae from common envelope evolution]{Bipolar Planetary Nebulae from Outflow Collimation by Common Envelope Evolution}
\author[Y. Zou et al.]{Yangyuxin Zou,$^{1}$\thanks{yzou5@ur.rochester.edu}
Adam Frank,$^{1}$
\newauthor
Zhuo Chen,$^{2}$
Thomas Reichardt,$^{3,4}$
Orsola De Marco,$^{3,4}$ 
\newauthor 
Eric G. Blackman,$^{1}$
Jason Nordhaus,$^{5,6}$
Bruce Balick,$^{7}$
\newauthor
Jonathan Carroll-Nellenback,$^{1}$
Luke Chamandy,$^{1}$
Baowei Liu,$^{1}$\\
$^{1}$Department of Physics and Astronomy, University of Rochester, Rochester, NY 14627, USA\\
$^{2}$Department of Physics, University of Alberta, Edmonton, AB T6G 2E1, Canada\\
$^{3}$Department of Physics and Astronomy, Macquarie University, Sydney, NSW 2109, Australia\\
$^{4}$Astronomy, Astrophysics and Astrophotonics Research Centre, Macquarie University, Sydney, NSW 2109, Australia\\
$^{5}$National Technical Institute for the Deaf, Rochester Institute of Technology, Rochester, NY 14623, USA\\
$^{6}$Center for Computational Relativity and Gravitation, Rochester Institute of Technology, Rochester, NY 14623, USA\\
$^{7}$Department of Astronomy, University of Washington, Seattle, WA 98195, USA 
}
\date{Accepted XXX. Received YYY; in original form ZZZ}
\pubyear{2020}
\begin{document}
\label{firstpage}
\pagerange{\pageref{firstpage}--\pageref{lastpage}}
\maketitle
\begin{abstract}
The morphology of bipolar planetary nebulae (PNe) can be attributed to interactions between a fast wind from the central engine and dense toroidal shaped ejecta left over from common envelope (CE) evolution.  
Here we use the 3-D hydrodynamic AMR code {\tt AstroBEAR} to study the possibility that bipolar PN outflows can emerge collimated even from an uncollimated spherical wind in the aftermath of a CE event.
The output of a single CE simulation via the SPH code {\tt PHANTOM} serves as the initial conditions.  Four cases of winds, all with high enough momenta to account for observed high momenta preplanetary nebula outflows, are injected spherically from the region of the CE binary remnant into the ejecta.  We compare cases with two different momenta and cases with no radiative cooling versus application of optically thin emission via a cooling curve to the outflow. 
Our simulations show that in all cases highly collimated bipolar outflows result from deflection of the spherical wind via the interaction with the CE ejecta. Significant asymmetries between the top and bottom lobes are seen in all cases. The asymmetry is strongest for the lower momentum case with radiative cooling.
While real post CE winds may be aspherical, our models show that collimation via "inertial confinement" will be strong enough to create jet-like outflows even beginning with maximally uncollimated drivers. 
Our simulations reveal detailed shock structures in the shock focused inertial confinement (SFIC) model and develop a lens-shaped inner shock that is a new feature of SFIC driven bipolar lobes.

\end{abstract}
\section{Introduction}
Planetary nebulae (PNe) are remnants of the late stages of stellar evolution of low and intermediate mass stars.  They are created by the interaction between a fast wind ejected during the post-asymptotic giant branch (AGB) evolutionary state of a star and slow moving, previously ejected AGB material. 

We begin with a short review of the history of the problem as some of the explicit issues this paper addresses have been outstanding in the community for some time.
The earliest version of this paradigm called the Interacting Stellar Wind model (ISW) was first described by \citet{Kwok+1978}. 
According to the ISW model the slow-wind was considered to be a spherical outflow such as the superwind phase of an AGB star ($v \sim 10~\mathrm{km~s^{-1}}$).  
As the star evolves from the AGB phase to the PN nucleus, the speed of the line-driven winds would then rise up to $v\sim1000~\mathrm{km~s^{-1}}$, 
sweeping the slow wind material into a dense shell while also forming a hot bubble of shocked fast wind material.  
The bright rims of PNe are then the ionized shells of swept up AGB material.
The ISW model has proven itself highly successful at
explaining the properties of spherical or mildly elliptical PNe \citep{Schonberner+2014}.

Most PNe, especially young PNe, are not spherical, and a generalization of this model relied on "inertial confinement" of the spherical fast wind by an aspherical AGB slow wind \citep{Balick1987}.  Analytic solutions \citep{Icke1988} and later simulations \citep{Icke+1992a, Icke+1992b, Frank+1993, Frank+Mellema1994a, Frank+Mellema1994b} verified the ability of this Generalized Interacting Stellar Wind (GISW) model to capture observed PN morphologies. 
In the GISW  model, the variation from mildly elliptical to strongly bipolar butterfly-shaped PNe is depended on the pole-to-equator density contrast in the AGB slow wind ($e = \rho_\mathrm{p}/\rho_\mathrm{e}$), as well as the specifics of its aspherical morphology.

After the introduction of the GISW model, other shaping processes were considered, including the role of magnetic fields \citep{Garcia-Segura+1999,2001Natur.409..485B,Garcia-Segura+2005,2007MNRAS.376..599N,BalickFrankIII}, clumps \citep{Steffen+Lopez2004} and the possibility that the fast winds are collimated close to the star \citep{Lee+Sahai2003}. New observations also showed that the PN shaping process may begin when the central star is still in the post-AGB or pre-PN (PPN) phases, before nebular ionization occurs \citep{Bujarrabal+2001}. 

The importance of collimated outflows as the driving source of PPN and PN was emphasized in a series of papers that attempted to strongly link observational characteristics of bipolar lobes to numerical models of winds and/or jets \citep{BalickFrankI,BalickFrankII, Akashi+Soker2018, Rechy-Garcia+2019, Estrella-Trujillo+2019, Schreier+2019}.
From these results it appears that applications of the GISW model must be capable of taking a wind from the central engine of any geometry and converting it into highly collimated flow that then drives highly collimated PPN or PN bipolar lobes. 

In spite of these advances, the GISW model often remains a background aspect of newer scenarios for PN formation.  Many MHD models, for example, still require a strong pole-to-equator density contrast to produce very tight-waisted nebulae (i.e. the bipolar lobes converge to narrow width region near the source).
Thus a pre-existing toroidal density distribution still figures centrally in our understanding of the PN shaping process.

One question which has not yet been fully addressed in GISW models of PNe is the origin of asphericity in the AGB slow wind.
Single star models for the production of a pole-to-equator density contrast have been proposed \citep{Bjorkman+Cassinelli1993} but most of these rely on rotation of the central star. The required rotation rates for these models have proven to be difficult to square with observations \citep{Garcia-Segura+2014}. 
Binary stars, on the other hand, offer a natural source of angular momentum that can be tapped to create toroidal-shaped flows, i.e a slow wind that has most of the material confined around the equatorial plane, or a small pole-to-equator density $\rho_p/\rho_e$. Binary star models for AGB wind shaping (and non-spherical PN production) have long been part of the PN literature 
\citep{Fabian+Hansen1979, Soker+Livio1989, Morris1981, Morris1990, Livio1993, Soker1992, Soker1997, Nordhaus+Blackman2006}. 
More recent 3-D simulation work has shown how detached binary interactions can produce density distributions that can shape PPN and PN flows \citep{Chen+2017}. 

Throughout the extensive discussion of binary stars and PNe, Common envelope (CE) evolution has been seen as a principle means of generating high pole-to-equator density contrast toroidal flows \citep{Livio+Soker1988,Nordhaus+Blackman2006}. 
CE interaction occurs when a more compact companion plunges into the envelope of a red giant branch (RGB) or AGB star \citep{Paczynski1976, Ivanova+2013}. 
The release of gravitational energy during the rapid orbital decay unbinds most of the AGB envelope creating an expanding toroidal flow that later becomes the aspherical slow wind needed for GISW models \citep{Nordhaus+Blackman2006}.
The ability to numerically simulate CE interaction to the point where the expanding envelope could serve as input for GISW models has, in the past, been hampered by the complexity of CE 3-D flows.
Over the last decade, however, a number of new simulation platforms have been developed for CE evolution including smooth particle methods (SPH) \citep{Passy+2012, Nandez+2014, Nandez+Ivanova2016, Ivanova+Nandez2016, Reichardt+2019}, adaptive mesh refinement (AMR) fixed mesh methods \citep{Ricker+Taam2012, Staff+2016, Kuruwita+2016,  Iaconi+2017, Iaconi+2018, Chamandy+2018} as well as moving mesh methods \citep{Ohlmann+2016,Prust+Chang2019}.

In this paper we use an SPH CE simulation as the input for GISW models. Our goal is to explore the capacity of CE ejecta to serve as the source for strongly collimated bipolar flows in PPNe. 
This question was also explored in \citet{Garcia-Segura+2018} who used AMR CE simulations as input. That work showed that PN collimation would occur but their results were restricted to 2.5 D axisymmetric models. 
Our studies, which build on those of \citet{Frank+2018}, are fully 3-D and, as such, we are particularly interested in the development of asymmetric features in the bipolar outflows.

The organization of the paper is as follows. In Section 2, we describe the methods and model. In Section 3, we present our results and explore the role of shock focused inertial confinement in generating bipolar outflows. In Section 4, we characterize and discuss asymmetries that develop as well as momentum budget questions. We conclude in Section 5.

\begin{table}
    \centering
    \caption{Summary of the four models. }
    \label{tab:models}
    \begin{tabular}{c|c|c|c|c}
        \hline 
        Model & A & B & C & D  \\
        \hline
        Cooling & No & No & Yes & Yes \\
        Momentum & High & Low & High & Low\\
        \hline 
    \end{tabular}
\end{table}

\section{Simulation setup}\label{sec:setup}

The initial condition in our simulation comes from a single 
CE simulation \citep{Iaconi+2017, Reichardt+2019} produced by the smoothed 
particle hydrodynamic (SPH) code {\tt PHANTOM}
\citep{Price+2017, Price+2018}. 
The model used here is the 1.1 million particle simulation in \citet[][rendered in fig. 1, 2, and the top two rows in fig. 14 of their paper]{Reichardt+2019}.
The simulation involves a $0.88~\msun$ RGB star with a $0.392~\msun$ core, and a $0.6~\msun$ compact companion. 
Thus the final binary system of the two stellar cores has a total mass of about $1~\msun$. 
The SPH simulation lasted for $\sim14$ years of physical time, and the binary orbital separation had been approximately constant for the last $\sim2000$ days of that period.  
The orbit-averaged final separation is about $20~\rsun$. 

Our simulation uses the AMR code {\tt AstroBEAR} \citep{Carroll-Nellenback+2013}. 
We compare simulations of jets with two different values of momenta and cases with no radiative cooling versus application of optically thin emission via a cooling curve to the outflow. Table~\ref{tab:models} summarizes the four models presented in this work. 
Each model runs on 120 cores. Wall clock time for models A, B, and C evolving to the scale presented here is about 5 days, and 10 days for model D. 

The AMR simulation domain is a cubic box with side length of $128\,000\,\rsun$. The coarsest grid cells have a length of $1000\,\rsun$ on each side. There are 7 levels of AMR, with each finer grid having a refinement level of 2 over the previous. Hence, the finest grid has cells of $\sim7.8\,\rsun$ per side. The data for these large and small grids, along with the 4th level grid, were obtained by interpolating the density, internal energy and velocity from the SPH CE simulation to 3D grids. This interpolation was done with the SPH visualisation tool, \textsc{splash} \citep{Price2007}. The mapping is not strictly conservative, the gas quantities are interpolated and sampled at each point on a specified grid.

After mapping the initial condition onto the grids, a point particle of $1~\msun$ is placed at the center of the domain to represent the binary system. The point particle interacts with the gas only by its gravity and has a softening radius of $39~\rsun$ (5 times the size of a 7th level grid cell). Self-gravity of the gas is not included. Note that the final binary separation of the SPH simulation is within this gravitational softening radius. Thus we do not resolve the evolution of the binary any further or consider its effect on the fast wind in our simulation except for its gravity.

A spherical fast wind is initialized within a radius of $46.9~\rsun$ (6 times the size of a 7th level grid cell, hereafter $r_{\mathrm{FW}}$) around the central particle. Within that radius, density and temperature are set to fixed initial values, and are updated at each time step. This fast wind condition is initialized at the beginning of our simulation, but with zero outflow speed until the time we ``turn on" the outflow by giving it a uniform radial speed in all directions. 

We note that the choice of a spherical fast wind is likely an oversimplification. As has been discussed in the literature, the wind may form from the primary core or via a disk around the primary, the secondary, or the binary \citep{2001ApJ...546..288B,Matt+2006,2019MNRAS.490.1179G}.  

During initialization, a floor temperature of 20 K and floor density of $4\times 10^{-18}~\mathrm{g\,cm^{-3}}$ have been applied to remove noise from mapping of the SPH data to the AMR grids. After the first time step, the allowed minimum values in the simulation are lowered for temperature to 10 K, and for density to $1\times 10^{-22}~\mathrm{g\,cm^{-3}}$. 

\begin{figure}
    \centering
    \includegraphics[height = 6.4 cm]{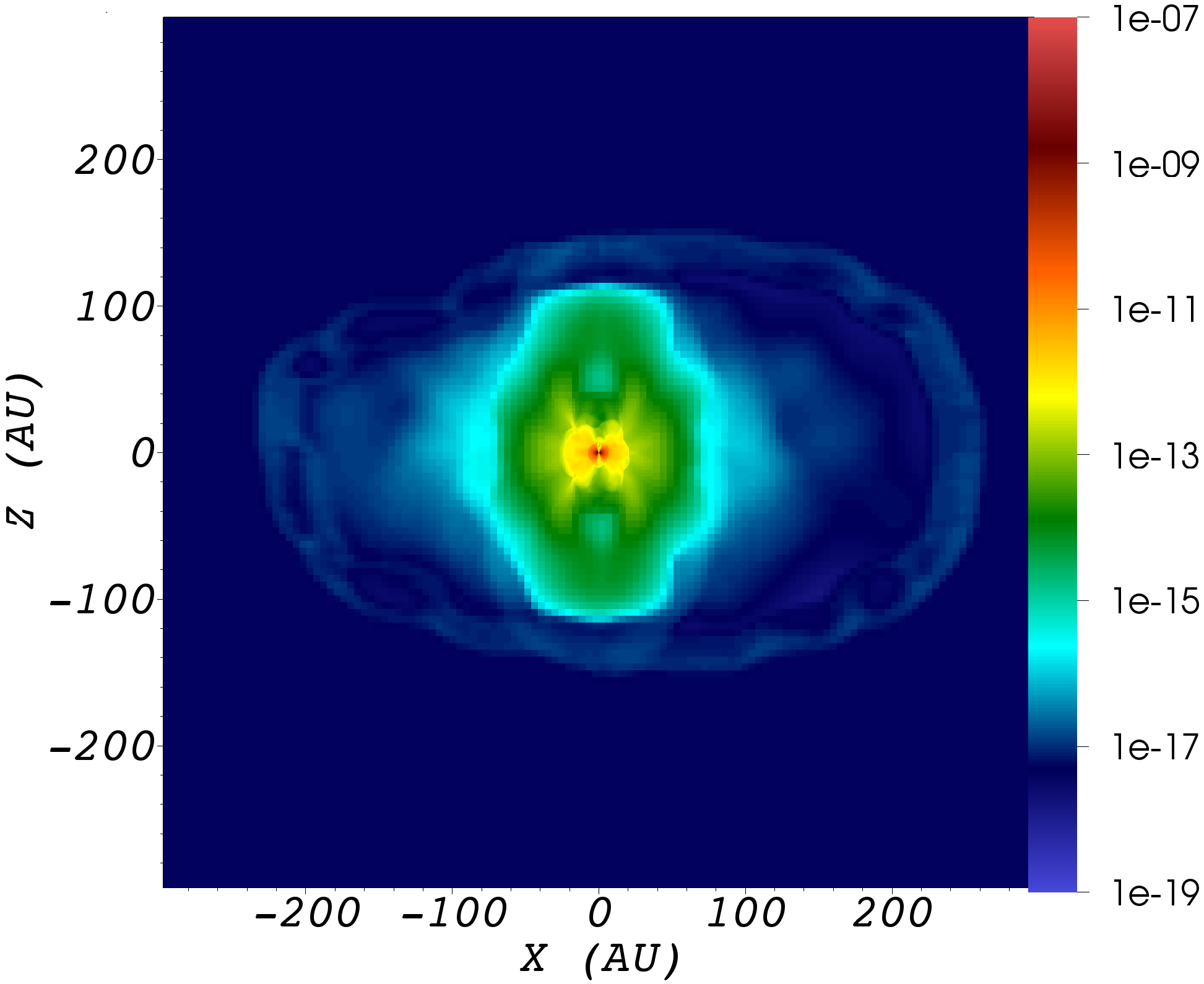}
     \includegraphics[height = 6.4 cm ]{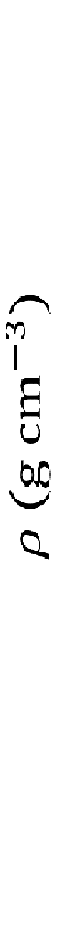}\\
    \includegraphics[height = 6.4 cm]{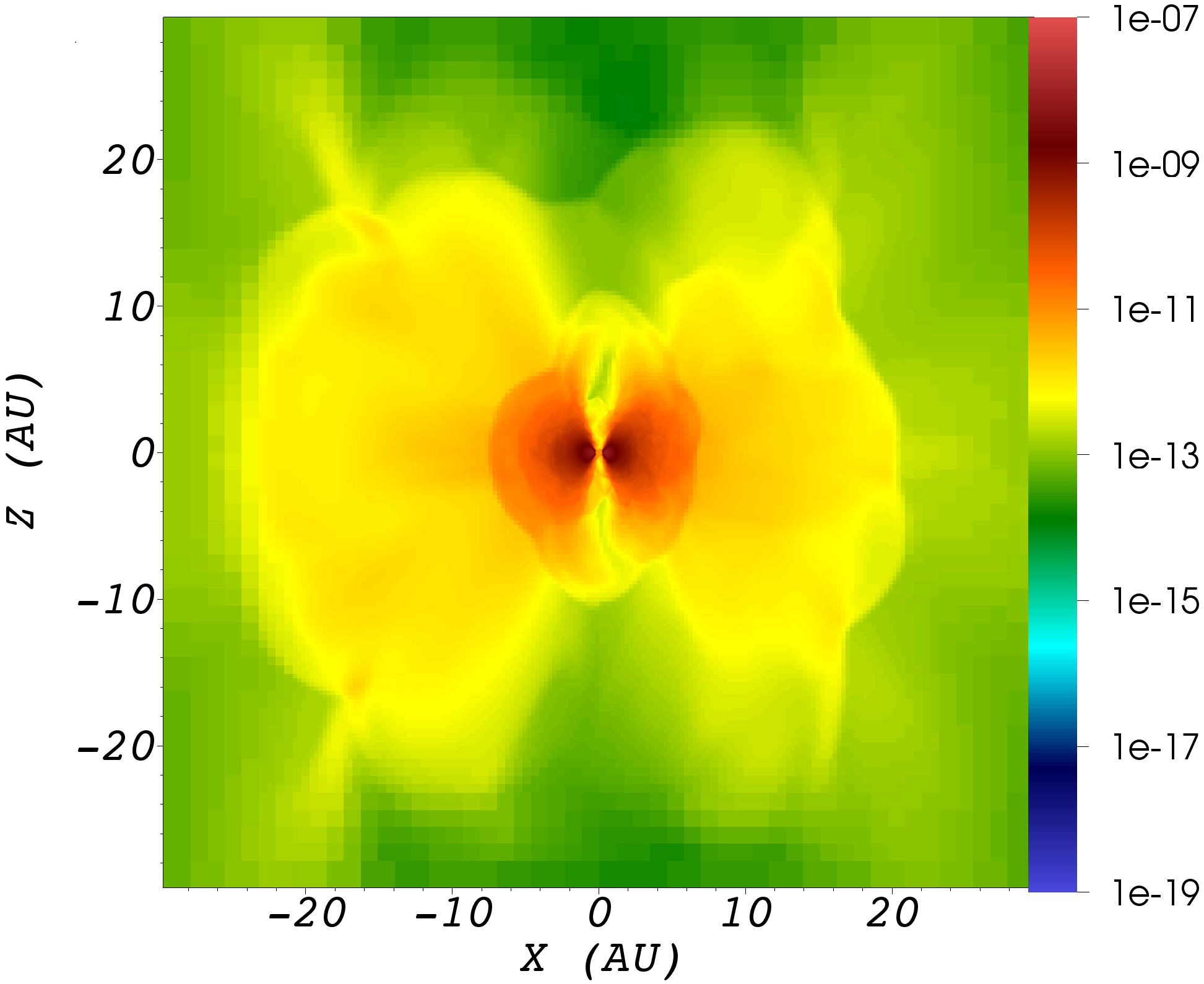}
     \includegraphics[height = 6.4 cm ]{time-evolutions/rho-short.png}
    \caption{Gas density of the CE ejecta before the fast wind turned on (2980 days after the simulation starts). Upper panel shows the entire simulation domain. The snapshot shown here is a vertical slice taken at the middle of the domain with the binary orbit in the x-y plane. The x- and z-axis are distance in $\mathrm{AU}$, and the color bar shows log scaled density in unit of $\mathrm{g\,cm^{-3}}$. Lower panel is a zoom-in to the central binary remnant. Note the density contrast between the relatively low density ``funnels'' bounded by high density equatorial walls. Minimum pole to equator contrast is $\rho_\mathrm{p}/\rho_\mathrm{e}=1.67\times10^{-4}$ at radius $r\sim 0.4~\mathrm{AU}$.}
    \label{fig:IW}
\end{figure}

The first part of the simulation is a quiescent period of evolution to let the CE ejecta relax on the AMR grids. 
During this 3000-day quiescent phase, the ejecta expands by about 4 times in length scale without significant change of the morphology\footnote{This behaviour independently confirms the studies of \citet{Iaconi+2019} who demonstrated that in the simulation of \citet{Reichardt+2019} the gas has, by the end of their simulation, reached homologous expansion.}.
This also ensures that, later in the simulation, bipolar lobes of the nebulae grow into regions already swept by the expanding ejecta, reducing possible artifacts from the SPH to AMR mapping.
Fig.~\ref{fig:IW} shows the density distribution before turning on the fast wind. 
The fast wind boundary allows material to pass both ways, but whatever comes within the radius $r_{\mathrm{FW}}\sim0.2~\mathrm{AU}$ will be removed from the simulation on the next time step. The dense regions immediately adjacent to the fast wind boundary condition show fallback towards the point source with approximately $0.19 ~\msun$ passing through the boundary during the entire quiescent period.
This infalling material reflects the fact that our choice of fast wind boundary is arbitrary and some gas in the densest inner regions would still remain bound to the binary perhaps in an accretion disk which may be the source of the wind \citep{Kuruwita+2016}. 
We note that the choice of fast wind boundary modifies the amount of CE material that falls back. In another test run where the fast wind boundary is doubled, $r_{\mathrm{FW}}\sim 0.4~\mathrm{AU}$, the total fallback is about $0.23~\msun$. These changes do not have a significant effect on our results however as a test run that has no quiescent period at all shows a morphology of the bipolar lobes that is qualitatively the same (see Fig.~\ref{fig:noIW-rho}).

Parameters for the fast wind injection are chosen with the problem of momentum excess for PPN in consideration. We chose a radial speed reasonable for a post-AGB wind, and set the total mechanical luminosity ($L_{\mathrm{mech}}=\dot{M}_{\mathrm{FW}}\cdot v_{\mathrm{FW}}^2$) be comparable with the radiative luminosity of such a star. Therefore, all models have high enough momenta in the fast wind to account for observed high momenta PPN outflows. We discuss the momentum budget questions later in Section \ref{sec:momentum}.
Table~\ref{tab:params} summarises parameters
related to the two momentum fluxes used in the simulation. 
The fast wind is turned on instantaneously after the quiescent
period, with a radial speed of $300~\mathrm{km\,s^{-1}}$. 
For each momentum flux, we run two models comparing the case of
no radiative cooling (models A and B) versus applying the Dalgarno-McCray cooling curve \citep{Dalgarno+McCray1972} to material above $30\,000~\mathrm{K}$ (models C and D).
This critical temperature is chosen to let cooling happen mostly in material interior to the bipolar outflow but not so much in the ejecta. We note our choice is arbitrary for the purpose of comparative study, while Dalgarno-McCray cooling becomes significant above $12\,000~\mathrm{K}$. 

Refinement of the grid is controlled by a combination of layers of fixed cubic boxes and an narrow rectangular region (referred to as a prism) whose length in z increases. When mapping the SPH data, the base level grid covers the entire domain, while the 4th level grid occupies a cubic box of $\sim37~\mathrm{AU}$ per side, and the 7th level grid occupies a cubic box of $\sim7~\mathrm{AU}$ per side, both centred around the point particle. Intermediate levels are placed around (by the AMR routines) so that finer grids are nested within coarser grids, and the maximum refinement changes by one level between regions of different resolution. This cubic region of higher refinement (level $\ge4$), though a small fraction of the entire simulation domain, contains the entire CE ejecta in its initial condition and later on the central PN nucleus. 
The other part of the control is the prism with a square base equal to the 7th level box, but its height along the z-axis grows, leading ahead the extent of the outflow material. A critical outflow density is set to capture the extent of the outflow material. Refinement is forced to the 7th level within this prism, and gradually reduced outside by the AMR. A snapshot with grid patches can be found in Appendix~\ref{apx:mesh}.


    \begin{table}
    \centering
	\caption{Simulation's input parameters.}
	\label{tab:params}
	\begin{tabular}{ccc} 
		\hline 
        & High-  & Low-  \\
        & momentum & momentum \\
        \hline 
        Models & A, C & B, D\\
        $\rho_{\mathrm{FW}}\,\,(\mathrm{g\,cm^{-3}})$ & $1\times10^{-11}$ & $5\times10^{-13}$ \\
        $v_{\mathrm{FW}}\,\,(\mathrm{km\,s^{-1}})$ & 300 & 300 \\
        Quiescent period (days) & $3\,000$ & $3\,000$  \\
        $\dot{M}_{\mathrm{FW}}\,\,(\mathrm{M_{\odot}~yr^{-1}})$ & $6.42\times10^{-4}$ & $3.21\times10^{-5}$ \\
        $T_{\mathrm{FW}}\,\,(\mathrm{K})$ & $30\,000$ & $30\,000$ \\
        $L_{\mathrm{mech}}\,\,(\mathrm{L_{\odot}})\simeq L_\mathrm{AGB}$ & $4.75\times10^{3}$ & $2.38\times10^{2}$ \\
        $(\dot{M}_{\mathrm{FW}}\cdot v_\mathrm{FW})/( L_{\mathrm{mech}}/c)$ & 
        $2000$ &   $100$\\
        \hline 
	\end{tabular}
    \end{table}

\section{Results}\label{sec:results}

Here we present a description of the results from our four numerical models. We first describe the two adiabatic models in Sections \ref{sec:A} and \ref{sec:B}. In Section \ref{sec:hot-bubble} we elaborate on the detailed structure of the shocks in the outflow. Then, in Sections \ref{sec:C} and \ref{sec:D}, we point out features that change when the system is allowed to cool via optically thin radiative emission. For each model, we choose four snapshots to represent the time evolution. The times are chosen such that the lobes are of comparable size for each model even if this means the snapshots are taken at different times across models. The snapshots are selected to match the physical height of the top lobe in model D, because it is the slowest evolving model and we stop our simulation at $7\,000$ days. Note that we have run models A and B to a larger physical scale (but less time) and we do not see significant change in the morphology. Thus the  expansion of the bipolar outflows eventually becomes homologous. In Section \ref{sec:asymmetry} we address the asymmetries in the models. 

Before we begin discussion of the results we remind the reader of the basic shock structure in GISW models. As in any colliding flow, two shocks form as the fast wind interacts with the previously deposited slow moving material. 
An outer shock, moving radially outward from the central wind source, is created as slow wind material is accelerated and swept up via the ram pressure of the interior fast wind. 
In addition, an inner shock forms moving back to the fast wind source via the deceleration of the fast wind. The inner shock converts fast wind kinetic energy into thermal energy of post-shock fast wind material (this energy will be radiated away if cooling is present). If both the fast and slow wind are spherical, the two shocks will be spherical.  
If, however, the slow wind is aspherical, then the inner shock can take on a convex semi-elliptical shape. This distortion of the inner shock is important because the radial fast wind streamlines no longer encounter the shock normally but instead strike it at an angle. 
Streamlines passing through a shock at an acute angle are refracted away from the shock normal.  In this way an aspherical slow wind in GISW models leads to a ``shock focused inertial confinement'' of the fast wind.  This leads to well-collimated jets \citep{Icke+1992a, Icke+1992b}.  
The stronger the pole-to-equator contrast ($e$) in the slow wind, the more effective is the shock focusing of fast wind material into bipolar jets \citep{Frank1999}. The formation of jets and bipolar outflows from shock focused inertial confinement (SFIC) has been studied in a variety of contexts from PN to star formation \citep{Frank1994, Mellema1994, Mellema1995, Mellema+Frank1995}. In Section \ref{sec:hot-bubble} we will use this basic understanding of shock structure to interpret our models.

The collimator for our flows is the highly aspherical CE ejecta (Fig.~\ref{fig:IW}).  In particular note the presence of what we will call ``funnels'' above and below the central core deep in the CE ejecta. These are relatively low density polar regions bounded by high density equatorial walls and form due to the angular momentum transfer during the CE orbital inspiral of the companion.  The pole to equator contrast at the end of the quiescent phase is $e = \rho_\mathrm{p}/\rho_\mathrm{e} \simeq 0.126 $ at radius $r \sim 0.2~\mathrm{AU}$ (against the fast wind radius), and the minimum is $e=1.67\times10^{-4}$ at radius $r\sim 0.4~\mathrm{AU}$.

\subsection{Model A: Adiabatic, high-momentum outflow}
\label{sec:A}
\begin{figure*}
    \includegraphics[height = 9 cm]{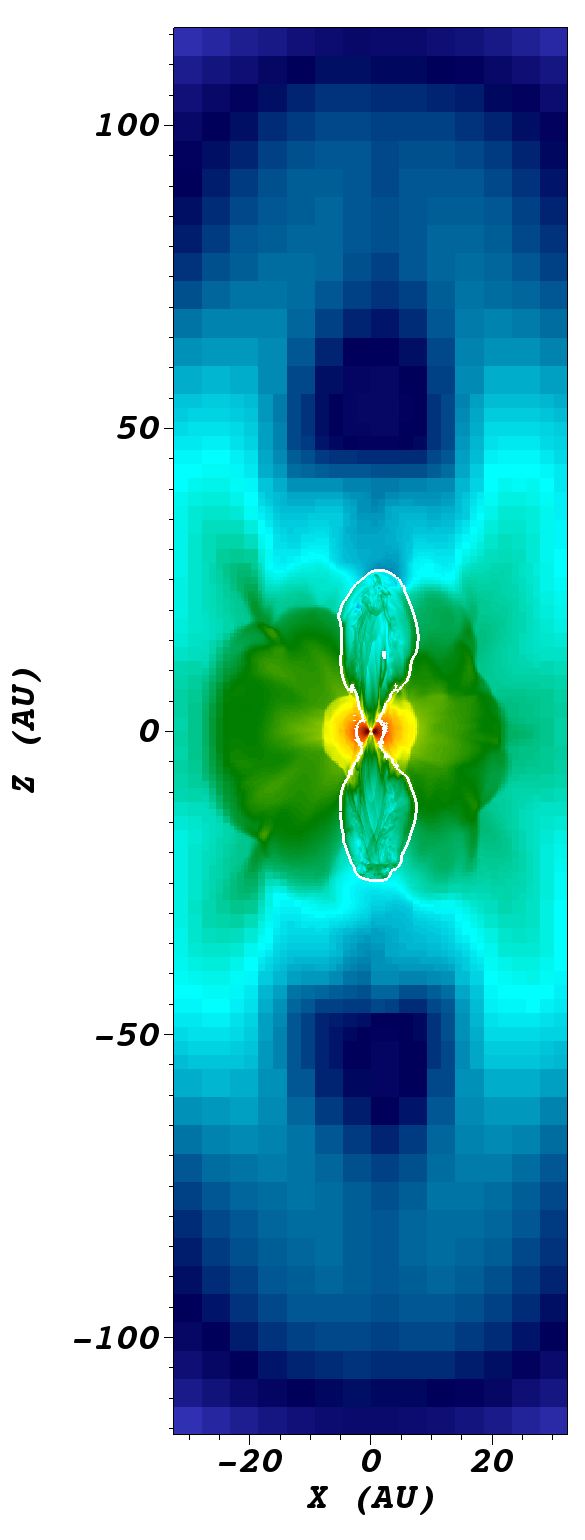}
    \includegraphics[height = 9 cm]{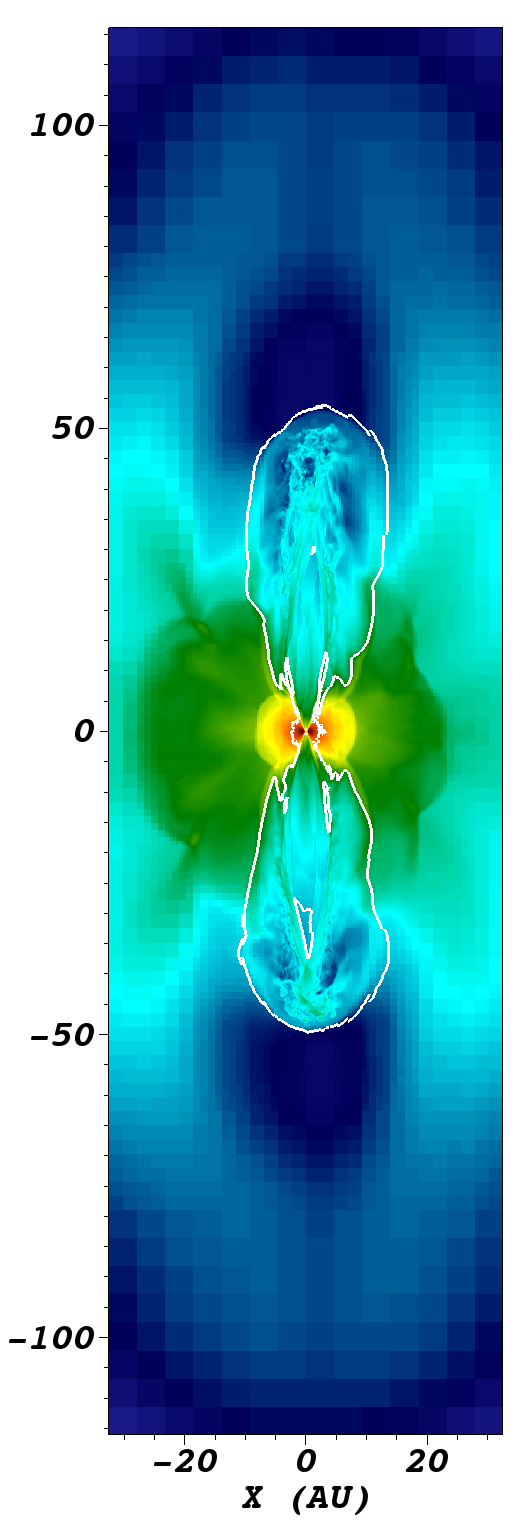}
    \includegraphics[height = 9 cm]{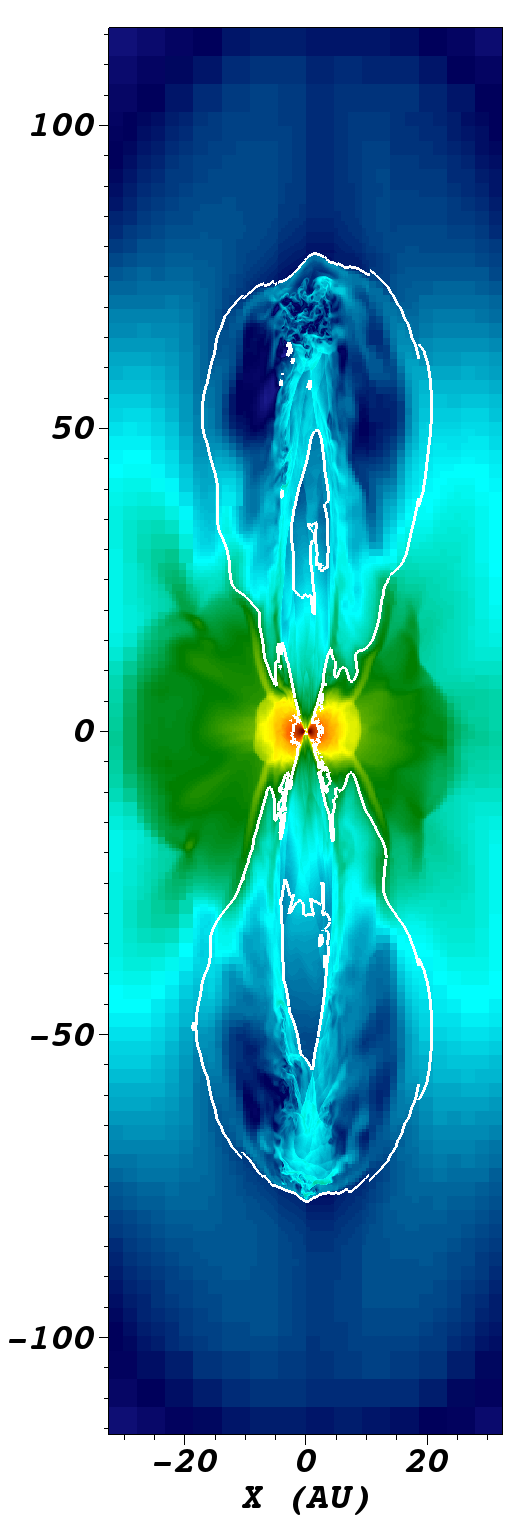}
    \includegraphics[height = 9 cm]{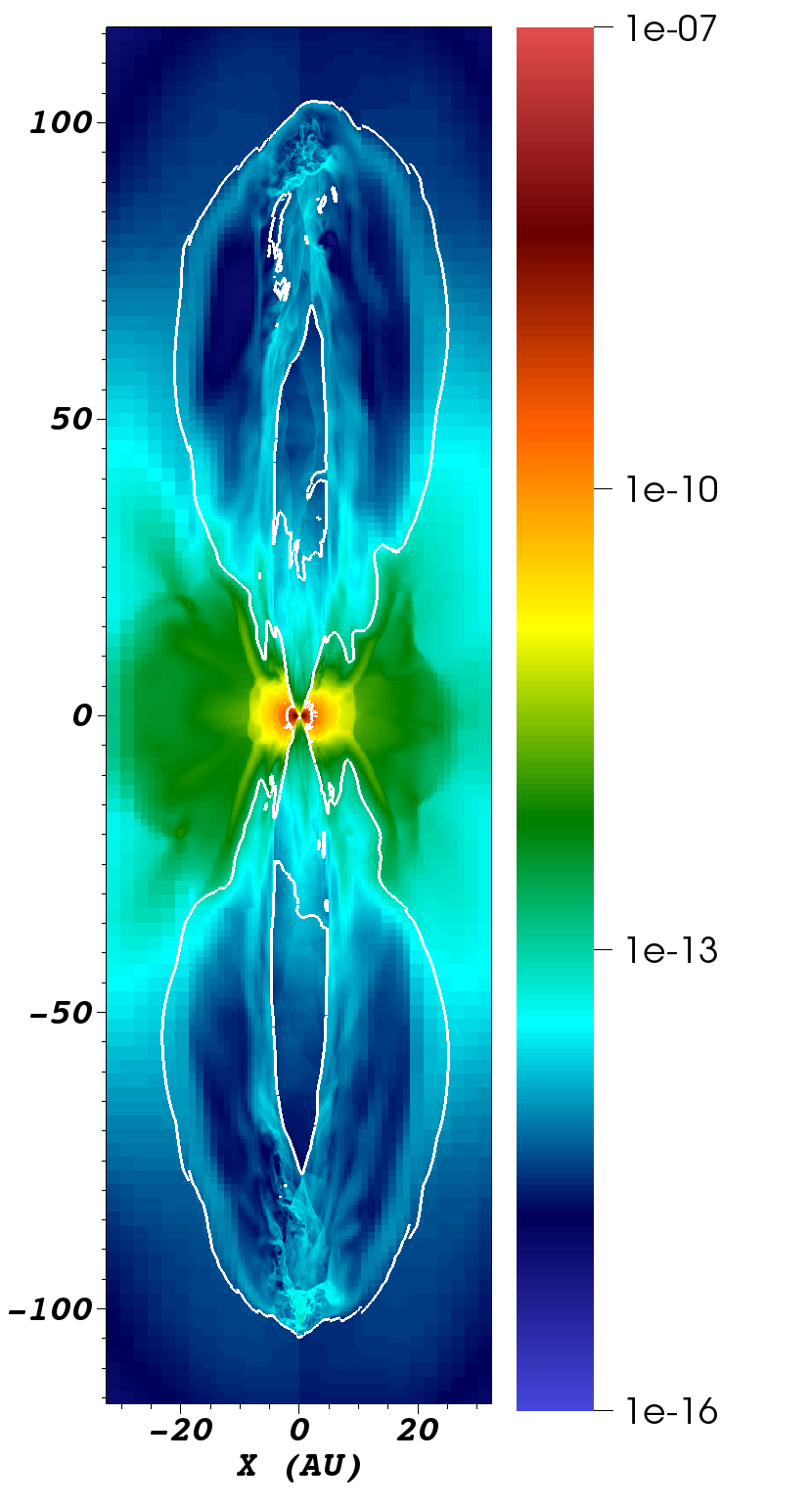}
    \includegraphics[height = 9 cm]{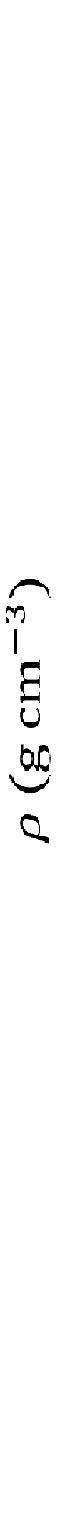}
    \caption{Model A: density ($\mathrm{g\,cm^{-3}}$, in log scale) plots of the high-momentum, no radiative cooling outflow taken at 3260, 3480, 3700, 3920 days in the simulation (260, 480, 700, 920 days after the fast wind turned on). White contours mark constant temperature of $10\,000~\mathrm{K}$. Note the outer contour coincides  
    with the outward-moving bow shock of the bipolar outflow, while the inner contour indicates a lens-shaped shock moving backward into the fast wind material. More details on the shock structures in Section \ref{sec:hot-bubble}. }
    \label{fig:A-rho}
\end{figure*}

\begin{figure*}
    \includegraphics[height = 9 cm]{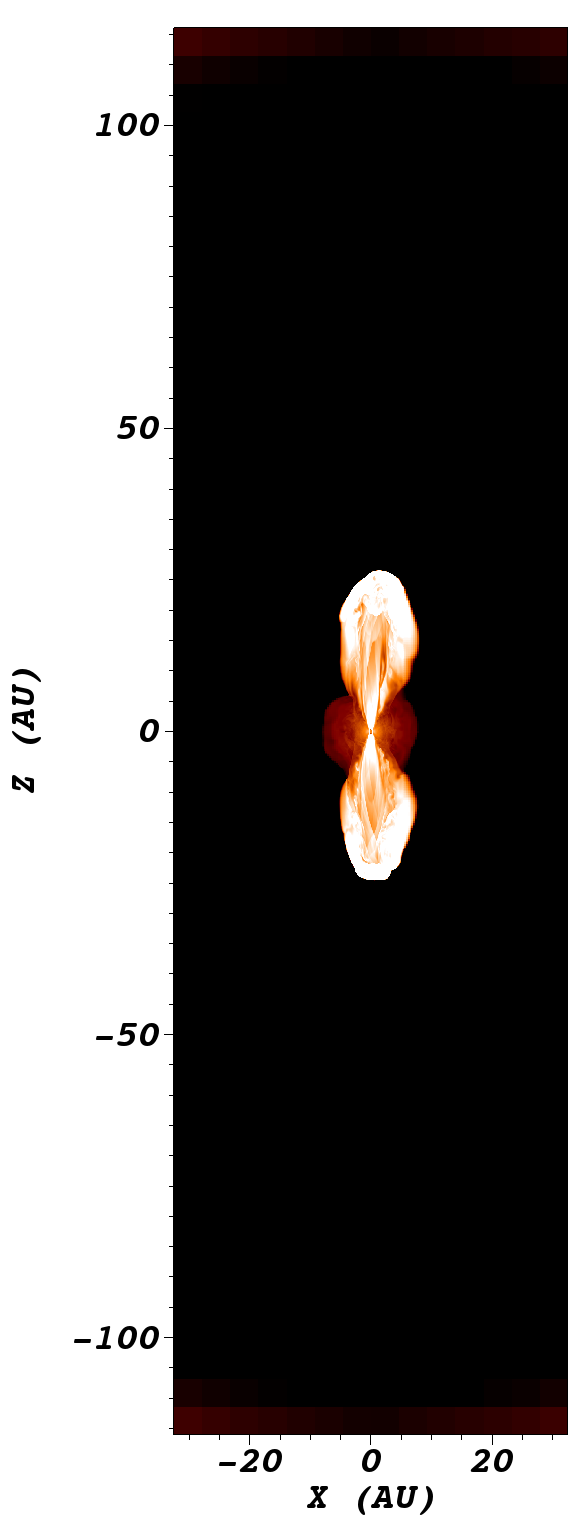}
    \includegraphics[height = 9 cm]{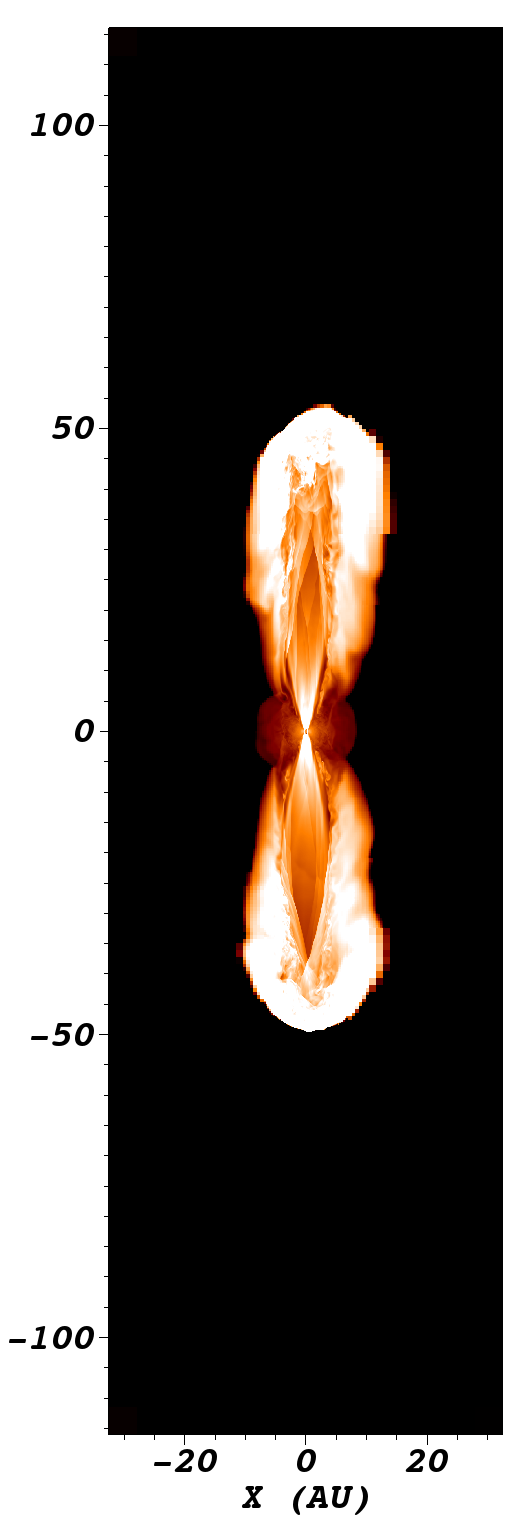}
    \includegraphics[height = 9 cm]{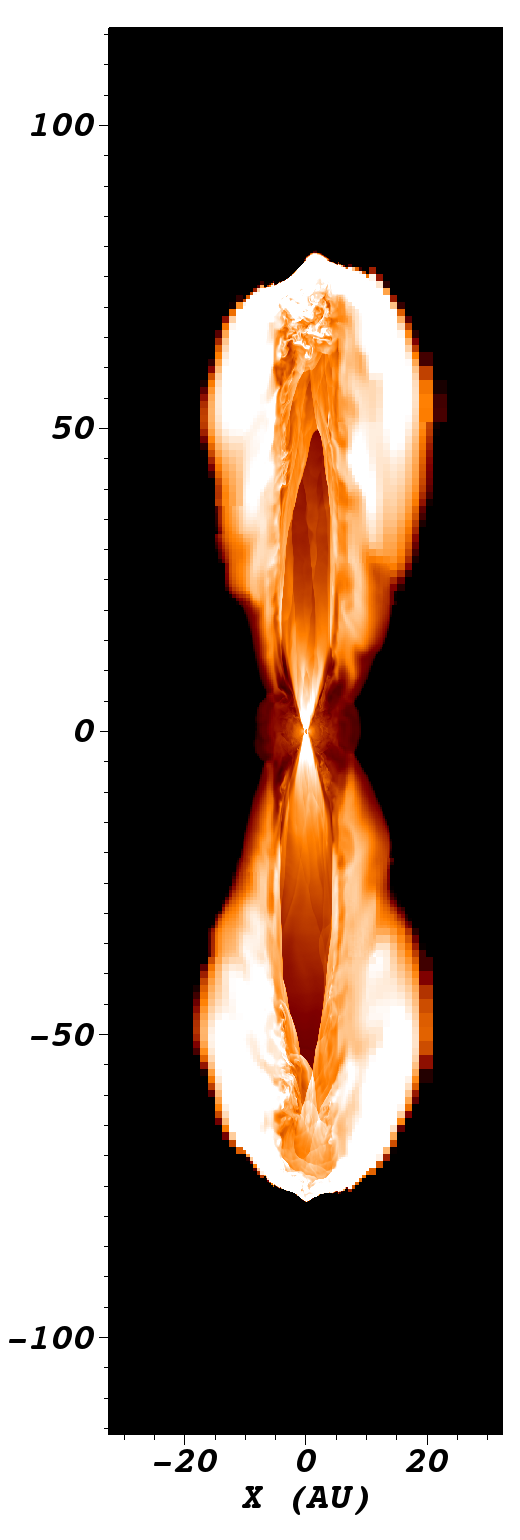}
    \includegraphics[height = 9 cm]{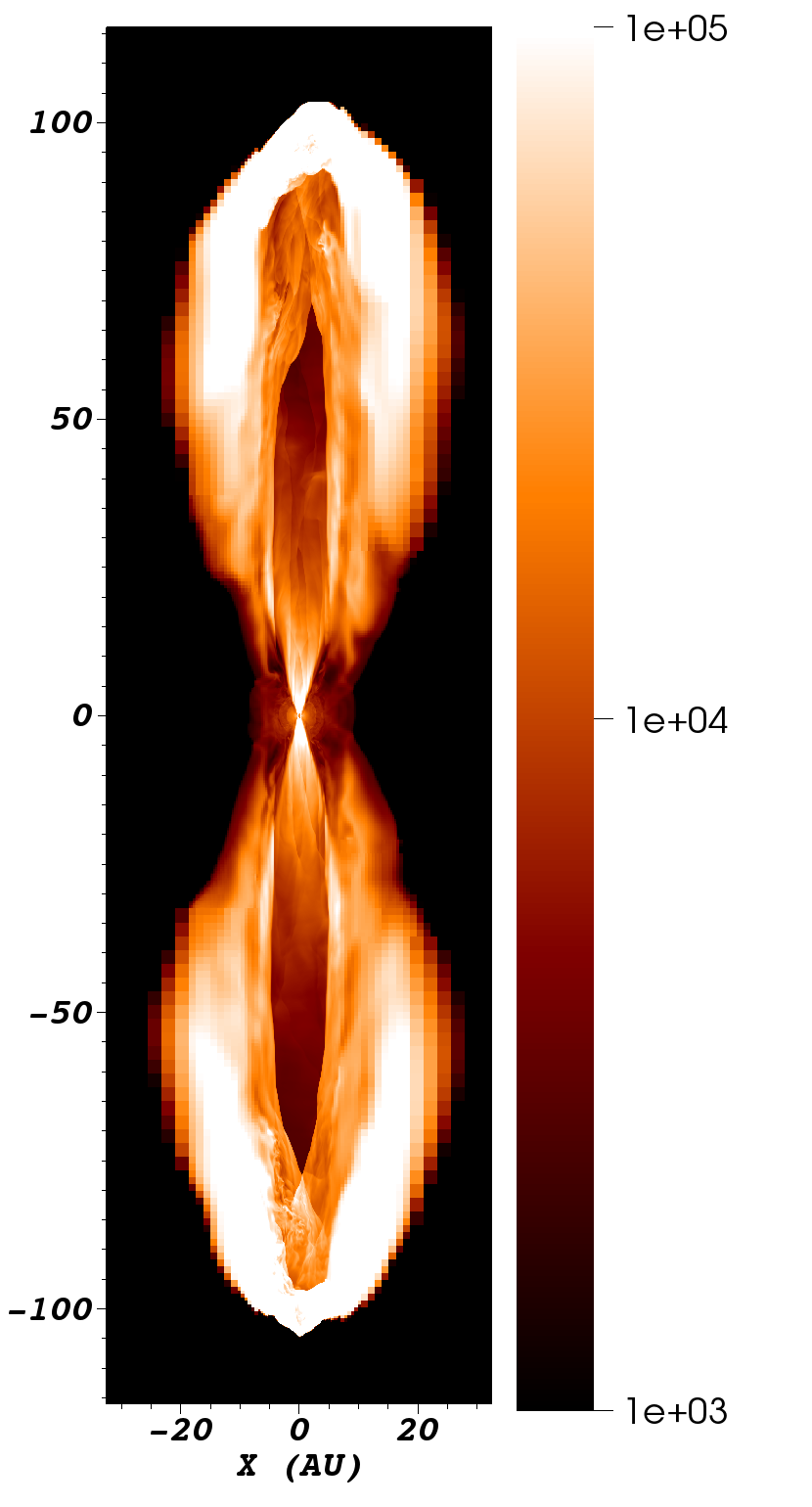}
    \includegraphics[height = \columnwidth]{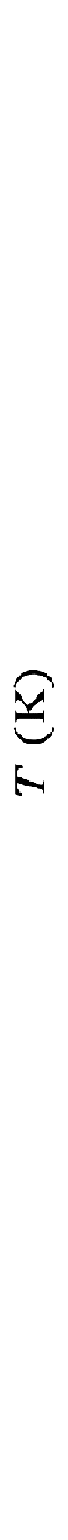}
    \caption{Model A: temperature (K, in log scale) plots of the high-momentum, no radiative cooling outflow taken at the same time as the above density plots. The hot ($\sim\mathrm{10^5~K}$) region defining the outer rim of the bipolar lobes is where the outer shock drives into the CE ejecta. A pair of low temperature, lens-shaped regions behind the elliptical shock are fast wind material that cools via geometric dilution. Note that near the base of the outflow, colder material from the CE ejecta is dragged into the lobes. This region helps define the funnel in the CE ejecta which is responsible for the overall collimation of the bipolar lobes. }
    \label{fig:A-Temp}
\end{figure*}

Model A has a high-density outflow of $6.42\times10^{-4} \msun \mathrm{yr}^{-1}$ and evolves adiabatically. Fig.~\ref{fig:A-rho} shows the density evolution and Fig.~\ref{fig:A-Temp} shows the temperature evolution over 920 days after the onset of the fast wind. All frames are sliced at the middle of the y-axis. The PN core (i.e. the binary) is at the center of the frames (coordinate $x,y,z=0$) and the binary orbit is in the x-y plane. 
White contours, representing temperature of $10\,000~\mathrm{K}$, are plotted over in the density plots as this is a good indication of the outline of the bipolar outflow. 

The first thing to notice is that although the fast wind is injected spherically into the simulation grids, two highly collimated lobes emerge shortly after we turn on the fast wind.
On the leading edge of the lobes is a shell of hot CE ejecta (AGB) material swept up by the outward going bow shock, represented by the light blue regions just behind the outer temperature contour in Fig.~\ref{fig:A-rho}. 
Interior to this shell is a tenuous hot bubble of gas at temperature around $10^5~\mathrm{K}$ and filled with shocked ejecta material. At early times, this hot bubble has almost uniform density around $10^{-13}~\mathrm{g\,cm^{-3}}$, and later grows into a  hollower cavity with density below  $10^{-15}~\mathrm{g\,cm^{-3}}$. 

We see the SFIC process produces collimated jets that run from the central PN nucleus up and down along the polar axis. Fig.~\ref{fig:A-Temp} shows abrupt temperature changes across a pair of lens-shaped inner shocks going backward into the fast wind. Kinetic energy is  converted into thermal energy in the shocked fast wind material, which produces turbulence at the far end of the jets. Interior to the lens-shaped region, the jet cools with distance from the PN nucleus due to geometrical dilution. 
We note that in later stages of the simulation, the bipolar lobes grow laterally extending beyond the refinement region defined by the prism described in Section \ref{sec:setup}. For example, in Fig.~\ref{fig:A-rho}, the sides of the bipolar lobes in light blue are in a region with 4 levels of AMR refinement. This lack of resolution might have suppressed turbulence (compare the density maps in Fig.~\ref{fig:mesh}). 

When we use a tracer (not shown in figures) to track the fast wind, we find that the material originally injected near the equatorial plane has been redirected by the funnel in the dense CE ejecta into the collimated jets. 
To explore the ability for the CE ejecta to redirect the wind we performed an additional test run.  In this case the wind was confined to the equatorial plane rather than being spherically symmetric.  Even in this case, using a maximally uncollimated wind we find the interaction with CE ejecta again drives bipolar lobes (see Appendix \ref{apx:xjet}).
 
The presence of asymmetries is an important result of our models, as PPN/PN observations often show significant differences between lobes in bipolar flows on small scales.  On larger scales we see the upper lobe has a more pointed tip than the lower lobe, while the lower lobe is slightly wider at its midpoint. On a smaller scales, the detailed structures of the filaments are visibly different between the lobes.  
Similar asymmetries also present in an orthogonal cut through the y-z plane. We quantify the asymmetries later in Section \ref{sec:asymmetry}.

Note that near the base of the bipolar lobes we see cold ejecta material being dragged into the lobes.  This region helps define the funnel in the CE ejecta which is responsible for the overall collimation of the bipolar lobes. The entrainment 
of collimating material into the lobes is an effect that was seen even in early SFIC studies \citep{Icke+1992a}.

\subsection{Model B: Adiabatic, low-momentum outflow}
\label{sec:B}
\begin{figure*}
    \includegraphics[height = 9 cm]{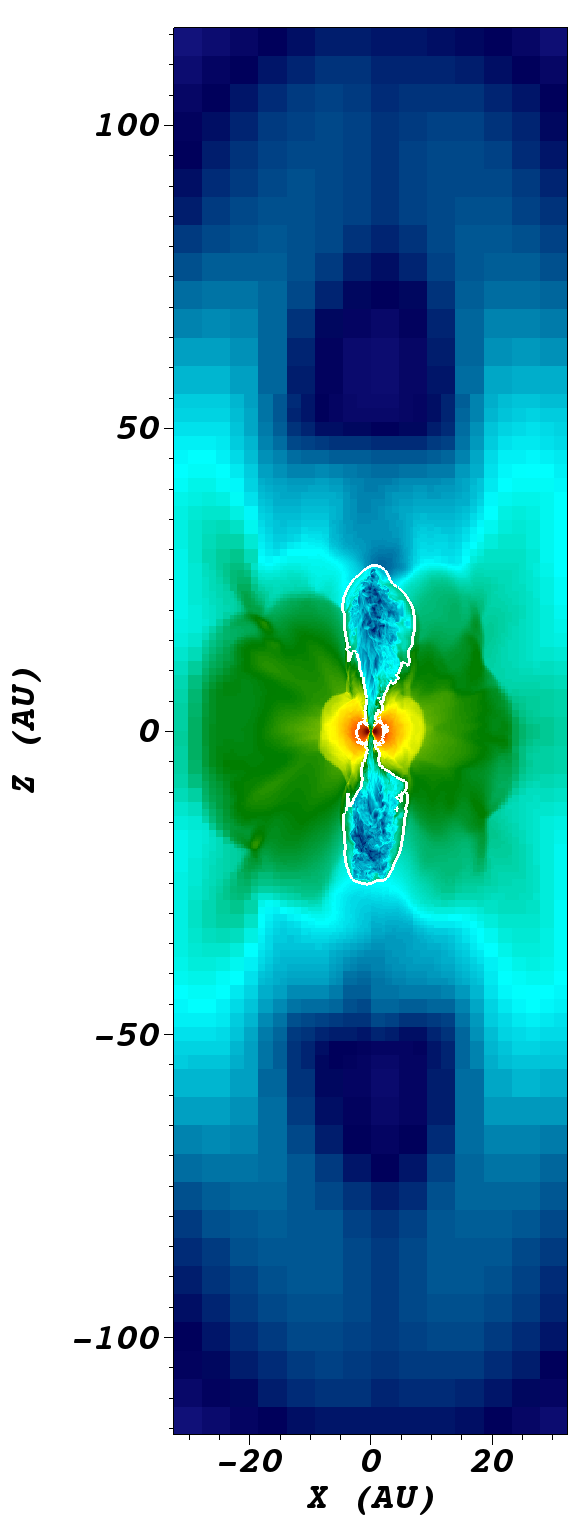}
    \includegraphics[height = 9 cm]{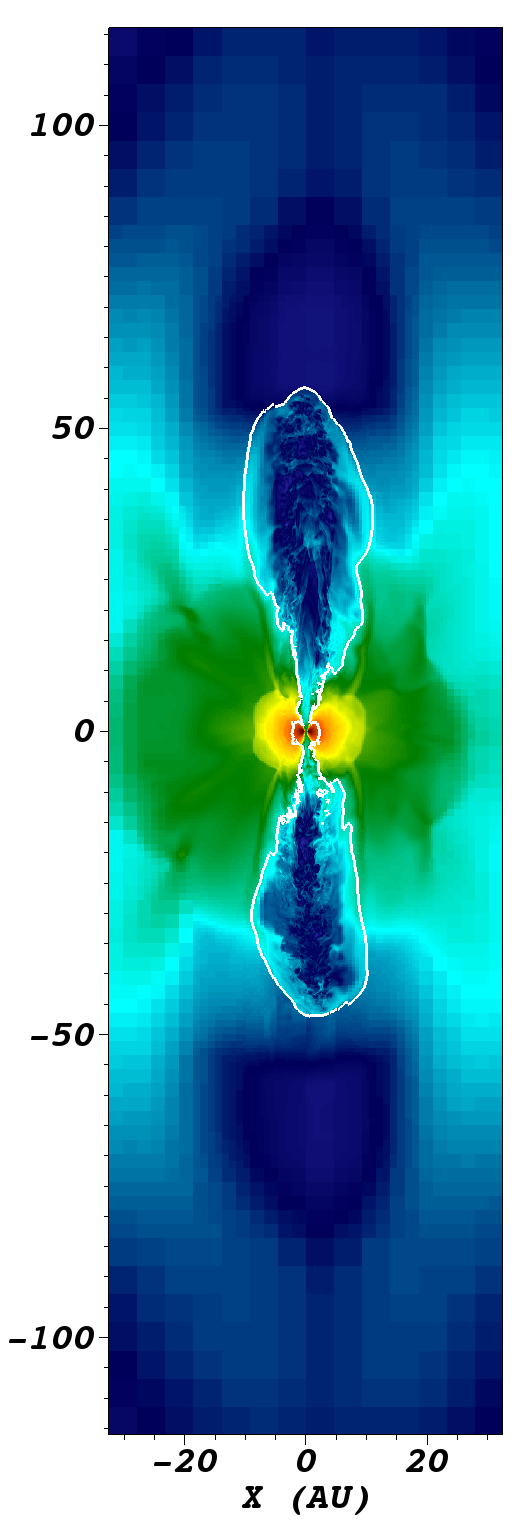}
    \includegraphics[height = 9 cm]{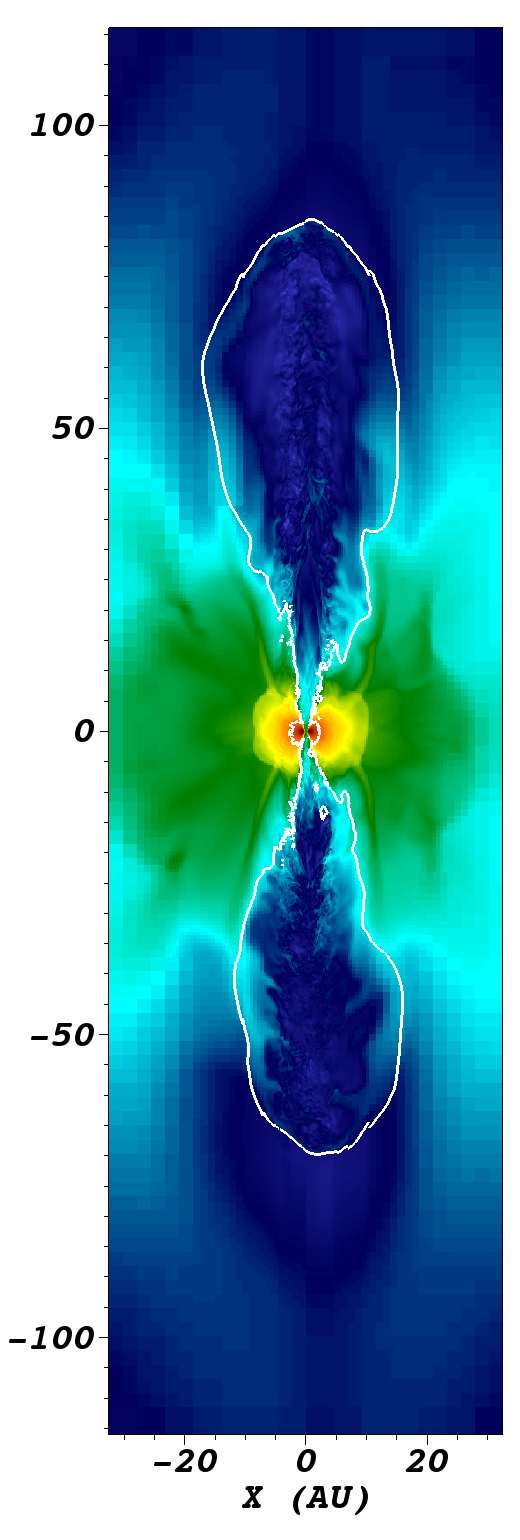}
    \includegraphics[height = 9 cm]{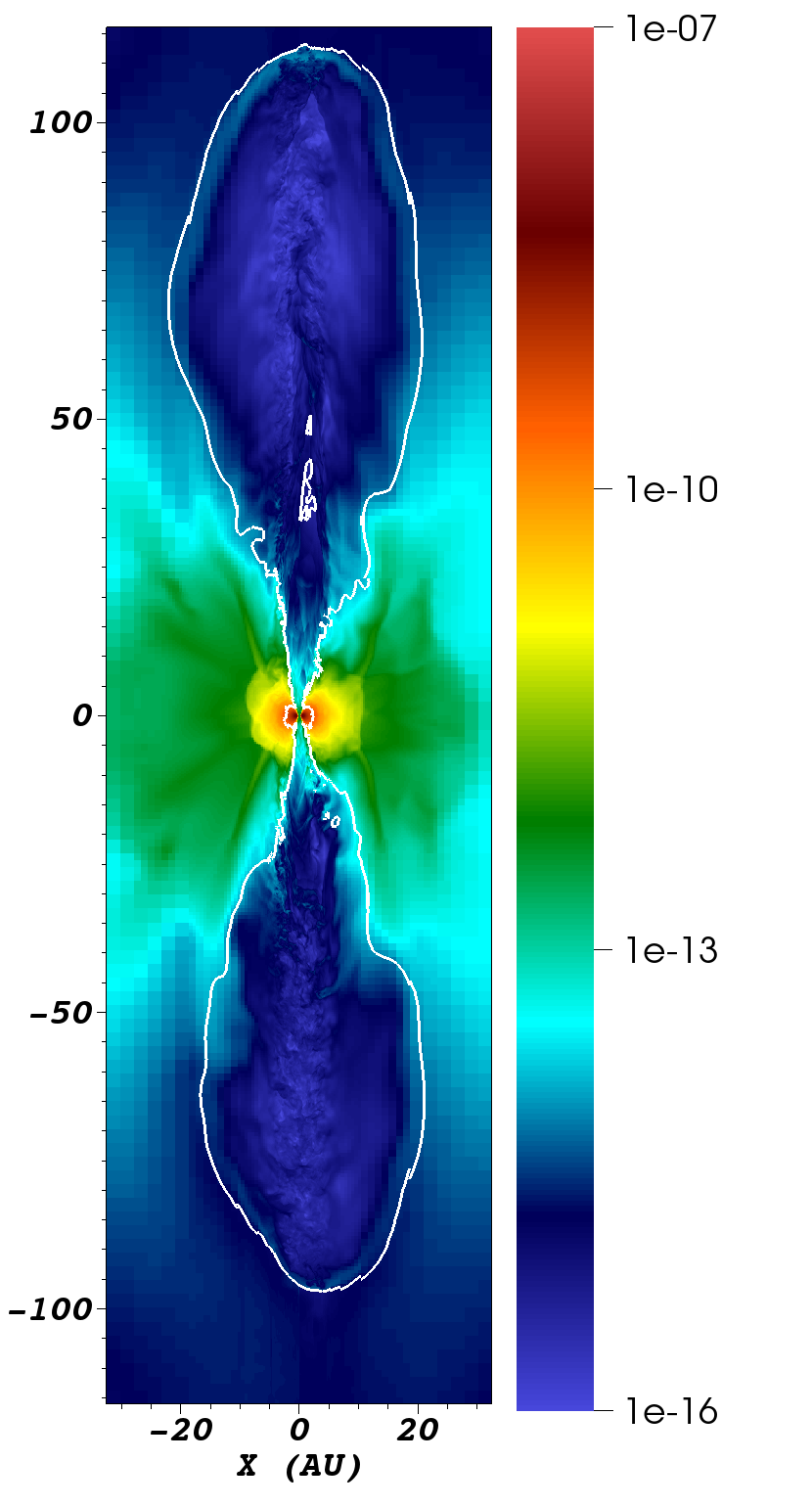}
    \includegraphics[height = 9 cm]{time-evolutions/rho-new.png}
    \caption{Model B: density ($\mathrm{g\,cm^{-3}}$, in log scale) plots of the low-momentum, no radiative cooling outflow taken at 3640, 4060, 4380, 4780 days (640, 1060, 1380, 1780 days after the fast wind turned on). White contours mark constant temperature of $10\,000~\mathrm{K}$. It takes much longer for the low-momentum outflow to evolve to similar physical scale compared to the high-momentum case. }
    \label{fig:B-rho}
\end{figure*}

\begin{figure*}
    \includegraphics[height = 9 cm]{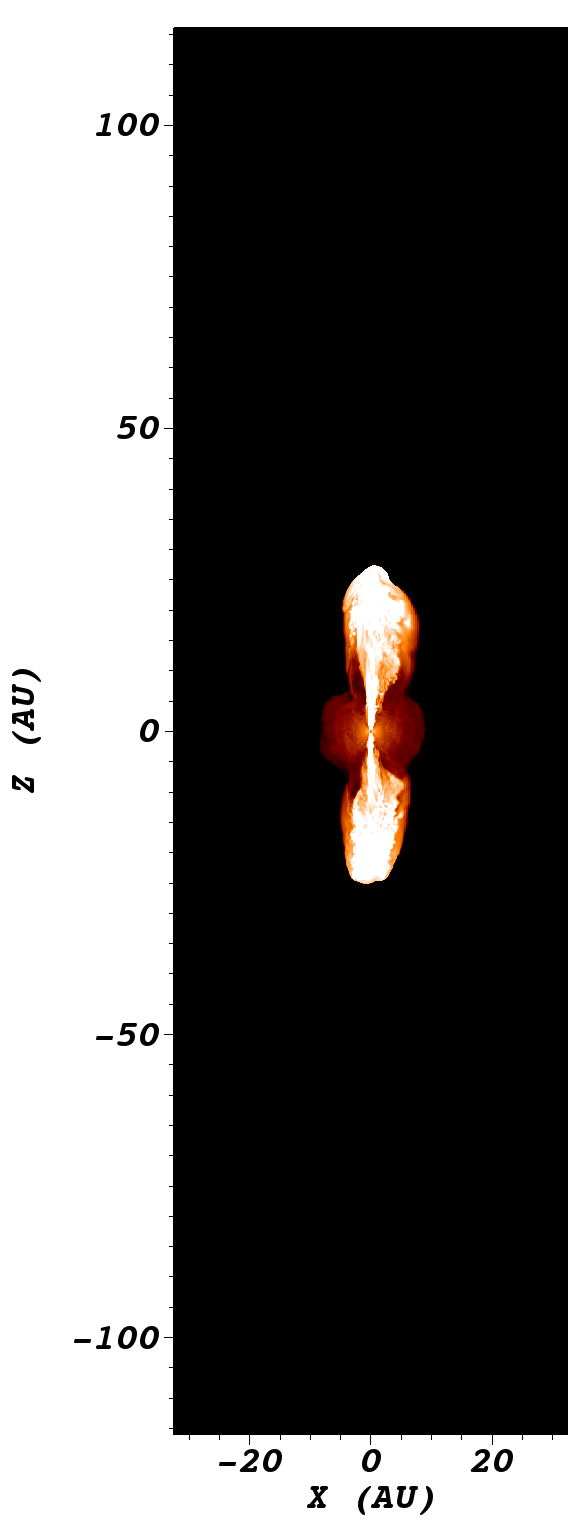}
    \includegraphics[height = 9 cm]{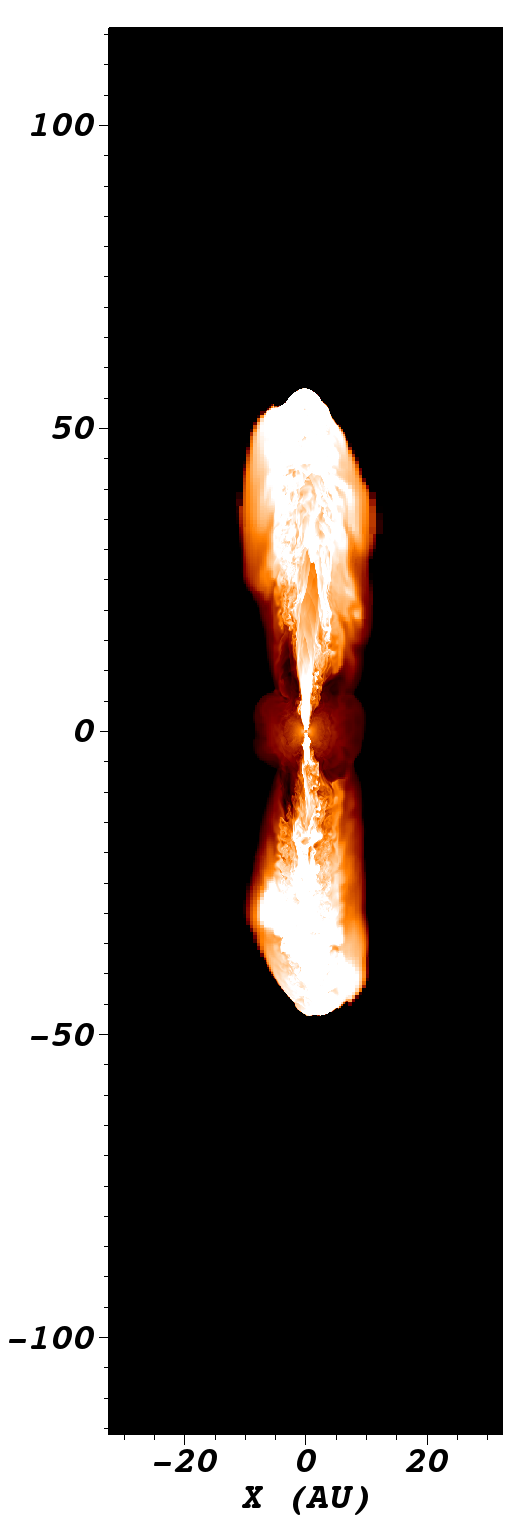}
    \includegraphics[height = 9 cm]{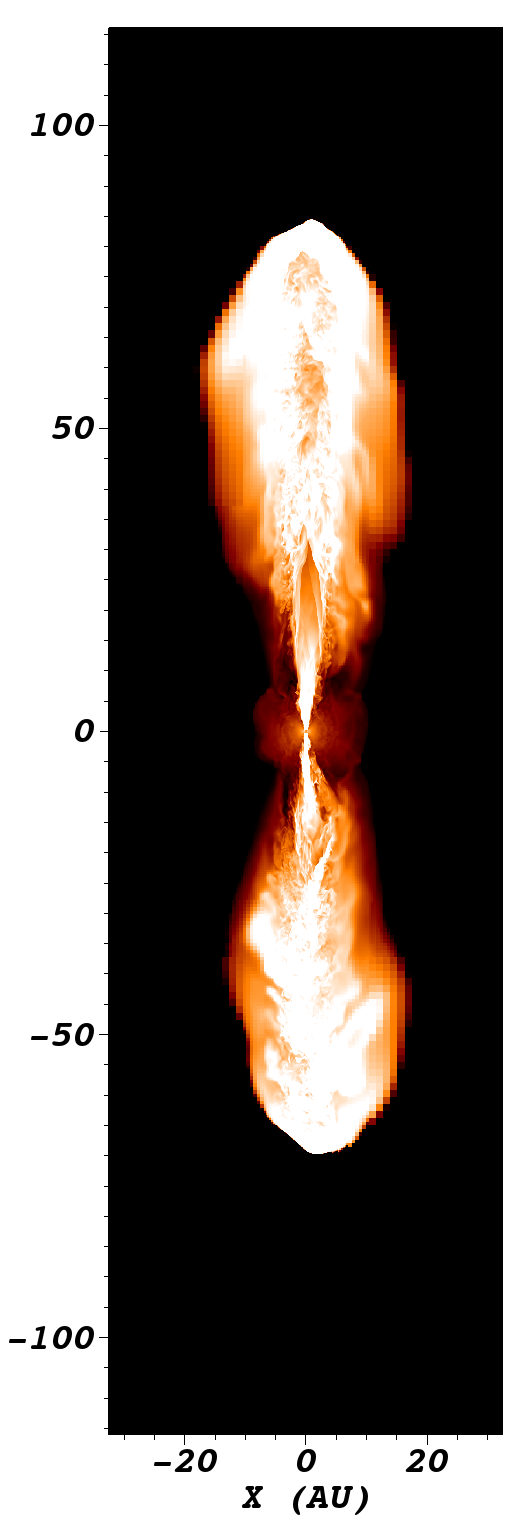}
    \includegraphics[height = 9 cm]{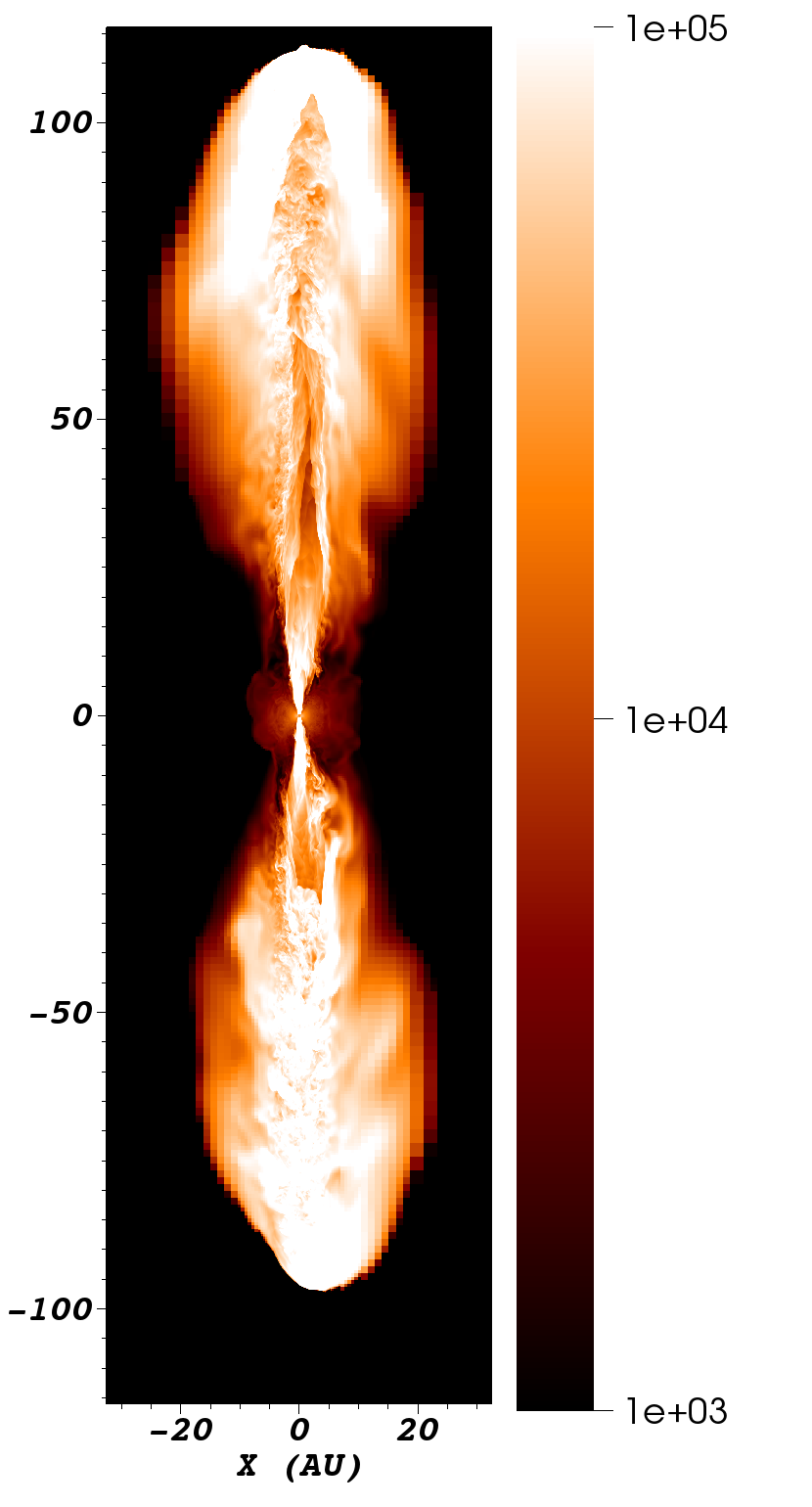}
    \includegraphics[height = 9 cm]{time-evolutions/T-new.png}
    \caption{Model B: temperature (K, in log scale) plots of the low-momentum, no radiative cooling outflow taken at the same time as Fig.~\ref{fig:B-rho}. Note that the inner shock in the top lobe has been pushed back closer to the center and has an irregular shape. }
    \label{fig:B-Temp}
\end{figure*}

Model B has an adiabatically evolving, low-density outflow with mass loss rate of $3.21\times 10^{-5} \msun \mathrm{yr}^{-1}$. Other parameters are set the same as in model A. Fig.~\ref{fig:B-rho} shows the density evolution and Fig.~\ref{fig:B-Temp} shows the temperature evolution on the same scale as for the previous section.

It takes longer for the lobes in model B to evolve to a similar size compared to model A, as would be expected given the lower value of the injected momentum. The first snapshot (leftmost) of model B is taken 380 days later than the first snapshot shown for model A, and the last snapshot (rightmost) is 860 days later than the last snapshot for model A. 

In terms of morphology, once again we see ambient material piles up just after the bow shock, but in thinner layers compared to model A.  The bow shock is aligned quite well to the $10\,000~\mathrm{K}$ temperature contour in Fig.~\ref{fig:B-rho}. 
Inside the bow shock, the lobes appear more homogeneous in this case, both in density and in temperature. Features that have been pointed out in Section \ref{sec:A}, such as the central pillars of jet, and the lens-shaped inner shocks, are still present but exhibit a different morphology.  The inner shock in the top lobe has been pushed back closer to the center and has an irregular shape like a candle flame, which can still be identified as the abrupt temperature change from less than $10^4$ K to about $10^5$ K in Fig.~\ref{fig:B-Temp}. This is to be expected given the lower momentum injection rate.  Note that the inner shock in the bottom lobe, however, can hardly be recognized in the very turbulent flow.
The jets are also narrower and quickly become more turbulent as they propagate outward.  

Large-scale asymmetry is more apparent in model B. The bottom lobe has a smaller volume compared to the top one. Starting from the second panel shown in Fig.~\ref{fig:B-rho} and \ref{fig:B-Temp}, the leading edge of the bottom lobe has turned slightly to the right of the polar axis (the side of positive x). Thus the funnel in the CE ejecta, which is the driver of the collimation, is not symmetric.  The low momentum runs, with their lower ram pressure, are more sensitive to the density distribution in the funnel as we will discuss in Section \ref{sec:asymmetry}. We also note that the enhanced asymmetry between the high and low momentum runs will be more exaggerated when we consider the impact of cooling.  

\subsection{Shocks in the GISW model}
\label{sec:hot-bubble}

\begin{figure*}
    \centering
    \includegraphics[height=8cm]{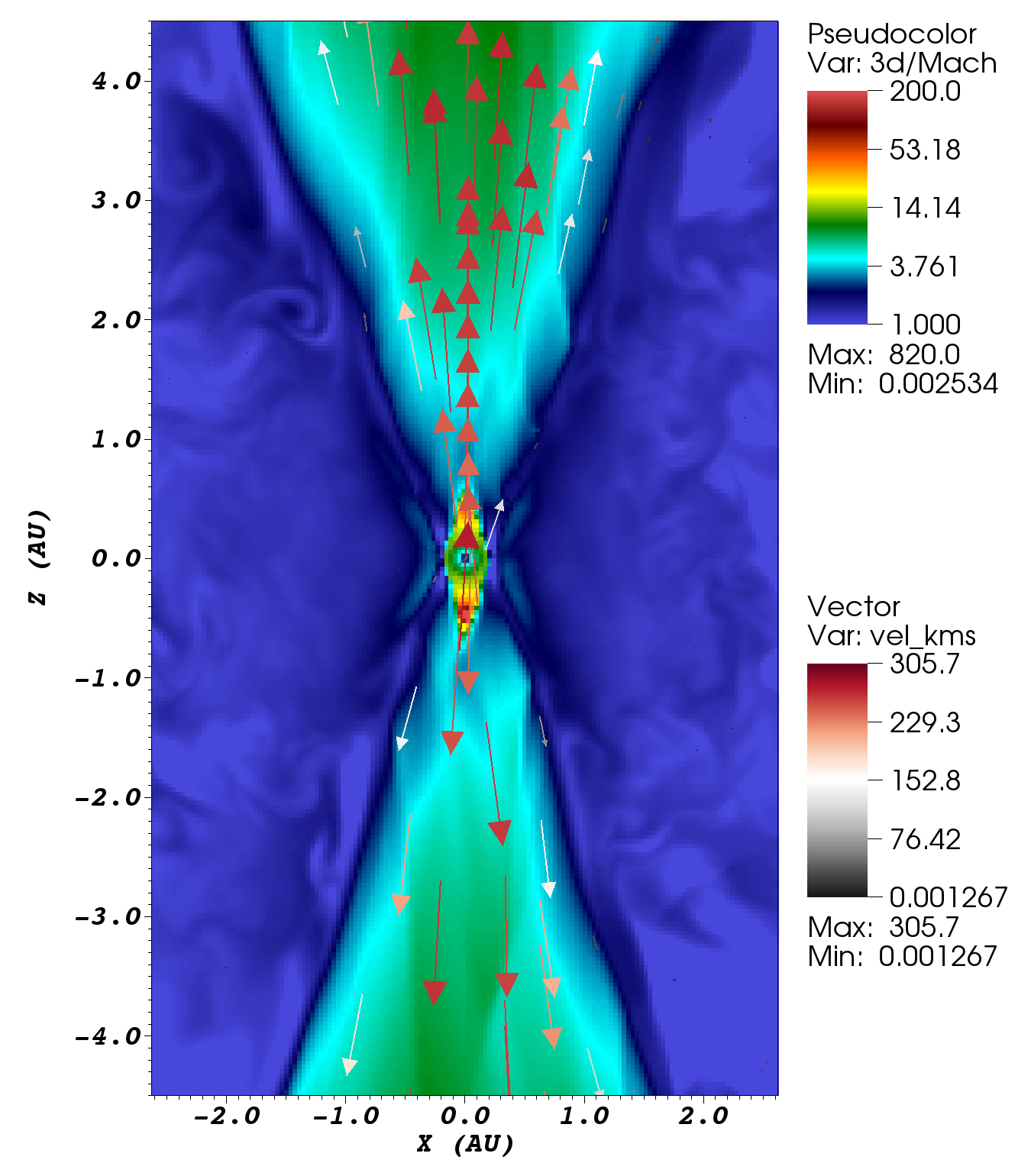}
    \includegraphics[height=8cm]{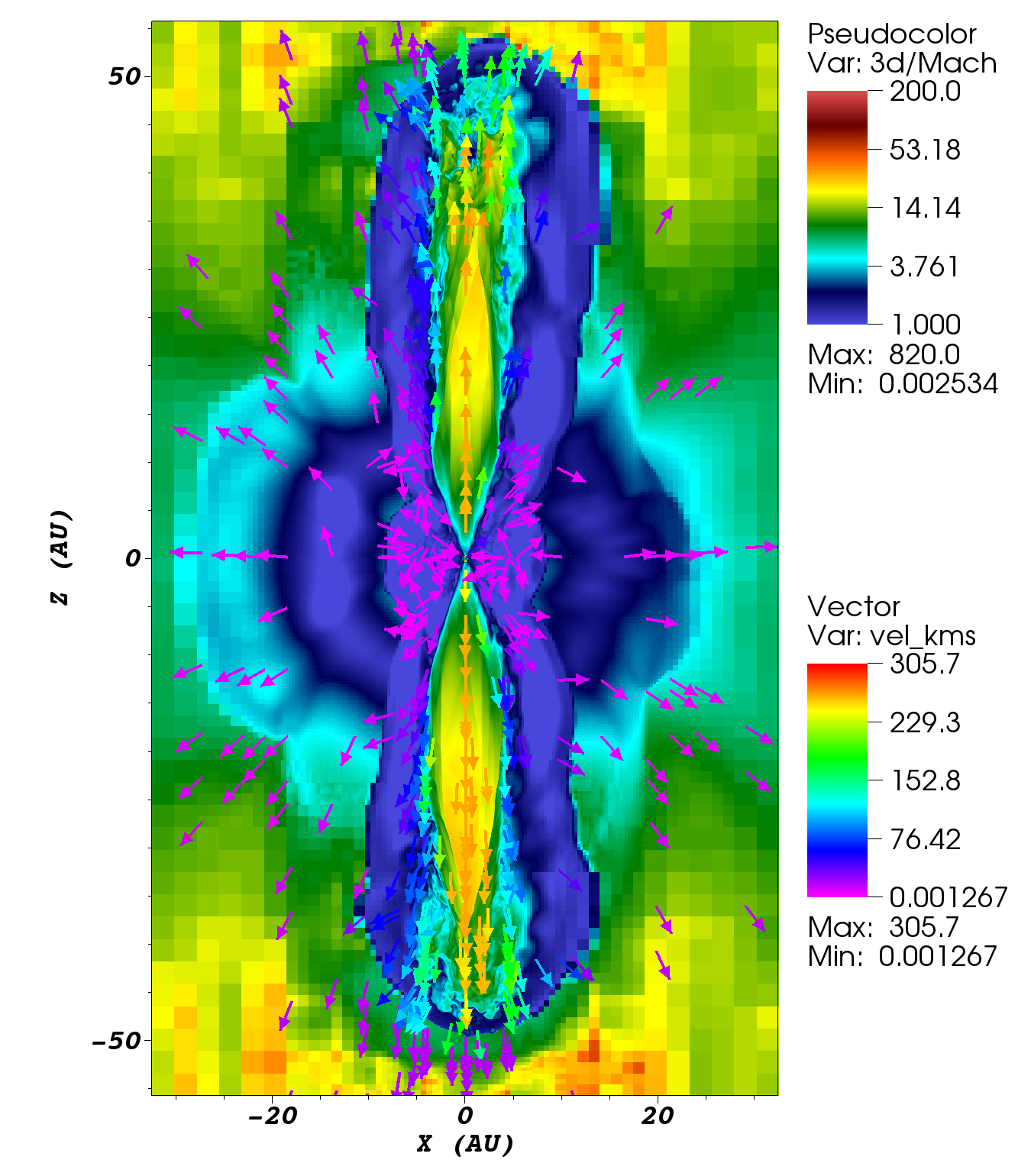}
    \caption{Pseudo-color Mach number in the central region of model A at 3480 days (same time as the second panel shown in Fig.~\ref{fig:A-rho}). Vectors are gas velocity in $\mathrm{km\,s^{-1}}$. Left panel zooms in around the central source to show the elliptical inner shock. Arrow length scales with speed magnitude. Right panel captures the entire PPN lobes and has a scale 10 times larger than the left. Arrows in the right panel have uniform length to show the slow expansion of the CE ejecta together with the high speed outflow.}
    \label{fig:shocks}
\end{figure*}

As discussed at the beginning of Section \ref{sec:results}, the classical GISW model involves the development of three discontinuities: an outer shock moving into, and accelerating, the ambient medium; an inner shock moving back towards the central engine and decelerating the fast wind; a contact discontinuity separating the two shocks. When the AGB/slow wind material is aspherical, these discontinuities all become aspherical as well.  Furthermore, shock focused inertial confinement (SFIC) occurs when the inner shock takes on an elliptical morphology which collimates the post-shock fast wind flow. While the basic physics of these shocks has been understood since the studies of Icke, Balick, Frank \& Mellema \citep{Icke+1992a, Icke+1992b, Mellema+Frank1995, Frank+Mellema1996}, the use of AMR methodologies in {\tt AstroBEAR} in the current work allows us to explore the evolution of the flows for much longer times as well as see the inner shock dynamics at much higher resolution.  As we shall see the high density contrasts of CE ejecta produce new behaviors not seen in previous studies.   

In Fig.~\ref{fig:shocks} we present a pseudo-color image Mach number with velocity vectors tracing the flow pattern. While the right panel shows the entire bipolar flow, the left panel zooms in around the central nucleus on a scale 10 times smaller. Both panels are taken from model A at 480 days after the start of the outflow (the second panel in Fig.~\ref{fig:A-rho} and \ref{fig:A-Temp}). 

We begin with the left panel which shows the flow close to the fast wind source. Note first the low Mach number CE ejecta (dark blue, $\mathcal{M}\sim 2$).  
The initial funnel in the CE ejecta has now become the base of the bipolar flow. Within the funnel and surrounding the central wind source we see the elliptical inner shock which goes backwards to the central source and shocks the fast wind material that passed through. This shock is quite small with a semi-minor axis length of just $\sim0.2~\mathrm{AU}$ and an eccentricity of about 0.95. 
The position of this elliptical shock front remains stationary during the entire  simulation. 
Within the elliptical shock we see highly supersonic, freely expanding fast wind material whose Mach number is increasing with distance from the central source, while the physical speed of the gas ($v_{\text{gas}}>250~\mathrm{km\,s^{-1}}$) stays close to the initial speed  of the injected spherical wind ($v_{\mathrm{FW}}=300~\mathrm{km\,s^{-1}}$).  Such an increase in Mach number is to be expected as the fast wind temperature decreases with radius as $T \propto r^{2(1-\gamma)}$ ($T\sim r^{-4/3}$ for monatomic ideal gas) due to geometrical dilution. 

Immediately beyond the elliptical shock, the Mach number in the post-shock fast wind material drops by a factor of 10, as the light blue region around the yellow/red ellipse.  Note that while the Mach number drops as would be expected at a shock, it remains above unity indicating the flow is not subsonic ($\mathcal{M}\sim 2$). Thus, as expected from SFIC theory, the oblique inner shock leads to the production of a flow which is both supersonic and collimated.  

The relatively lower Mach number of the immediate post-shock fast wind also means it has a relatively high thermal energy content.  Thus as this collimated flow expands along the z direction, its thermal energy will lead to a lateral expansion of the flow.  This geometric expansion will, once again, lead to a gradual increase in the Mach number.  Thus, after the initial collimation at the small scale inner shock, we see the development of a large-scale diverging supersonic flow which fills the lobes.  It is the interaction of flow with both the CE ejecta and down-stream fast wind material that creates the lens-shaped shocks discussed in Sections \ref{sec:A} and \ref{sec:B}.  
\textit{The development of the lens-shaped shocks is a new feature of SFIC driven bipolar lobes and could only have been seen by tracking the flows with high resolution (AMR).}  In addition it can only form in a high density contrast environment such as that created by CE ejecta.

We now focus on the right panel of Fig.~\ref{fig:shocks} which shows the entire bipolar flow. Once again the spherical dark blue region is the slowly expanding, low Mach number CE ejecta. The outer parts of the lobes are also in dark blue, and a steep jump in Mach number from around 2 to 10 (dark blue to dark green) demarcates the bow shock separating the undisturbed CE ejecta and material which has been accelerated by the outer shock. The yellow-green pillars running up the center of the lobes represent fast wind collimated by the small-scale inner shock whose Mach number has increased with the flow's expansion. This material has high velocity ($\sim260~\mathrm{km\,s^{-1}}$) and relatively high Mach numbers ($\mathcal{M}\ge 15$).

To summarize we see three layers of shocks in the bipolar lobes. From the outer boundary and moving inward, these are: the bow shock or outer/forward shock, the lens-shaped inner/backward shock, and the small scale elliptical shock around the central region, which is also a backward shock.

\subsection{Model C: cooling, high-momentum outflow}
\label{sec:C}

\begin{figure*}
    \includegraphics[height = 9 cm]{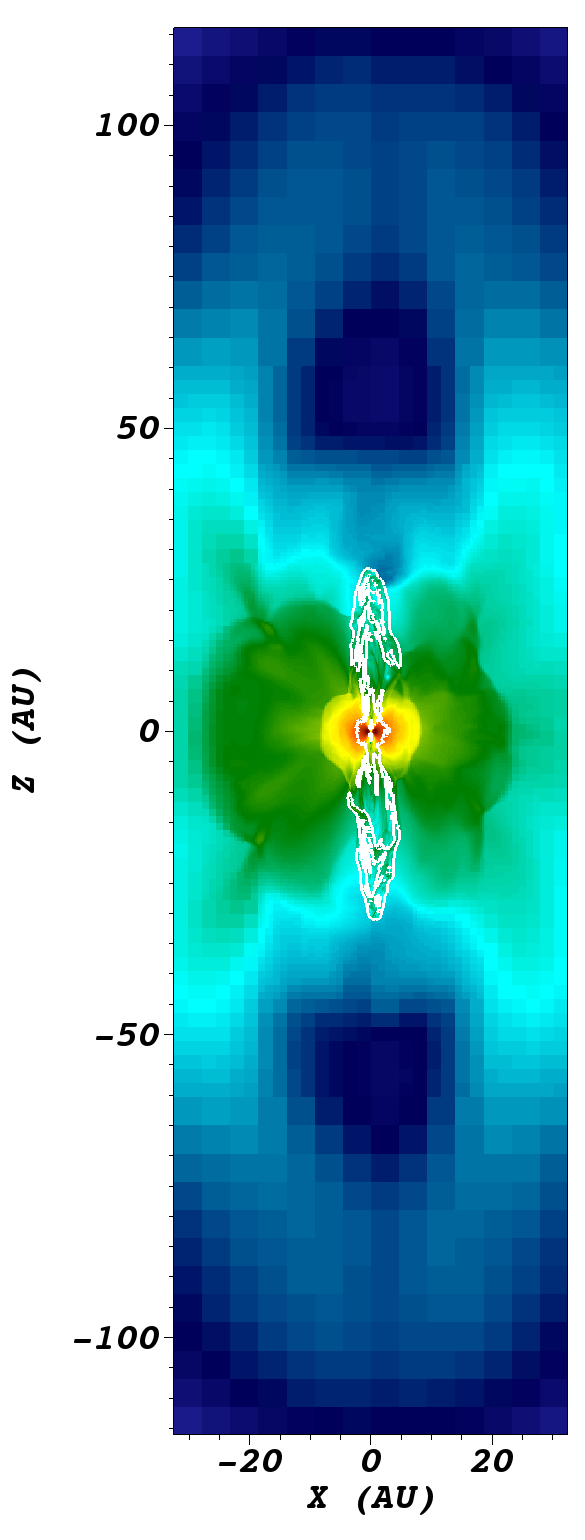}
    \includegraphics[height = 9 cm]{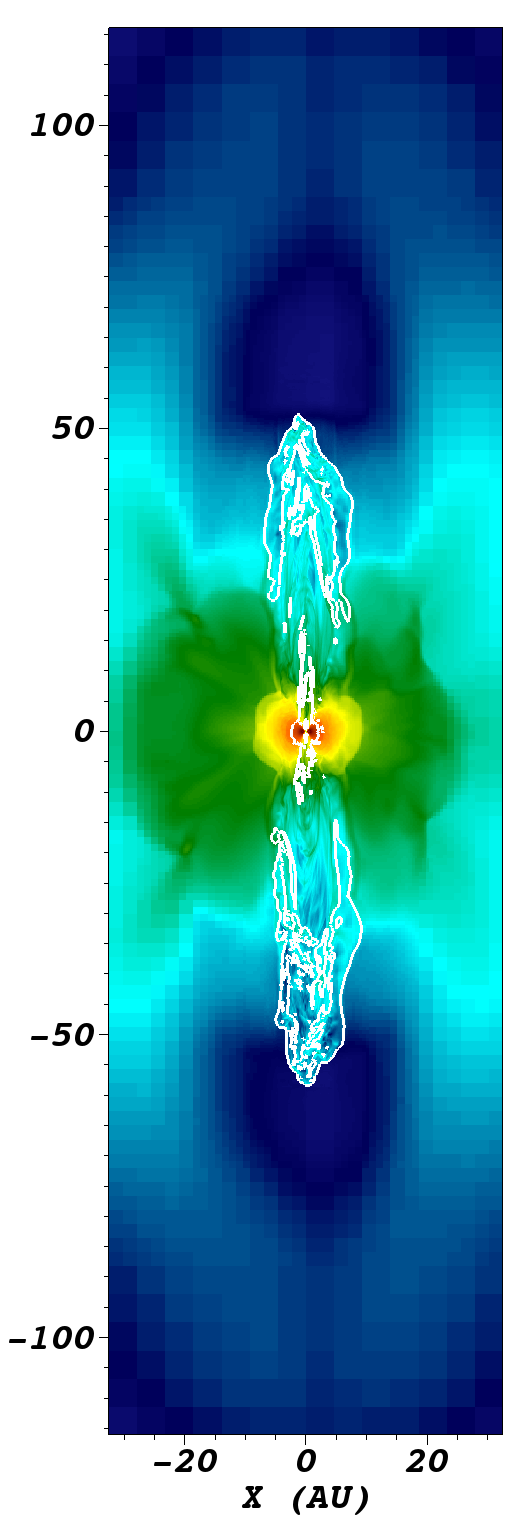}
    \includegraphics[height = 9 cm]{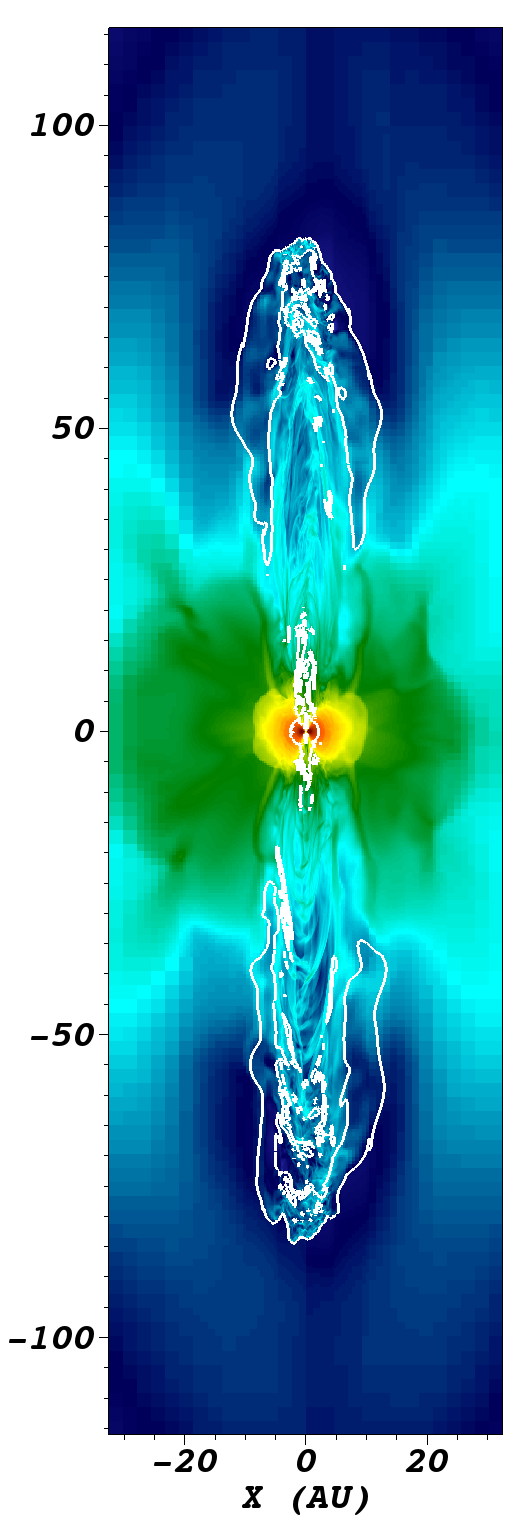}
    \includegraphics[height = 9 cm]{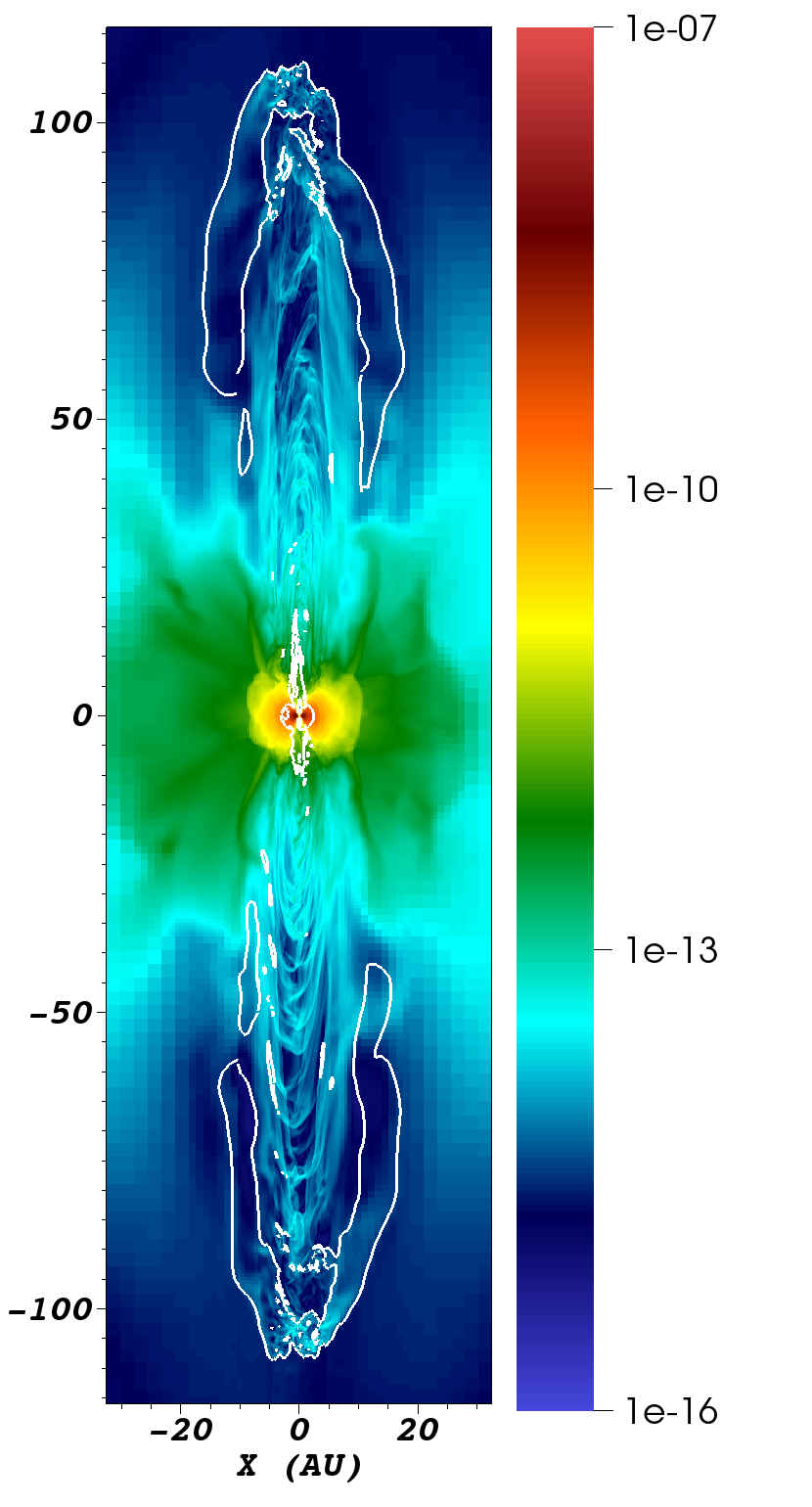}
    \includegraphics[height = 9 cm]{time-evolutions/rho-new.png}
    \caption{Model C: density ($\mathrm{g\,cm^{-3}}$, in log scale) evolution of the high-momentum outflow with radiative cooling. Panels, from left to right, are taken at 3460, 3840, 4200, 4600 days (460, 840, 1200, 1600 days after the fast wind turned on). White contours mark constant temperature of $10\,000~\mathrm{K}$. Note the highly turbulent flow in the central jets.}
    \label{fig:C-rho}
\end{figure*}

\begin{figure*}
    \includegraphics[height = 9 cm]{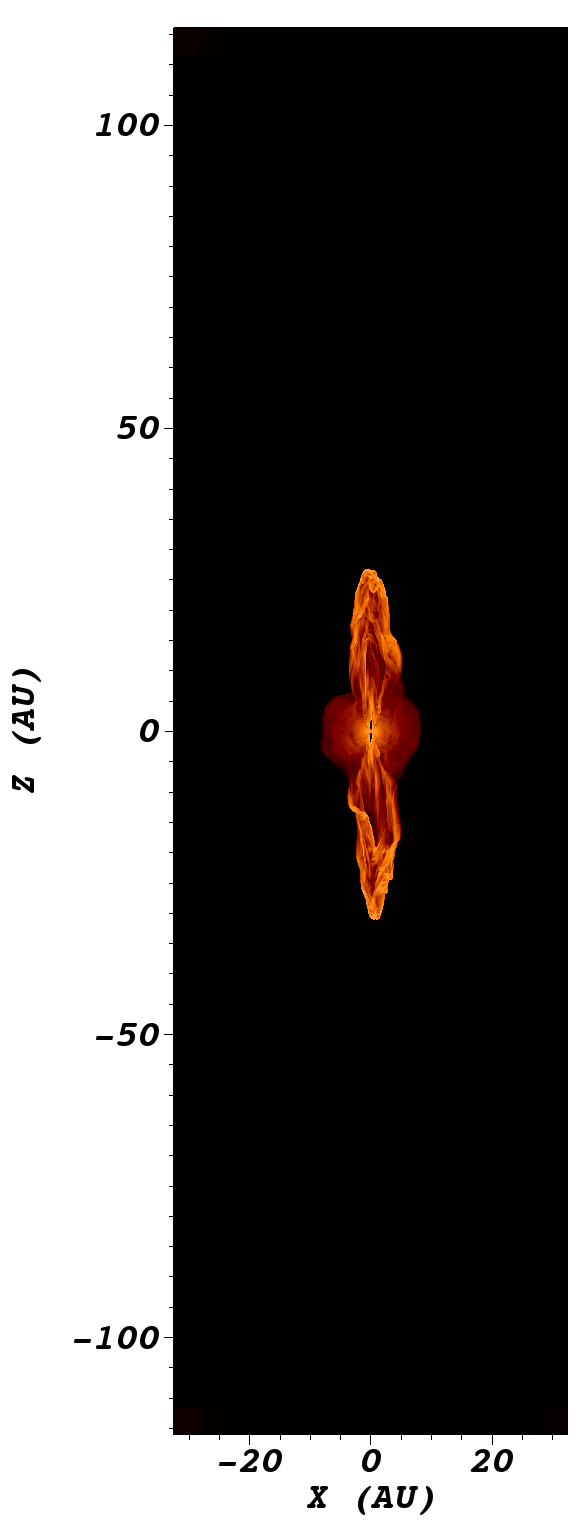}
    \includegraphics[height = 9 cm]{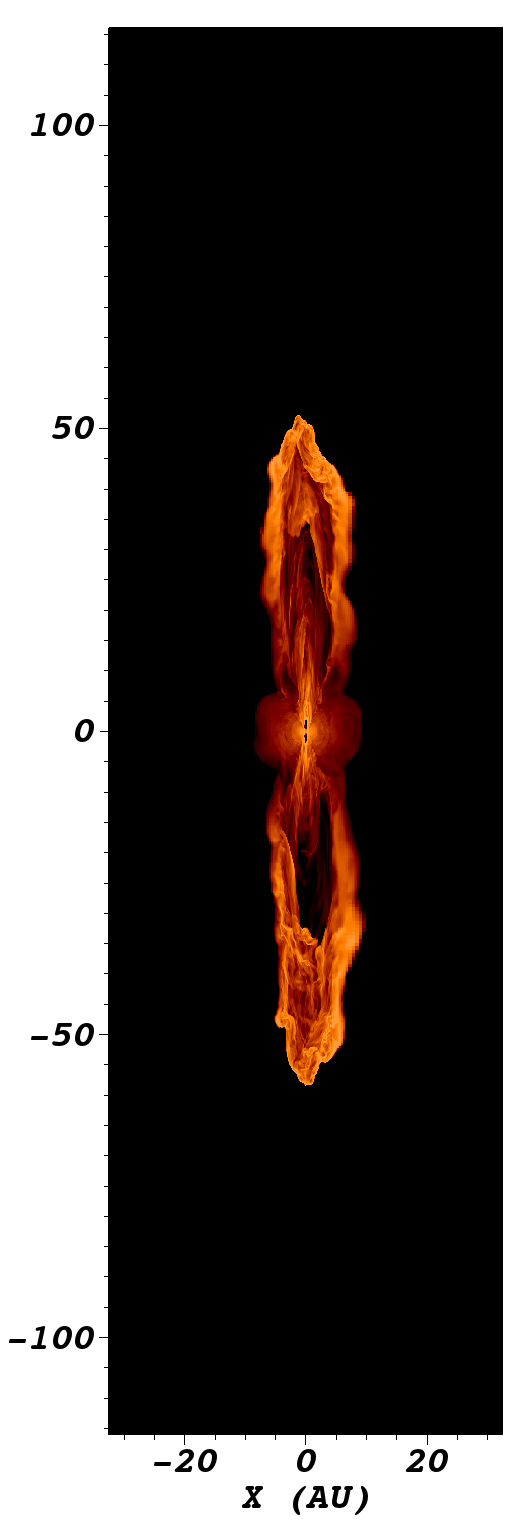}
    \includegraphics[height = 9 cm]{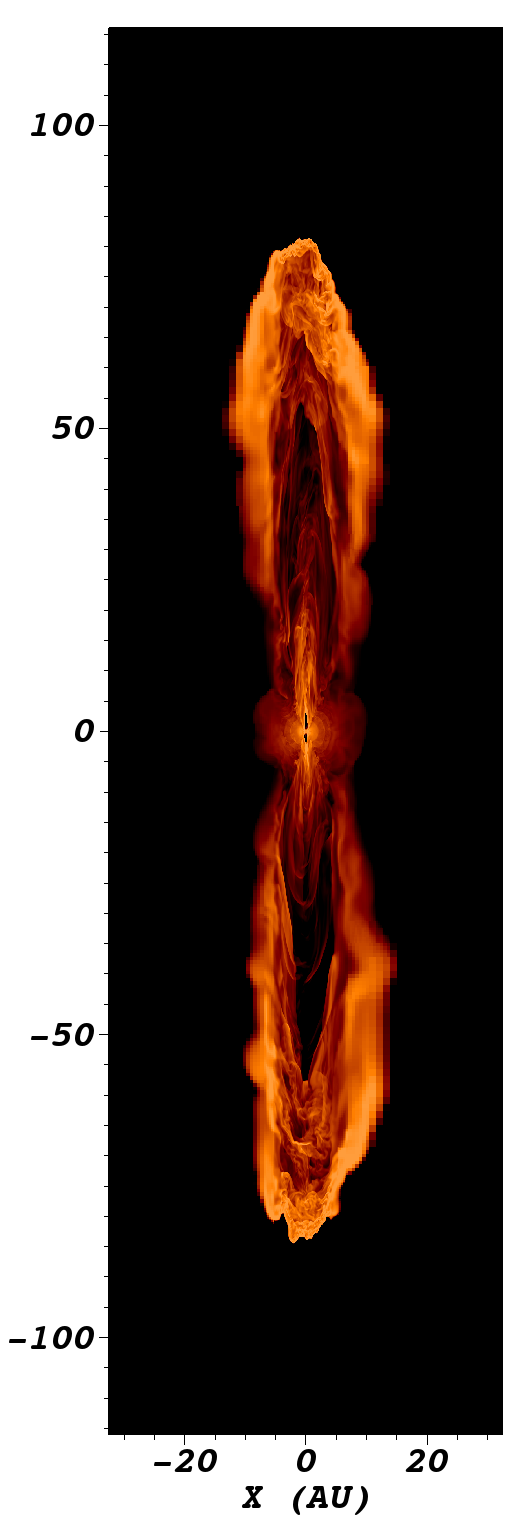}
    \includegraphics[height = 9 cm]{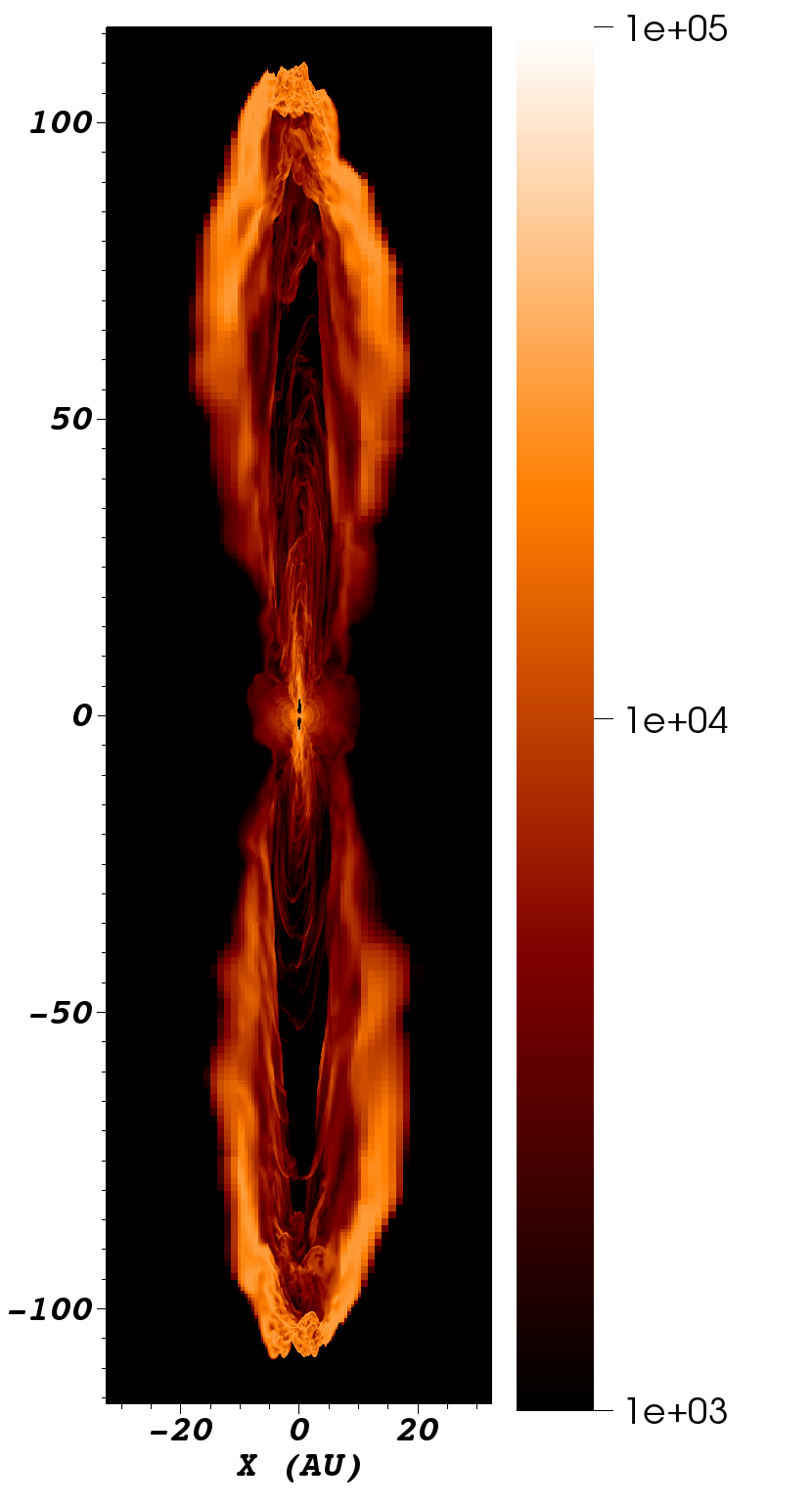}
    \includegraphics[height = 9 cm]{time-evolutions/T-new.png}
    \caption{Model C: temperature (K, in log scale) evolution of the high-momentum outflow with radiative cooling, taken at the same 
    times as Fig.~\ref{fig:C-rho}. Central jets have quickly been cooled to $\sim10^3$ K, and the inner shock is no longer distinguishable in this case.}
    \label{fig:C-Temp}
\end{figure*}

Energy loss or "radiative cooling" in GISW models leads to the expectation that thermal energy gained behind shocks will be lost, leading to decreased pressure support for these regions of the flow.  Thus rather than extended post-shock shells we expect to see thin, cold dense regions immediately following the shocks. Such thin cold "slabs" are prone to a variety of instabilities including thin-shell modes \citep{Vishniac+1989}.

To investigate the role of radiative cooling in our simulations in model C we apply the Dalgarno-McCray cooling curve in all grids 
with $ T> 3\times10^4~\mathrm{K}$ to the condition of model A. Fig.~\ref{fig:C-rho} and \ref{fig:C-Temp} are the density and temperature evolving 1600 days after starting the outflow. While we do not apply the cooling curve below the floor temperature of 30,000 K, colder material can still cool down adiabatically by geometrical expansion. 

Inspection of Fig.~\ref{fig:C-rho} and \ref{fig:C-Temp} demonstrates that the cooling does create thinner, colder and denser post shock regions. Compared with Model A we see the bipolar lobes have a reduced width which occurs as lateral expansion is inhibited with the loss of thermal energy to optically thin radiation. The bipolar shell is also clearly fragmented as is the interior pillar of the jet formed from the elliptical inner shock.  It is of particular interest to note that the degree of asymmetry between the two lobes is greater with cooling (Model C) than without (Model A).  This is likely due to the fact that the cooling triggers instabilities which are sensitive to small-scale details of the interaction of fast wind with the CE ejecta, the latter of which is inherently 3-D and asymmetric, a point we return to in the next section. 

\subsection{Model D: cooling, low-momentum outflow}
\label{sec:D}
\begin{figure*}
    \includegraphics[height = 9 cm]{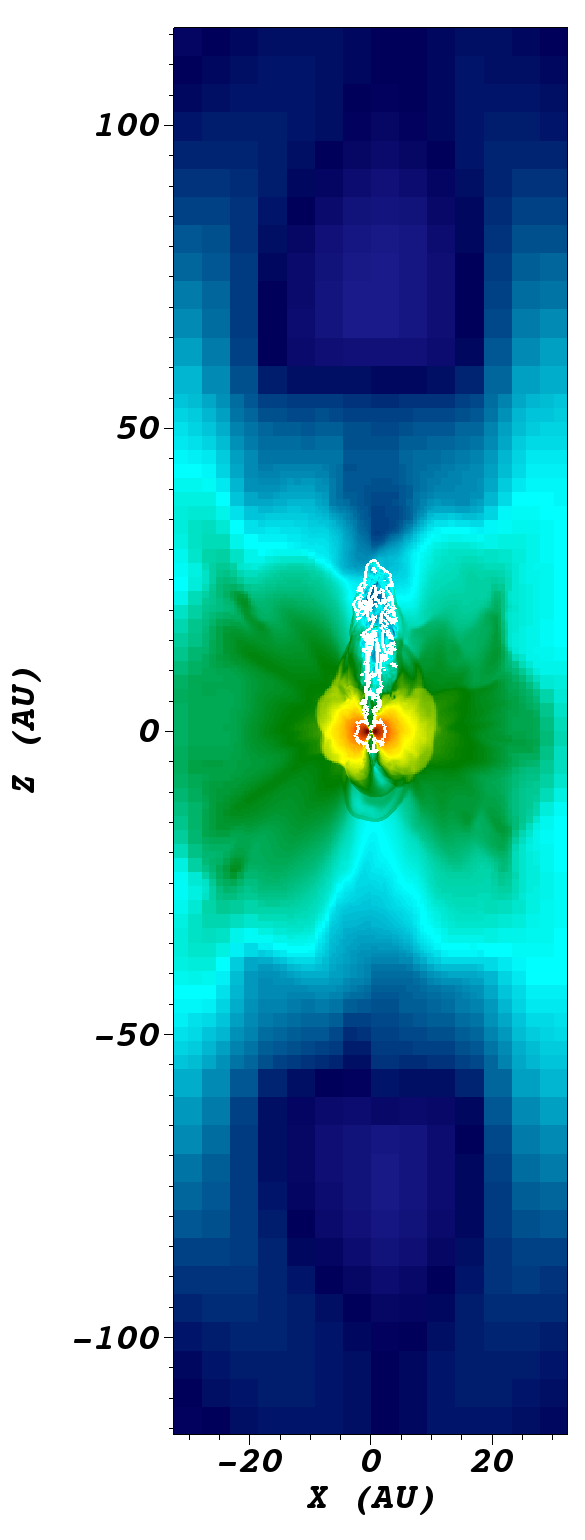}
    \includegraphics[height = 9 cm]{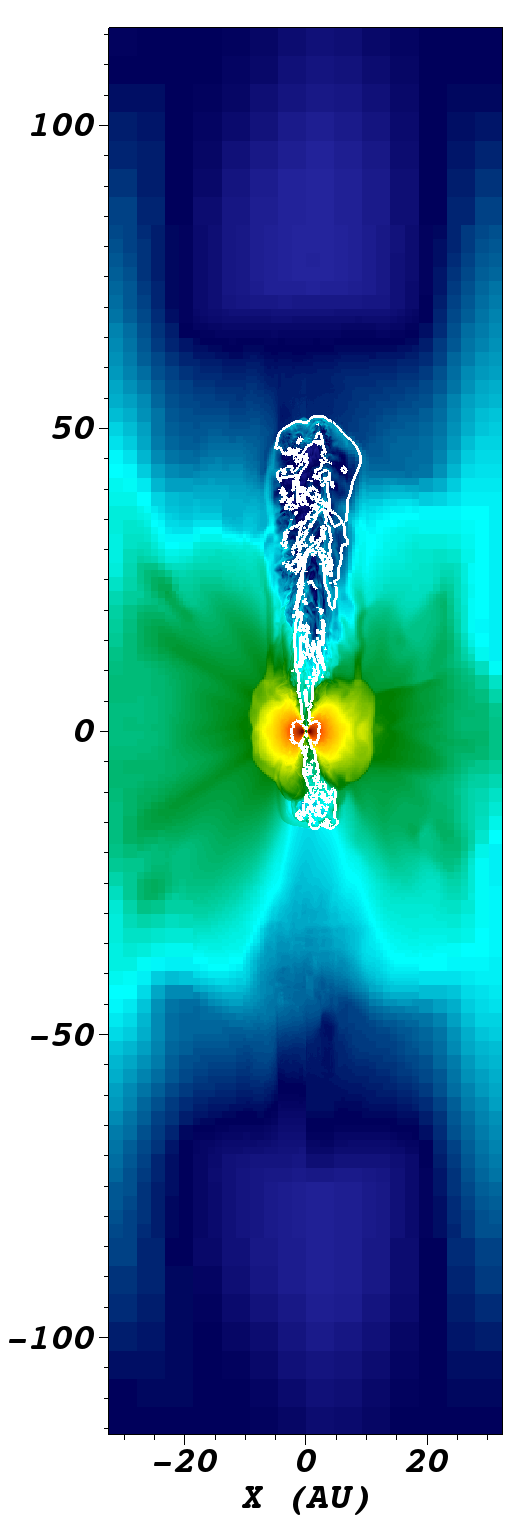}
    \includegraphics[height = 9 cm]{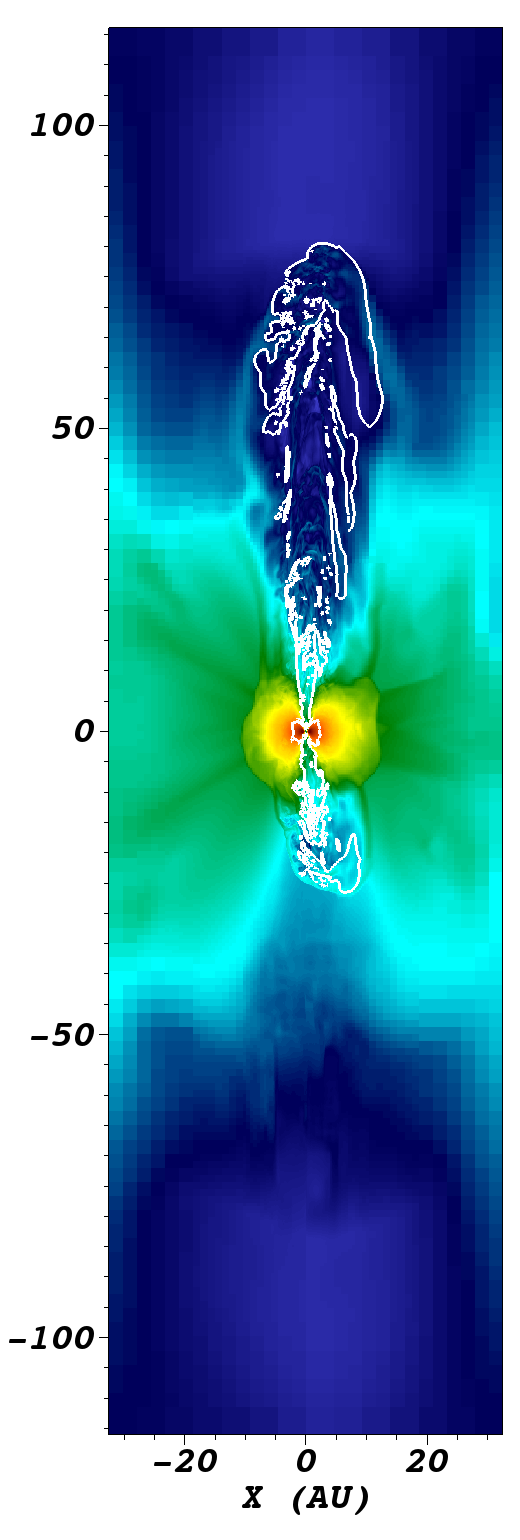}
    \includegraphics[height = 9 cm]{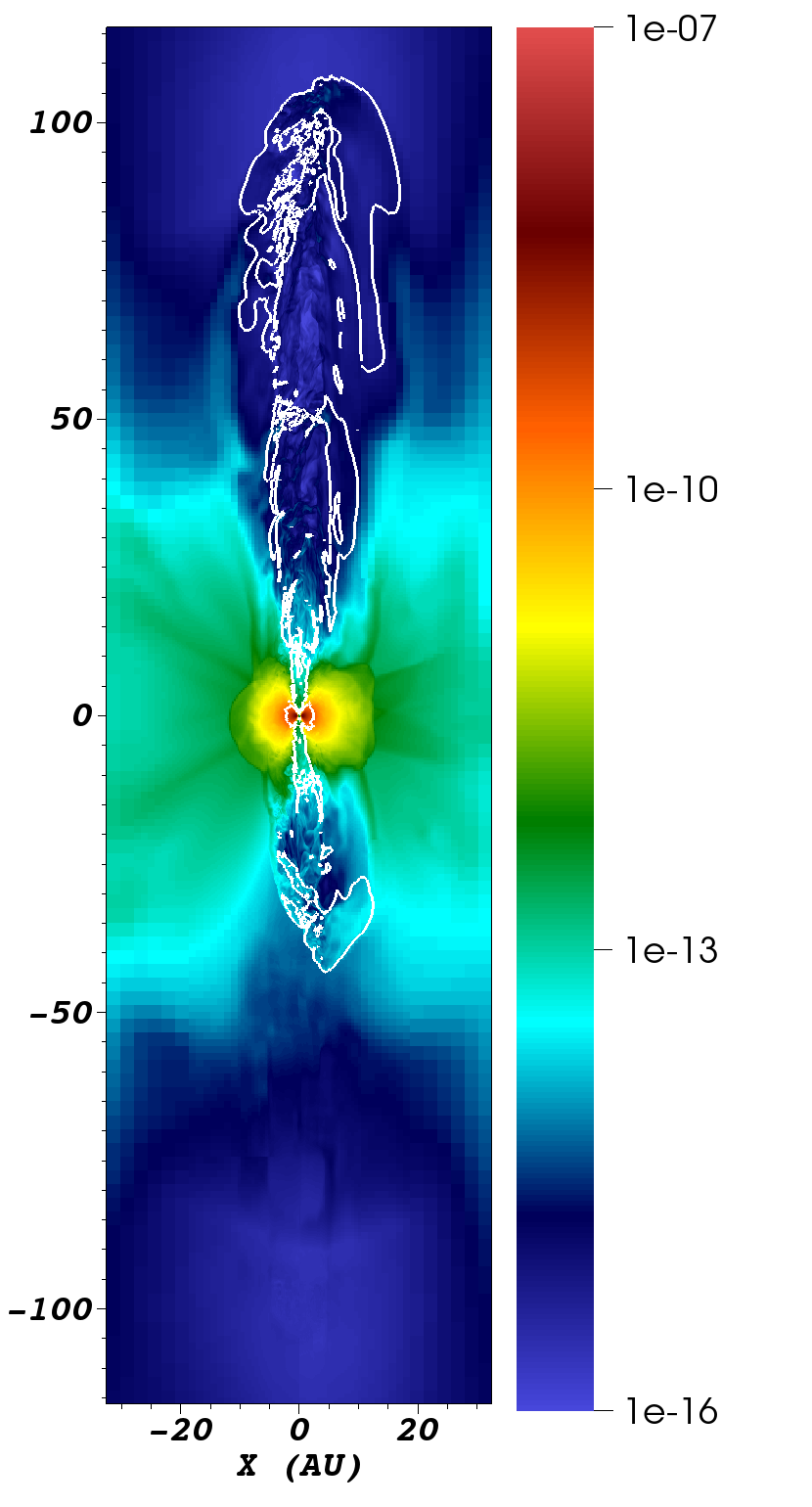}
    \includegraphics[height = 9 cm]{time-evolutions/rho-new.png}
    \caption{Model D: density ($\mathrm{g\,cm^{-3}}$, in log scale) evolution of the low-momentum outflow with radiative cooling. Panels, from left to right, are taken at 4760, 5580, 6400, 7000 days (1760, 2580, 3400, 4000 days after the fast wind turned on). White contours mark constant temperature of $10\,000~\mathrm{K}$. The bottom lobe has hardly made way out of the CE ejecta for over 2000 days after injecting the fast wind.}
    \label{fig:D-rho}
\end{figure*}

\begin{figure*}
    \includegraphics[height = 9 cm]{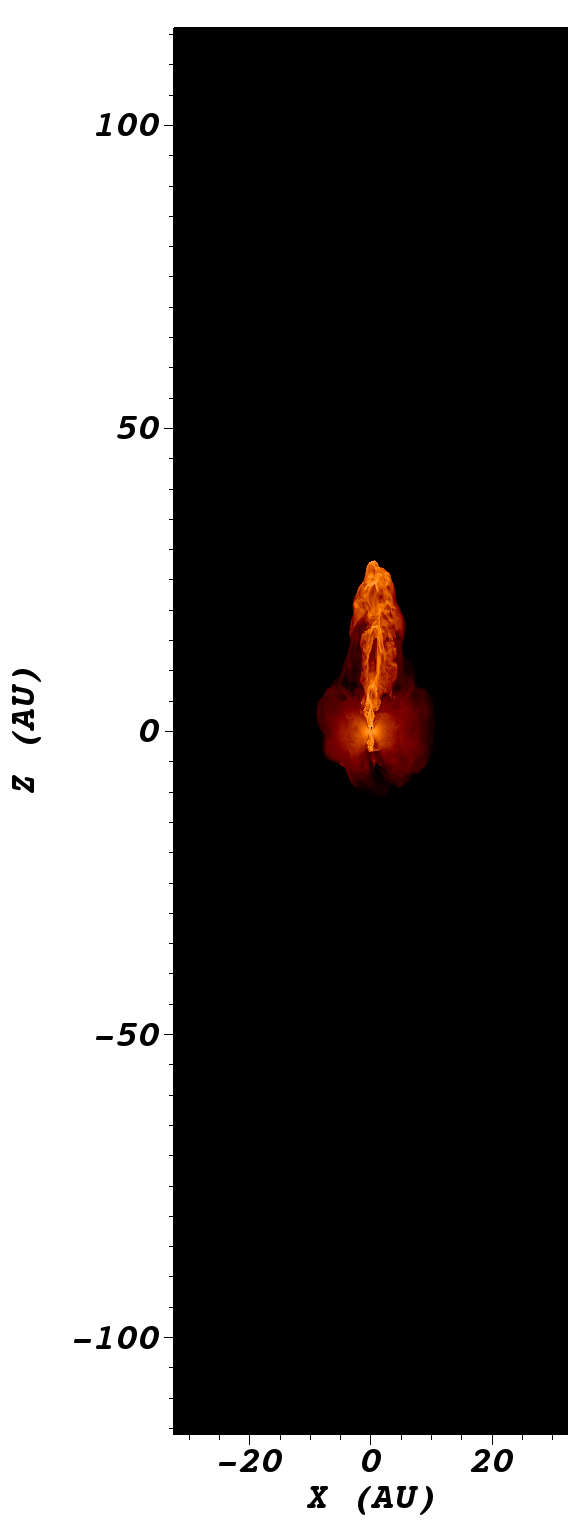}
    \includegraphics[height = 9 cm]{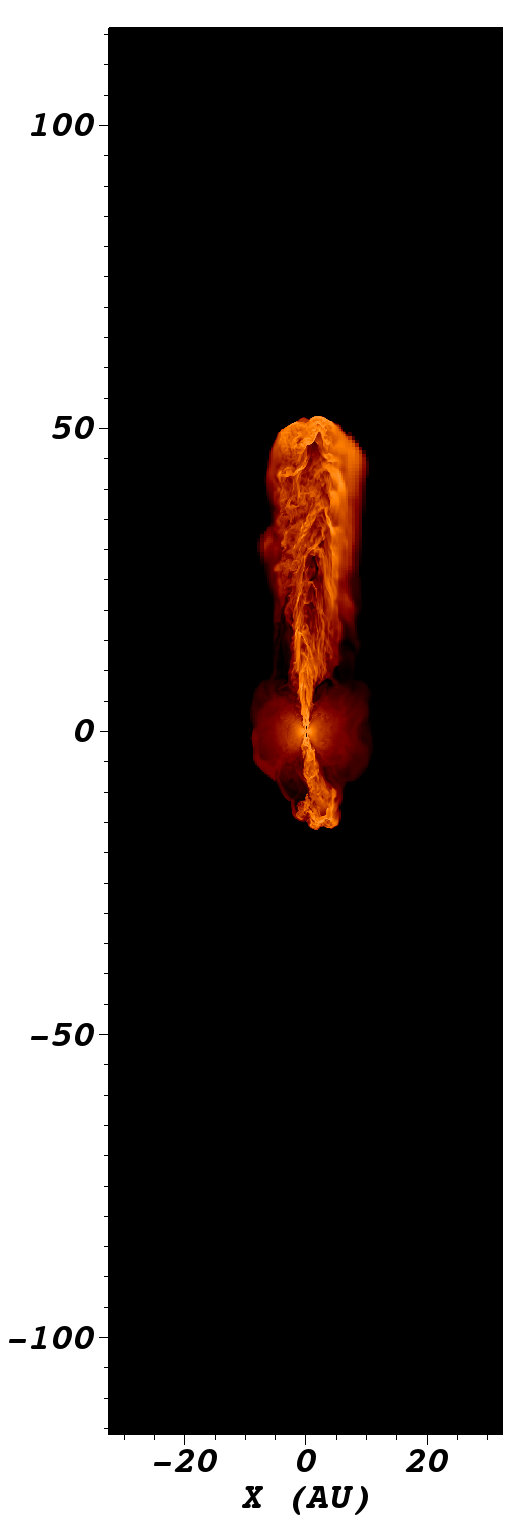}
    \includegraphics[height = 9 cm]{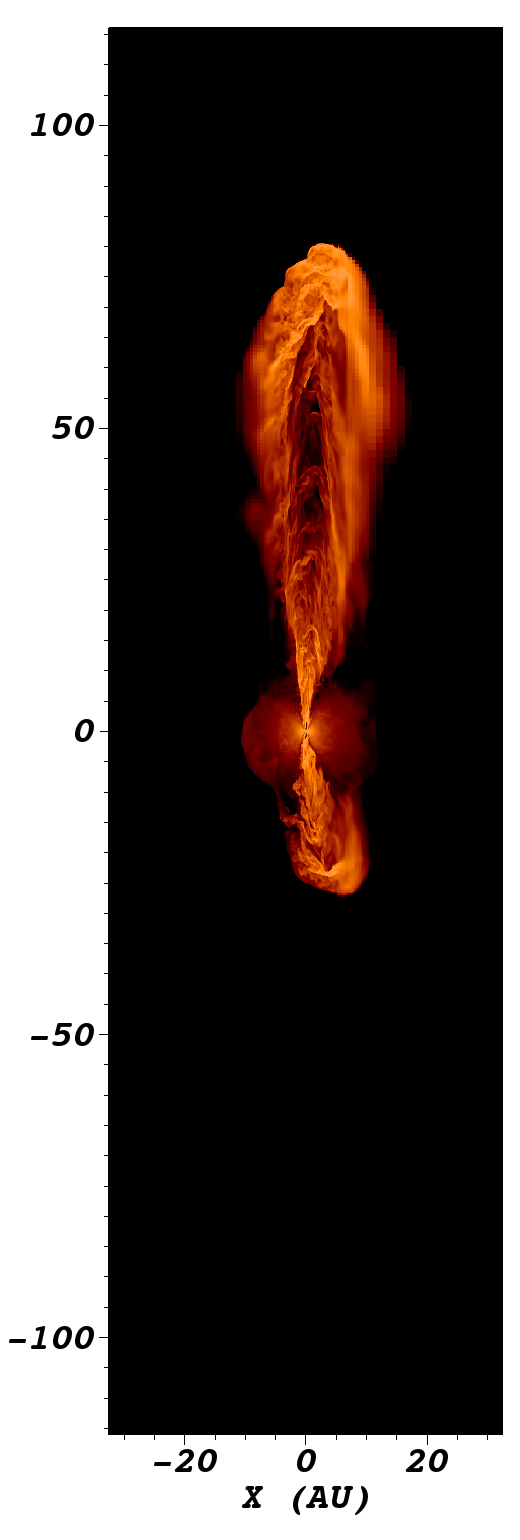}
    \includegraphics[height = 9 cm]{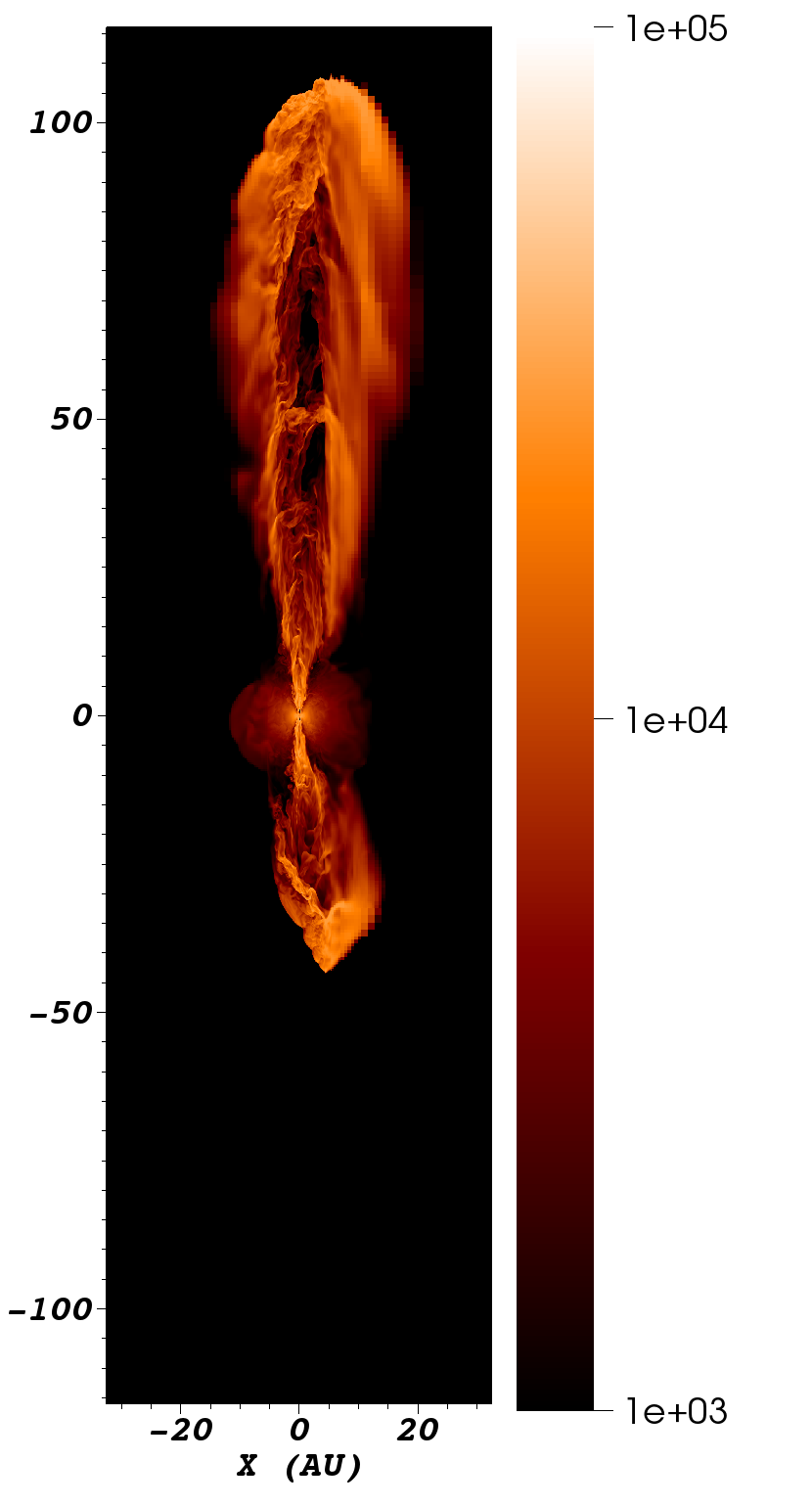}
    \includegraphics[height = 9 cm]{time-evolutions/T-new.png}
    \caption{Model D: temperature (K, in log scale) evolution of the low-momentum outflow with radiative cooling, taken at the same times as Fig.~\ref{fig:D-rho}.}
    \label{fig:D-Temp}
\end{figure*}

Large scale asymmetry becomes most apparent in model D, where we apply cooling to the low-momentum flow from model B. Fig.~ \ref{fig:D-rho} and \ref{fig:D-Temp} show evolution in density and temperature respectively for 4000 days after the outflow is turned on. The outflow to the negative z direction has been choked for more than 1500 days after starting, so the bottom lobe is only about one-third as long compared to the top lobe in its z-extent. Also, we see the the bottom lobe has been redirected to the right side of the polar axis.  While this was apparent in model B, it is also observed here to a greater extent especially in the central two panels where almost the entire lobe has been shunted to the positive-x side of the top lobe's nominal symmetry axis. 

Spatial evolution of the top lobe in model D is also the slowest. It takes 4000 days for the top lobe to reach the same height in the positive and negative z-directions, while the other three models take less than 2000 days. This is as expected since for model D we have both a low momentum injection rate and the presence of radiative cooling.

While the scale of these simulations is still smaller than most observed PPN, its worth noting the similarity between the morphology of this model and OH\,231.8+04.2 \footnote{The morphology of OH\,231.8+04.2 is tantalisingly similar to the asymmetries we simulate. However it should be pointed out that this nebula contains an AGB Mira and a distant companion, making it a far cry from a classic post-CE system. For an alternative model of this system, which however cannot account for the asymmetry of the nebula, see \citet{Staff+2016}.} which shows bipolar lobes of distinctly unequal size \citep{SanchezContreras2018}.  In addition a central "disk" (dark lane) is apparent in the nebula. Finally structures within the lobe termed the "spine" and "skirt" by \cite{BalickFrankI} appear to find analogues with the jets and lens shaped shock seen in our models.

We note that we also carried out runs with no quiescent period and once again the bottom lobe  develops much later compared to the top lobe.  Thus the strong asymmetry between top and bottom occurs regardless of the inclusion of a quiescent period between CE interaction and fast wind injection.

\section{Discussion}\label{sec:disc}
\subsection{Quantification of the asymmetries}\label{sec:asymmetry}

\begin{figure*}
	\includegraphics[width = 8.8 cm]{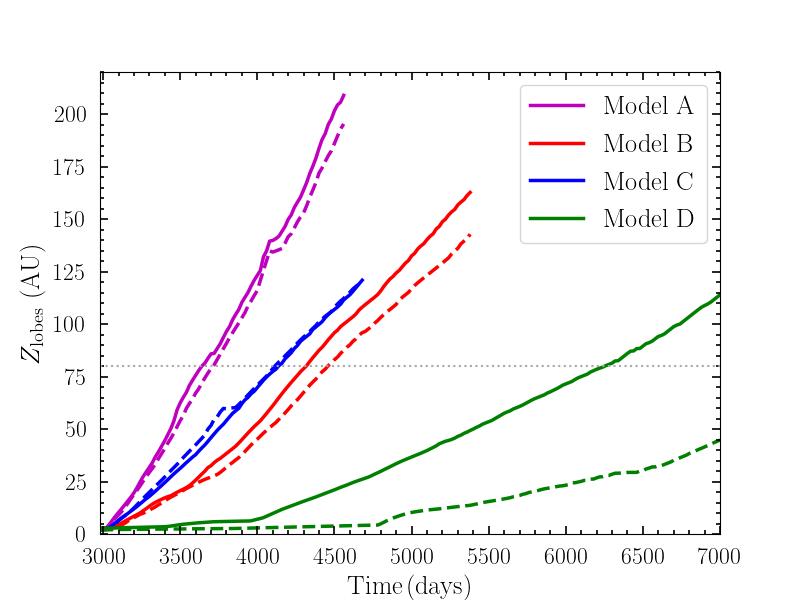}
	\includegraphics[width = 8.8 cm]{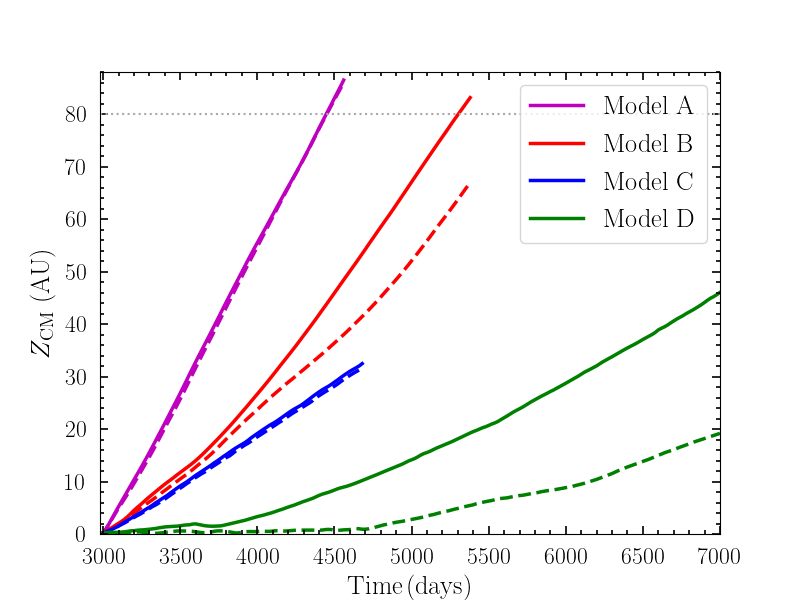}
    \caption{Extents of the lobes in z direction. (a) left: absolute length of the lobes; (b) right: position of the center of mass. For both panels, solid lines are the top lobes and dashed lines are the bottom lobes (in absolute value). 
    Note that once the slopes flatten and the lobe expansion will appear to be homologous.}
    \label{fig:tracer_z_flip}
\end{figure*}
\begin{figure*}
	\includegraphics[width = 8.8 cm]{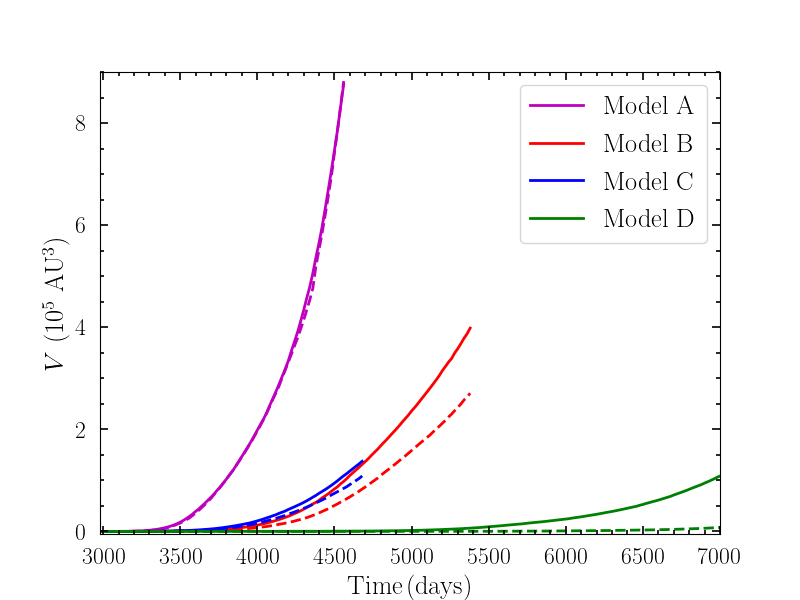}
	\includegraphics[width =8.8 cm]{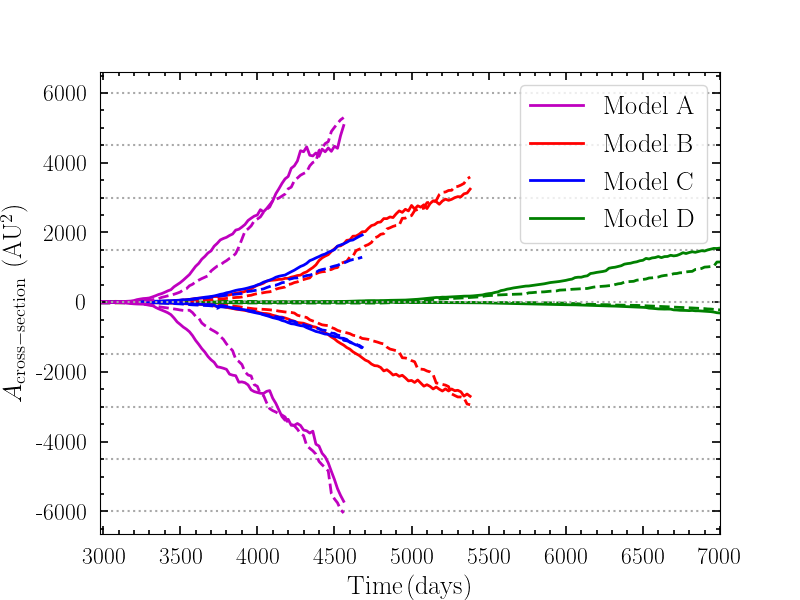}
    \caption{(a): Volume of the lobes, solid lines for top lobes and dashed lines for bottom lobes; (b): cross section of the lobes taking at $1/3$ (dashed lines) and $2/3$ (solid lines) heights of the lobes, with positive y-axis for the top lobes and negative y-axis for the bottom lobes.}
    \label{fig:shape}
\end{figure*}

In this section we try to quantify the degree of asymmetry using parameters of length, volume, and cross section associated with the lobes. 

Instead of the white temperature contour used previously in the density plots, here we define the lobes as regions filled with material from the fast wind outflow. 
We use tracers to keep track of material in the outflow. 
Tracers act as a label attached to the injected fast wind material.
They are attached to the outflow gas density in a $1:1$ ratio, where there is higher density, there are more tracers. 
A minimum tracer density at $10^{-15}~\mathrm{g\,cm^{-3}}$ distinguishes the boundary between outflow material and ambient material. 
Visually, this threshold density makes an envelope around the central jets inside the PN lobes, but within the white temperature contour in the density plots. 

Spatial extents of the lobes are shown in Fig.~\ref{fig:tracer_z_flip} using: (a) the extremum z-coordinates; (b) position of the center-of-mass along the z-direction. For each model, the solid line represents the top lobe and the dashed lined represents the bottom lobe. 
Comparing the momentum flux under the same cooling assumption, (A to B, and C to D), we see the low momentum cases not only lag the high momentum cases, but their bottom lobes also suffer a stronger effect compared to the top lobes. On the other hand, adding a cooling term does not affect the top-to-bottom asymmetry much besides that it slows down the spatial extension (compare model A to C, and B to D).

The left panel in Fig.~\ref{fig:shape} shows the integrated volume in each lobe. 
Similarly, the higher the momentum flux, or the more energetic (no radiative cooling) the outflow is, the larger the lobes will be.    
The right panel in Fig.~\ref{fig:shape} picks up more asymmetries on the small scales. It is the measurement of cross section of the lobes taken at the points of $1/3$ (dashed lines) and $2/3$ (solid lines) of their absolute z-extents (Fig.~\ref{fig:tracer_z_flip} left). 
Positive side on the y-axis is for the top lobes, and negative side is for the bottom lobes. 

A plausible hypothesis for the development of the asymmetries between top and bottom lobes is that while our wind is purely spherical, the CE substrate calculated via the SPH simulation is not expected to be top-bottom symmetric.  While some symmetry breaking will be due to the numerical algorithm, the underlying physics of the CE interaction is that of a highly turbulent 3-D flow which should not be expected to maintain top-bottom symmetry.  As we have emphasized above, the flows associated with CE interaction are inherently fully 3-D.  In particular, the binary stars` movement through the CE gas sets up strong shearing shocks that lead to Kelvin-Helmholtz instabilities which drive local regions of turbulence.  The complexity of the flows after repeatedly being shocked with the subsequent driving of unstable modes during each orbit creates local flow patterns that need not be axi-symmetric or top-bottom symmetric.  We note that such perturbations are seen not just in SPH simulations but also in AMR grid-based simulations (\citet{Chamandy+2018}, fig. 4 and  \citet{MacLeod+2018}, fig. 1). In addition even before the CE phase begins we can see the RGB and AGB stars, which will be the primary stars in CE interaction, will be convective. This means asymmetries should be expected to present at the start of CE evolution and therefore those asymmetries would also be expected in the ejecta into which our spherical wind propagates. Thus while the initial perturbations driving instabilities may have numerical seeds, the development of turbulent, asymmetric local flow conditions is to be expected. Therefore given the expected asymmetry in CE ejecta, it will manifest as a difference in wind obstruction between top and bottom and this will lead to asymmetries in the bipolar lobe evolution.

\subsection{Angular Distribution of  Linear Momentum} \label{sec:momentum}

In this paper, we have  injected a wind with momentum outflow rates well in excess of that which can be supplied by radiation alone (see last row of Table~\ref{tab:params}),
even though the mechanical energy associated with this outflow  
is about the luminosity of a post-AGB wind (see the second to last row of Table~\ref{tab:params}).
These values are chosen as input values because the bipolar outflows in PPN are  observed to have very strong constraints of momentum 
\citep{Bujarrabal+2001,Sahai+2008,Sahai+2015,Sahai+2017}.
Such outflows require another energy source besides radiation pressure which likely means extracting gravitational binding energy from within the binary, such as via accretion \citep{Blackman+Lucchini2014}.
Our present study addresses whether such outflows, even if highly uncollimated at the base, can emerge collimated.

In this context it is interesting to quantify the angular distribution of momentum that results from our 
collimation mechanism.
By injecting the isotropic fast wind at fixed density and speed, we are inputting momentum to the simulation at a rate of 
$\dot{P} = 4\pi r^2_{\mathrm{FW}} \rho_{\mathrm{FW}} v^2_{\mathrm{FW}}$, 
uniformly distributed around the sphere with radius $r_{\mathrm{FW}}\sim0.2\mathrm{AU}$. 
Most of this input momentum ($>90\%$) gets absorbed by the CE ejecta through inelastic collision between the fast wind material and the ejecta, but some is redirected into the collimated jets and reinforces them. 
We compare the ratio between the momentum in the collimated jets and the inputting momentum from the same solid angle $\Omega$ subtended by the jet as: 

\begin{equation}
     \frac{\text{Jet momentum flux}}{\text{Input momentum flux}}
    = \frac{\langle\rho_j v_j^2 \,\hat{z}\cdot\hat{r}\rangle\big|_{\Omega}}
    {\frac{1}{2}\, \rho_{\mathrm{FW}}v^2_{\mathrm{FW}}\,(\Omega/4\pi)}.
\end{equation}

We measure the jet momentum at a sphere with radius 1 AU from the central source,  
where the outflow has been strongly collimated, and the elliptical hot bubble shock is interior to this region.
Here, $\Omega$ is estimated from the traced outflow geometry, which is almost a cone shape.
The numerator is computed by integrating 
$( \rho_j v_j v_{j,\hat{z}})$ 
over the meshes along the sphere, then averaged over the area of the cone base $(\Omega r_{\mathrm{FW}}^2)$.  
The denominator depends on the input density and speed, as well as the geometry. The factor of $1/2$ in the denominator comes from isolating the $\hat{z}$ component of the velocity. 

For the two models without radiative cooling, we find the open angle of the collimated jet (i.e. $\Omega$) does not vary noticeably with time. We perform the above calculation for both top and bottom jets at four times throughout the evolution (same times as shown in Fig.~ \ref{fig:A-rho} to \ref{fig:B-Temp}). 
In the high momentum model A, the collimated jet subtends a solid angle $\Omega\sim0.37\pi$. Then for a single jet, this momentum ratio averages between 3.05 to 3.25.  Given the spherical input momentum for the high momentum case in Table~\ref{tab:params}, this gives momentum per unit time of the jet to be $\sim500$ times $L_{mech}/c$.

The factor of focusing is stronger for less input momentum.  In the low momentum case, model B, we measure a smaller $\Omega\sim0.2\pi$, and a higher momentum ratio between 6.1 to 7.8.  
Given the input spherical  momentum for the low momentum case in Table~\ref{tab:params}, this yields about 35 to 40 times $L_{mech}/c$ in each jet.

The calculations of this section have constrained the ratio of momentum  in the collimated jet compared to just "cutting out" that section of the solid angle of the initial spherically imposed momentum.
The ratios that we find are greater than unity but less than 10. As such,  matching the extremely high observed momenta  in PPN outflows  \citep{Bujarrabal+2001,Sahai+2008,Sahai+2015,Sahai+2017}
still requires an imposed isotropic momentum well in excess of what isotropic radiation from a star can provide.  That is why we chose such high momentum inputs for our isotropic values in the first place, and  assumed that there is some other way to supply this needed momentum.  Extraction of gravitational potential energy is likely needed, and one pathway is via accretion \citep{Blackman+Lucchini2014}. 
 

\section{Conclusions}

In this paper we have shown that by beginning with full 3D CE simulation as an input, it is possible to drive strongly bipolar outflows from a maximally uncollimated (i.e. a spherical) wind from the post-AGB star or remnant binary system.

We find that CE ejecta provide the high pole-to-equator density contrast environment needed in the classical GISW model. Three layers of shocks are fully resolved in the simulation.  These include the outer shock sweeping up the CE ejecta, the lens-shaped inner shock and, most importantly, the small scale, innermost, elliptical shock predicted by the GISW model. This shock focused inertial confinement shapes the spherical fast wind to become highly collimated bipolar outflow. 

Our simulations start with high enough momenta in all four cases to account for observed high momenta PPN outflows. We assume other source to supply such strong outflow such as accretion. We see significant focusing of the input momentum into the jet, and the fraction being redirected depends on the input momentum.

We also note that by showing that a wind from the central PPN/PN engine of any geometry can be converted into a jet, our results can be used as inputs for models of larger scale bipolar outflows that begin with jets as initial conditions \citep{BalickFrankI,BalickFrankII,BalickFrankIII}.  This is an important point as it was not always clear that GISW models could produce the tight-waisted nebula seen in some PPN/PN. The models presented in this chapter seem to indicate that GISW models can be the starting point for a wider variety of nebular models than previously thought

As was discussed in \citet{Frank+2018}, we see symmetry breaking in the bipolar outflows driven by our GISW models. This occurs because the morphology of the bipolar lobes is shaped by the inherent asymmetries in the central funnel of the CE ejecta. These asymmetries affect the kinematic and energetic properties of the outflows. Thus our work may offer explanation for objects like OH\,231.8+04.2 which show significant asymmetries in their lobes \citep{SanchezContreras2018}.  

It would also be important to carry out this simulation using CE ejecta using simulations where most of the ejecta are
formally unbound (for the problem of non-unbound common envelopes, see \citet{Reichardt+2020}). Finally, in order to
properly model PN and PPN, it would be opportune to carry out a simulation that uses an AGB CE simulation, rather than
an RGB one.

\section*{Acknowledgements}

This work used the computational and visualization resources in the Center for Integrated Research Computing (CIRC) at the University of Rochester and the computational resources of the Texas Advanced Computing Center (TACC) at The University of Texas at Austin, provided through allocation TG-AST120060 from the Extreme Science and Engineering Discovery Environment (XSEDE) \citep{xsede}, which is supported by National Science Foundation grant number ACI-1548562. 
Financial support for this project was provided by the Department of Energy grant DE-SC0001063, DE-SC0020432, and DE-SC0020434, the National Science Foundation grants AST-1515648 and AST-1813298, and the Space Telescope Science Institute grant HST-AR-14563.001-A. 
YZ acknowledges financial support from University of Rochester Frank J. Horton Graduate Research Fellowship.
ZC is grateful to the CITA National Postdoctoral Fellowship for financial support.
TR acknowledges financial support from Macquarie University Research Excellence Scholarship.
OD acknowledges support from the Australian Research Council Future fellowship scheme FT120100452. 
EB acknowledges additional support from  KITP UC Santa Barbara funded by NSF Grant PHY-1748958, and Aspen Center for Physics funded by NSF Grant PHY-1607611. 
JN acknowledges support from NASA HST-AR-15044.


\bibliographystyle{mnras}
\bibliography{refs} 

\begin{thebibliography}{}
\makeatletter
\relax
\def\mn@urlcharsother{\let\do\@makeother \do\$\do\&\do\#\do\^\do\_\do\%\do\~}
\def\mn@doi{\begingroup\mn@urlcharsother \@ifnextchar [ {\mn@doi@}
  {\mn@doi@[]}}
\def\mn@doi@[#1]#2{\def\@tempa{#1}\ifx\@tempa\@empty \href
  {http://dx.doi.org/#2} {doi:#2}\else \href {http://dx.doi.org/#2} {#1}\fi
  \endgroup}
\def\mn@eprint#1#2{\mn@eprint@#1:#2::\@nil}
\def\mn@eprint@arXiv#1{\href {http://arxiv.org/abs/#1} {{\tt arXiv:#1}}}
\def\mn@eprint@dblp#1{\href {http://dblp.uni-trier.de/rec/bibtex/#1.xml}
  {dblp:#1}}
\def\mn@eprint@#1:#2:#3:#4\@nil{\def\@tempa {#1}\def\@tempb {#2}\def\@tempc
  {#3}\ifx \@tempc \@empty \let \@tempc \@tempb \let \@tempb \@tempa \fi \ifx
  \@tempb \@empty \def\@tempb {arXiv}\fi \@ifundefined
  {mn@eprint@\@tempb}{\@tempb:\@tempc}{\expandafter \expandafter \csname
  mn@eprint@\@tempb\endcsname \expandafter{\@tempc}}}

\bibitem[\protect\citeauthoryear{{Akashi} \& {Soker}}{{Akashi} \&
  {Soker}}{2018}]{Akashi+Soker2018}
{Akashi} M.,  {Soker} N.,  2018, \mn@doi [\mnras] {10.1093/mnras/sty2479},
  \href {https://ui.adsabs.harvard.edu/abs/2018MNRAS.481.2754A} {481, 2754}

\bibitem[\protect\citeauthoryear{{Balick}}{{Balick}}{1987}]{Balick1987}
{Balick} B.,  1987, \mn@doi [\aj] {10.1086/114504}, \href
  {https://ui.adsabs.harvard.edu/\#abs/1987AJ.....94..671B} {94, 671}

\bibitem[\protect\citeauthoryear{{Balick}, {Frank}, {Liu}  \&
  {Huarte-Espinosa}}{{Balick} et~al.}{2017}]{BalickFrankI}
{Balick} B.,  {Frank} A.,  {Liu} B.,   {Huarte-Espinosa} M.,  2017, \mn@doi
  [\apj] {10.3847/1538-4357/aa77f0}, \href
  {https://ui.adsabs.harvard.edu/abs/2017ApJ...843..108B} {843, 108}

\bibitem[\protect\citeauthoryear{{Balick}, {Frank}, {Liu}  \&
  {Corradi}}{{Balick} et~al.}{2018}]{BalickFrankII}
{Balick} B.,  {Frank} A.,  {Liu} B.,   {Corradi} R.,  2018, \mn@doi [\apj]
  {10.3847/1538-4357/aaa772}, \href
  {https://ui.adsabs.harvard.edu/abs/2018ApJ...853..168B} {853, 168}

\bibitem[\protect\citeauthoryear{{Balick}, {Frank}  \& {Liu}}{{Balick}
  et~al.}{2019}]{BalickFrankIII}
{Balick} B.,  {Frank} A.,   {Liu} B.,  2019, \mn@doi [\apj]
  {10.3847/1538-4357/ab16f5}, \href
  {https://ui.adsabs.harvard.edu/abs/2019ApJ...877...30B} {877, 30}

\bibitem[\protect\citeauthoryear{{Bjorkman} \& {Cassinelli}}{{Bjorkman} \&
  {Cassinelli}}{1993}]{Bjorkman+Cassinelli1993}
{Bjorkman} J.~E.,  {Cassinelli} J.~P.,  1993, \mn@doi [\apj] {10.1086/172676},
  \href {https://ui.adsabs.harvard.edu/\#abs/1993ApJ...409..429B} {409, 429}

\bibitem[\protect\citeauthoryear{{Blackman} \& {Lucchini}}{{Blackman} \&
  {Lucchini}}{2014}]{Blackman+Lucchini2014}
{Blackman} E.~G.,  {Lucchini} S.,  2014, \mn@doi [\mnras]
  {10.1093/mnrasl/slu001}, \href
  {https://ui.adsabs.harvard.edu/\#abs/2014MNRAS.440L..16B} {440, L16}

\bibitem[\protect\citeauthoryear{{Blackman}, {Frank}, {Markiel}, {Thomas}  \&
  {Van Horn}}{{Blackman} et~al.}{2001a}]{2001Natur.409..485B}
{Blackman} E.~G.,  {Frank} A.,  {Markiel} J.~A.,  {Thomas} J.~H.,   {Van Horn}
  H.~M.,  2001a, \mn@doi [\nat] {10.1038/35054008}, \href
  {https://ui.adsabs.harvard.edu/abs/2001Natur.409..485B} {409, 485}

\bibitem[\protect\citeauthoryear{{Blackman}, {Frank}  \& {Welch}}{{Blackman}
  et~al.}{2001b}]{2001ApJ...546..288B}
{Blackman} E.~G.,  {Frank} A.,   {Welch} C.,  2001b, \mn@doi [\apj]
  {10.1086/318253}, \href
  {https://ui.adsabs.harvard.edu/abs/2001ApJ...546..288B} {546, 288}

\bibitem[\protect\citeauthoryear{{Bujarrabal}, {Castro-Carrizo}, {Alcolea}  \&
  {S{\'a}nchez Contreras}}{{Bujarrabal} et~al.}{2001}]{Bujarrabal+2001}
{Bujarrabal} V.,  {Castro-Carrizo} A.,  {Alcolea} J.,   {S{\'a}nchez Contreras}
  C.,  2001, \mn@doi [\aap] {10.1051/0004-6361:20011090}, \href
  {https://ui.adsabs.harvard.edu/\#abs/2001A&A...377..868B} {377, 868}

\bibitem[\protect\citeauthoryear{{Carroll-Nellenback}, {Shroyer}, {Frank}  \&
  {Ding}}{{Carroll-Nellenback} et~al.}{2013}]{Carroll-Nellenback+2013}
{Carroll-Nellenback} J.~J.,  {Shroyer} B.,  {Frank} A.,   {Ding} C.,  2013,
  \mn@doi [Journal of Computational Physics] {10.1016/j.jcp.2012.10.004}, \href
  {https://ui.adsabs.harvard.edu/\#abs/2013JCoPh.236..461C} {236, 461}

\bibitem[\protect\citeauthoryear{{Chamandy} et~al.,}{{Chamandy}
  et~al.}{2018}]{Chamandy+2018}
{Chamandy} L.,  et~al., 2018, \mn@doi [\mnras] {10.1093/mnras/sty1950}, \href
  {https://ui.adsabs.harvard.edu/\#abs/2018MNRAS.480.1898C} {480, 1898}

\bibitem[\protect\citeauthoryear{{Chen}, {Frank}, {Blackman}, {Nordhaus}  \&
  {Carroll-Nellenback}}{{Chen} et~al.}{2017}]{Chen+2017}
{Chen} Z.,  {Frank} A.,  {Blackman} E.~G.,  {Nordhaus} J.,
  {Carroll-Nellenback} J.,  2017, \mn@doi [\mnras] {10.1093/mnras/stx680},
  \href {https://ui.adsabs.harvard.edu/abs/2017MNRAS.468.4465C} {468, 4465}

\bibitem[\protect\citeauthoryear{{Dalgarno} \& {McCray}}{{Dalgarno} \&
  {McCray}}{1972}]{Dalgarno+McCray1972}
{Dalgarno} A.,  {McCray} R.~A.,  1972, \mn@doi [Annual Review of Astronomy and
  Astrophysics] {10.1146/annurev.aa.10.090172.002111}, \href
  {https://ui.adsabs.harvard.edu/abs/1972ARA&A..10..375D} {10, 375}

\bibitem[\protect\citeauthoryear{{Estrella-Trujillo},
  {Hern{\'a}ndez-Mart{\'\i}nez}, {Vel{\'a}zquez}, {Esquivel}  \&
  {Raga}}{{Estrella-Trujillo} et~al.}{2019}]{Estrella-Trujillo+2019}
{Estrella-Trujillo} D.,  {Hern{\'a}ndez-Mart{\'\i}nez} L.,  {Vel{\'a}zquez}
  P.~F.,  {Esquivel} A.,   {Raga} A.~C.,  2019, \mn@doi [\apj]
  {10.3847/1538-4357/ab12e1}, \href
  {https://ui.adsabs.harvard.edu/abs/2019ApJ...876...29E} {876, 29}

\bibitem[\protect\citeauthoryear{{Fabian} \& {Hansen}}{{Fabian} \&
  {Hansen}}{1979}]{Fabian+Hansen1979}
{Fabian} A.~C.,  {Hansen} C.~J.,  1979, \mn@doi [\mnras]
  {10.1093/mnras/187.2.283}, \href
  {https://ui.adsabs.harvard.edu/abs/1979MNRAS.187..283F} {187, 283}

\bibitem[\protect\citeauthoryear{{Frank}}{{Frank}}{1994}]{Frank1994}
{Frank} A.,  1994, \mn@doi [\aj] {10.1086/116850}, \href
  {https://ui.adsabs.harvard.edu/abs/1994AJ....107..261F} {107, 261}

\bibitem[\protect\citeauthoryear{{Frank}}{{Frank}}{1999}]{Frank1999}
{Frank} A.,  1999, \mn@doi [New Astronomy Reviews]
  {10.1016/S1387-6473(99)00005-6}, \href
  {https://ui.adsabs.harvard.edu/\#abs/1999NewAR..43...31F} {43, 31}

\bibitem[\protect\citeauthoryear{{Frank} \& {Mellema}}{{Frank} \&
  {Mellema}}{1994a}]{Frank+Mellema1994a}
{Frank} A.,  {Mellema} G.,  1994a, \aap, \href
  {https://ui.adsabs.harvard.edu/\#abs/1994A&A...289..937F} {289, 937}

\bibitem[\protect\citeauthoryear{{Frank} \& {Mellema}}{{Frank} \&
  {Mellema}}{1994b}]{Frank+Mellema1994b}
{Frank} A.,  {Mellema} G.,  1994b, \mn@doi [\apj] {10.1086/174450}, \href
  {https://ui.adsabs.harvard.edu/\#abs/1994ApJ...430..800F} {430, 800}

\bibitem[\protect\citeauthoryear{{Frank} \& {Mellema}}{{Frank} \&
  {Mellema}}{1996}]{Frank+Mellema1996}
{Frank} A.,  {Mellema} G.,  1996, \mn@doi [\apj] {10.1086/178099}, \href
  {https://ui.adsabs.harvard.edu/abs/1996ApJ...472..684F} {472, 684}

\bibitem[\protect\citeauthoryear{{Frank}, {Balick}, {Icke}  \&
  {Mellema}}{{Frank} et~al.}{1993}]{Frank+1993}
{Frank} A.,  {Balick} B.,  {Icke} V.,   {Mellema} G.,  1993, \mn@doi [\apj]
  {10.1086/186735}, \href
  {https://ui.adsabs.harvard.edu/\#abs/1993ApJ...404L..25F} {404, L25}

\bibitem[\protect\citeauthoryear{{Frank}, {Chen}, {Reichardt}, {De Marco},
  {Blackman}  \& {Nordhaus}}{{Frank} et~al.}{2018}]{Frank+2018}
{Frank} A.,  {Chen} Z.,  {Reichardt} T.,  {De Marco} O.,  {Blackman} E.,
  {Nordhaus} J.,  2018, \mn@doi [Galaxies] {10.3390/galaxies6040113}, \href
  {https://ui.adsabs.harvard.edu/\#abs/2018Galax...6..113F} {6, 113}

\bibitem[\protect\citeauthoryear{{Garc{\'\i}a-Segura}, {Langer},
  {R{\'o}{\.z}yczka}  \& {Franco}}{{Garc{\'\i}a-Segura}
  et~al.}{1999}]{Garcia-Segura+1999}
{Garc{\'\i}a-Segura} G.,  {Langer} N.,  {R{\'o}{\.z}yczka} M.,   {Franco} J.,
  1999, \mn@doi [\apj] {10.1086/307205}, \href
  {https://ui.adsabs.harvard.edu/\#abs/1999ApJ...517..767G} {517, 767}

\bibitem[\protect\citeauthoryear{{Garc{\'\i}a-Segura}, {L{\'o}pez}  \&
  {Franco}}{{Garc{\'\i}a-Segura} et~al.}{2005}]{Garcia-Segura+2005}
{Garc{\'\i}a-Segura} G.,  {L{\'o}pez} J.~A.,   {Franco} J.,  2005, \mn@doi
  [\apj] {10.1086/426110}, \href
  {https://ui.adsabs.harvard.edu/\#abs/2005ApJ...618..919G} {618, 919}

\bibitem[\protect\citeauthoryear{{Garc{\'\i}a-Segura}, {Villaver}, {Langer},
  {Yoon}  \& {Manchado}}{{Garc{\'\i}a-Segura}
  et~al.}{2014}]{Garcia-Segura+2014}
{Garc{\'\i}a-Segura} G.,  {Villaver} E.,  {Langer} N.,  {Yoon} S.~C.,
  {Manchado} A.,  2014, \mn@doi [\apj] {10.1088/0004-637X/783/2/74}, \href
  {https://ui.adsabs.harvard.edu/abs/2014ApJ...783...74G} {783, 74}

\bibitem[\protect\citeauthoryear{{Garc{\'\i}a-Segura}, {Ricker}  \&
  {Taam}}{{Garc{\'\i}a-Segura} et~al.}{2018}]{Garcia-Segura+2018}
{Garc{\'\i}a-Segura} G.,  {Ricker} P.~M.,   {Taam} R.~E.,  2018, \mn@doi [\apj]
  {10.3847/1538-4357/aac08c}, \href
  {https://ui.adsabs.harvard.edu/\#abs/2018ApJ...860...19G} {860, 19}

\bibitem[\protect\citeauthoryear{{Guidarelli}, {Nordhaus}, {Chamandy}, {Chen},
  {Blackman}, {Frank}, {Carroll-Nellenback}  \& {Liu}}{{Guidarelli}
  et~al.}{2019}]{2019MNRAS.490.1179G}
{Guidarelli} G.,  {Nordhaus} J.,  {Chamandy} L.,  {Chen} Z.,  {Blackman} E.~G.,
   {Frank} A.,  {Carroll-Nellenback} J.,   {Liu} B.,  2019, \mn@doi [\mnras]
  {10.1093/mnras/stz2641}, \href
  {https://ui.adsabs.harvard.edu/abs/2019MNRAS.490.1179G} {490, 1179}

\bibitem[\protect\citeauthoryear{{Iaconi}, {Reichardt}, {Staff}, {De Marco},
  {Passy}, {Price}, {Wurster}  \& {Herwig}}{{Iaconi}
  et~al.}{2017}]{Iaconi+2017}
{Iaconi} R.,  {Reichardt} T.,  {Staff} J.,  {De Marco} O.,  {Passy} J.-C.,
  {Price} D.,  {Wurster} J.,   {Herwig} F.,  2017, \mn@doi [\mnras]
  {10.1093/mnras/stw2377}, \href
  {https://ui.adsabs.harvard.edu/\#abs/2017MNRAS.464.4028I} {464, 4028}

\bibitem[\protect\citeauthoryear{{Iaconi}, {De Marco}, {Passy}  \&
  {Staff}}{{Iaconi} et~al.}{2018}]{Iaconi+2018}
{Iaconi} R.,  {De Marco} O.,  {Passy} J.-C.,   {Staff} J.,  2018, \mn@doi
  [\mnras] {10.1093/mnras/sty794}, \href
  {https://ui.adsabs.harvard.edu/abs/2018MNRAS.477.2349I} {477, 2349}

\bibitem[\protect\citeauthoryear{{Iaconi}, {Maeda}, {De Marco}, {Nozawa}  \&
  {Reichardt}}{{Iaconi} et~al.}{2019}]{Iaconi+2019}
{Iaconi} R.,  {Maeda} K.,  {De Marco} O.,  {Nozawa} T.,   {Reichardt} T.,
  2019, \mn@doi [\mnras] {10.1093/mnras/stz2312}, \href
  {https://ui.adsabs.harvard.edu/abs/2019MNRAS.489.3334I} {489, 3334}

\bibitem[\protect\citeauthoryear{{Icke}}{{Icke}}{1988}]{Icke1988}
{Icke} V.,  1988, \aap, \href
  {https://ui.adsabs.harvard.edu/\#abs/1988A&A...202..177I} {202, 177}

\bibitem[\protect\citeauthoryear{{Icke}, {Balick}  \& {Frank}}{{Icke}
  et~al.}{1992a}]{Icke+1992a}
{Icke} V.,  {Balick} B.,   {Frank} A.,  1992a, \aap, \href
  {https://ui.adsabs.harvard.edu/\#abs/1992A&A...253..224I} {253, 224}

\bibitem[\protect\citeauthoryear{{Icke}, {Mellema}, {Balick}, {Eulderink}  \&
  {Frank}}{{Icke} et~al.}{1992b}]{Icke+1992b}
{Icke} V.,  {Mellema} G.,  {Balick} B.,  {Eulderink} F.,   {Frank} A.,  1992b,
  \mn@doi [\nat] {10.1038/355524a0}, \href
  {https://ui.adsabs.harvard.edu/\#abs/1992Natur.355..524I} {355, 524}

\bibitem[\protect\citeauthoryear{{Ivanova} \& {Nandez}}{{Ivanova} \&
  {Nandez}}{2016}]{Ivanova+Nandez2016}
{Ivanova} N.,  {Nandez} J.~L.~A.,  2016, \mn@doi [\mnras]
  {10.1093/mnras/stw1676}, \href
  {https://ui.adsabs.harvard.edu/\#abs/2016MNRAS.462..362I} {462, 362}

\bibitem[\protect\citeauthoryear{{Ivanova} et~al.,}{{Ivanova}
  et~al.}{2013}]{Ivanova+2013}
{Ivanova} N.,  et~al., 2013, \mn@doi [\aapr] {10.1007/s00159-013-0059-2}, \href
  {https://ui.adsabs.harvard.edu/abs/2013A&ARv..21...59I} {21, 59}

\bibitem[\protect\citeauthoryear{{Kuruwita}, {Staff}  \& {De Marco}}{{Kuruwita}
  et~al.}{2016}]{Kuruwita+2016}
{Kuruwita} R.~L.,  {Staff} J.,   {De Marco} O.,  2016, \mn@doi [\mnras]
  {10.1093/mnras/stw1414}, \href
  {https://ui.adsabs.harvard.edu/abs/2016MNRAS.461..486K} {461, 486}

\bibitem[\protect\citeauthoryear{{Kwok}, {Purton}  \& {Fitzgerald}}{{Kwok}
  et~al.}{1978}]{Kwok+1978}
{Kwok} S.,  {Purton} C.~R.,   {Fitzgerald} P.~M.,  1978, \mn@doi [\apj]
  {10.1086/182621}, \href
  {https://ui.adsabs.harvard.edu/\#abs/1978ApJ...219L.125K} {219, L125}

\bibitem[\protect\citeauthoryear{{Lee} \& {Sahai}}{{Lee} \&
  {Sahai}}{2003}]{Lee+Sahai2003}
{Lee} C.-F.,  {Sahai} R.,  2003, \mn@doi [\apj] {10.1086/346265}, \href
  {https://ui.adsabs.harvard.edu/abs/2003ApJ...586..319L} {586, 319}

\bibitem[\protect\citeauthoryear{{Livio}}{{Livio}}{1993}]{Livio1993}
{Livio} M.,  1993, in {Weinberger} R.,  {Acker} A.,  eds,  IAU Symposium Vol.
  155, Planetary Nebulae. p.~279

\bibitem[\protect\citeauthoryear{{Livio} \& {Soker}}{{Livio} \&
  {Soker}}{1988}]{Livio+Soker1988}
{Livio} M.,  {Soker} N.,  1988, \mn@doi [\apj] {10.1086/166419}, \href
  {https://ui.adsabs.harvard.edu/\#abs/1988ApJ...329..764L} {329, 764}

\bibitem[\protect\citeauthoryear{{MacLeod}, {Ostriker}  \& {Stone}}{{MacLeod}
  et~al.}{2018}]{MacLeod+2018}
{MacLeod} M.,  {Ostriker} E.~C.,   {Stone} J.~M.,  2018, \mn@doi [\apj]
  {10.3847/1538-4357/aae9eb}, \href
  {https://ui.adsabs.harvard.edu/abs/2018ApJ...868..136M} {868, 136}

\bibitem[\protect\citeauthoryear{{Matt}, {Frank}  \& {Blackman}}{{Matt}
  et~al.}{2006}]{Matt+2006}
{Matt} S.,  {Frank} A.,   {Blackman} E.~G.,  2006, \mn@doi [\apj]
  {10.1086/507325}, \href
  {https://ui.adsabs.harvard.edu/\#abs/2006ApJ...647L..45M} {647, L45}

\bibitem[\protect\citeauthoryear{{Mellema}}{{Mellema}}{1994}]{Mellema1994}
{Mellema} G.,  1994, \aap, \href
  {https://ui.adsabs.harvard.edu/abs/1994A&A...290..915M} {290, 915}

\bibitem[\protect\citeauthoryear{{Mellema}}{{Mellema}}{1995}]{Mellema1995}
{Mellema} G.,  1995, \mn@doi [\mnras] {10.1093/mnras/277.1.173}, \href
  {https://ui.adsabs.harvard.edu/abs/1995MNRAS.277..173M} {277, 173}

\bibitem[\protect\citeauthoryear{{Mellema} \& {Frank}}{{Mellema} \&
  {Frank}}{1995}]{Mellema+Frank1995}
{Mellema} G.,  {Frank} A.,  1995, \mn@doi [\mnras] {10.1093/mnras/273.2.401},
  \href {https://ui.adsabs.harvard.edu/abs/1995MNRAS.273..401M} {273, 401}

\bibitem[\protect\citeauthoryear{{Morris}}{{Morris}}{1981}]{Morris1981}
{Morris} M.,  1981, \mn@doi [\apj] {10.1086/159317}, \href
  {https://ui.adsabs.harvard.edu/abs/1981ApJ...249..572M} {249, 572}

\bibitem[\protect\citeauthoryear{{Morris}}{{Morris}}{1990}]{Morris1990}
{Morris} M.,  1990, in {Mennessier} M.~O.,  {Omont} A.,  eds, From Miras to
  Planetary Nebulae: Which Path for Stellar Evolution?. p.~520

\bibitem[\protect\citeauthoryear{{Nandez} \& {Ivanova}}{{Nandez} \&
  {Ivanova}}{2016}]{Nandez+Ivanova2016}
{Nandez} J.~L.~A.,  {Ivanova} N.,  2016, \mn@doi [\mnras]
  {10.1093/mnras/stw1266}, \href
  {http://adsabs.harvard.edu/abs/2016MNRAS.460.3992N} {460, 3992}

\bibitem[\protect\citeauthoryear{{Nandez}, {Ivanova}  \& {Lombardi}}{{Nandez}
  et~al.}{2014}]{Nandez+2014}
{Nandez} J.~L.~A.,  {Ivanova} N.,   {Lombardi} J.~C. J.,  2014, \mn@doi [\apj]
  {10.1088/0004-637X/786/1/39}, \href
  {https://ui.adsabs.harvard.edu/abs/2014ApJ...786...39N} {786, 39}

\bibitem[\protect\citeauthoryear{{Nordhaus} \& {Blackman}}{{Nordhaus} \&
  {Blackman}}{2006}]{Nordhaus+Blackman2006}
{Nordhaus} J.,  {Blackman} E.~G.,  2006, \mn@doi [\mnras]
  {10.1111/j.1365-2966.2006.10625.x}, \href
  {https://ui.adsabs.harvard.edu/abs/2006MNRAS.370.2004N} {370, 2004}

\bibitem[\protect\citeauthoryear{{Nordhaus}, {Blackman}  \& {Frank}}{{Nordhaus}
  et~al.}{2007}]{2007MNRAS.376..599N}
{Nordhaus} J.,  {Blackman} E.~G.,   {Frank} A.,  2007, \mn@doi [\mnras]
  {10.1111/j.1365-2966.2007.11417.x}, \href
  {https://ui.adsabs.harvard.edu/abs/2007MNRAS.376..599N} {376, 599}

\bibitem[\protect\citeauthoryear{{Ohlmann}, {R{\"o}pke}, {Pakmor}  \&
  {Springel}}{{Ohlmann} et~al.}{2016}]{Ohlmann+2016}
{Ohlmann} S.~T.,  {R{\"o}pke} F.~K.,  {Pakmor} R.,   {Springel} V.,  2016,
  \mn@doi [\apj] {10.3847/2041-8205/816/1/L9}, \href
  {https://ui.adsabs.harvard.edu/\#abs/2016ApJ...816L...9O} {816, L9}

\bibitem[\protect\citeauthoryear{{Paczynski}}{{Paczynski}}{1976}]{Paczynski1976}
{Paczynski} B.,  1976, in {Eggleton} P.,  {Mitton} S.,   {Whelan} J.,  eds,
  IAU Symposium Vol. 73, Structure and Evolution of Close Binary Systems. p.~75

\bibitem[\protect\citeauthoryear{{Passy} et~al.,}{{Passy}
  et~al.}{2012}]{Passy+2012}
{Passy} J.-C.,  et~al., 2012, \mn@doi [\apj] {10.1088/0004-637X/744/1/52},
  \href {https://ui.adsabs.harvard.edu/\#abs/2012ApJ...744...52P} {744, 52}

\bibitem[\protect\citeauthoryear{{Price}}{{Price}}{2007}]{Price2007}
{Price} D.~J.,  2007, \mn@doi [\pasa] {10.1071/AS07022}, \href
  {https://ui.adsabs.harvard.edu/abs/2007PASA...24..159P} {24, 159}

\bibitem[\protect\citeauthoryear{{Price} et~al.,}{{Price}
  et~al.}{2017}]{Price+2017}
{Price} D.~J.,  et~al., 2017, {PHANTOM: Smoothed particle hydrodynamics and
  magnetohydrodynamics code} (\mn@eprint {ascl} {1709.002})

\bibitem[\protect\citeauthoryear{{Price} et~al.,}{{Price}
  et~al.}{2018}]{Price+2018}
{Price} D.~J.,  et~al., 2018, \mn@doi [\pasa] {10.1017/pasa.2018.25}, \href
  {https://ui.adsabs.harvard.edu/abs/2018PASA...35...31P} {35, e031}

\bibitem[\protect\citeauthoryear{{Prust} \& {Chang}}{{Prust} \&
  {Chang}}{2019}]{Prust+Chang2019}
{Prust} L.~J.,  {Chang} P.,  2019, \mn@doi [\mnras] {10.1093/mnras/stz1219},
  \href {https://ui.adsabs.harvard.edu/abs/2019MNRAS.486.5809P} {486, 5809}

\bibitem[\protect\citeauthoryear{{Rechy-Garc{\'\i}a}, {Pe{\~n}a}  \&
  {Vel{\'a}zquez}}{{Rechy-Garc{\'\i}a} et~al.}{2019}]{Rechy-Garcia+2019}
{Rechy-Garc{\'\i}a} J.~S.,  {Pe{\~n}a} M.,   {Vel{\'a}zquez} P.~F.,  2019,
  \mn@doi [\mnras] {10.1093/mnras/sty2758}, \href
  {https://ui.adsabs.harvard.edu/abs/2019MNRAS.482.1163R} {482, 1163}

\bibitem[\protect\citeauthoryear{{Reichardt}, {De Marco}, {Iaconi}, {Tout}  \&
  {Price}}{{Reichardt} et~al.}{2019}]{Reichardt+2019}
{Reichardt} T.~A.,  {De Marco} O.,  {Iaconi} R.,  {Tout} C.~A.,   {Price}
  D.~J.,  2019, \mn@doi [\mnras] {10.1093/mnras/sty3485}, \href
  {https://ui.adsabs.harvard.edu/\#abs/2019MNRAS.484..631R} {484, 631}

\bibitem[\protect\citeauthoryear{{Reichardt}, {De Marco}, {Iaconi}, {Chamandy}
  \& {Price}}{{Reichardt} et~al.}{2020}]{Reichardt+2020}
{Reichardt} T.~A.,  {De Marco} O.,  {Iaconi} R.,  {Chamandy} L.,   {Price}
  D.~J.,  2020, \mn@doi [\mnras] {10.1093/mnras/staa937}, \href
  {https://ui.adsabs.harvard.edu/abs/2020MNRAS.494.5333R} {494, 5333}

\bibitem[\protect\citeauthoryear{{Ricker} \& {Taam}}{{Ricker} \&
  {Taam}}{2012}]{Ricker+Taam2012}
{Ricker} P.~M.,  {Taam} R.~E.,  2012, \mn@doi [\apj]
  {10.1088/0004-637X/746/1/74}, \href
  {https://ui.adsabs.harvard.edu/\#abs/2012ApJ...746...74R} {746, 74}

\bibitem[\protect\citeauthoryear{{Sahai} \& {Patel}}{{Sahai} \&
  {Patel}}{2015}]{Sahai+2015}
{Sahai} R.,  {Patel} N.~A.,  2015, \mn@doi [\apjl]
  {10.1088/2041-8205/810/1/L8}, \href
  {https://ui.adsabs.harvard.edu/abs/2015ApJ...810L...8S} {810, L8}

\bibitem[\protect\citeauthoryear{{Sahai}, {Claussen}, {S{\'a}nchez Contreras},
  {Morris}  \& {Sarkar}}{{Sahai} et~al.}{2008}]{Sahai+2008}
{Sahai} R.,  {Claussen} M.,  {S{\'a}nchez Contreras} C.,  {Morris} M.,
  {Sarkar} G.,  2008, \mn@doi [\apj] {10.1086/587638}, \href
  {https://ui.adsabs.harvard.edu/abs/2008ApJ...680..483S} {680, 483}

\bibitem[\protect\citeauthoryear{{Sahai}, {Vlemmings}, {Gledhill}, {S{\'a}nchez
  Contreras}, {Lagadec}, {Nyman}  \& {Quintana-Lacaci}}{{Sahai}
  et~al.}{2017}]{Sahai+2017}
{Sahai} R.,  {Vlemmings} W.~H.~T.,  {Gledhill} T.,  {S{\'a}nchez Contreras} C.,
   {Lagadec} E.,  {Nyman} L.~{\r{A}}.,   {Quintana-Lacaci} G.,  2017, \mn@doi
  [\apjl] {10.3847/2041-8213/835/1/L13}, \href
  {https://ui.adsabs.harvard.edu/abs/2017ApJ...835L..13S} {835, L13}

\bibitem[\protect\citeauthoryear{{S{\'a}nchez Contreras}, {Alcolea},
  {Bujarrabal}, {Castro-Carrizo}, {Velilla Prieto}, {Santand er-Garc{\'\i}a},
  {Quintana-Lacaci}  \& {Cernicharo}}{{S{\'a}nchez Contreras}
  et~al.}{2018}]{SanchezContreras2018}
{S{\'a}nchez Contreras} C.,  {Alcolea} J.,  {Bujarrabal} V.,  {Castro-Carrizo}
  A.,  {Velilla Prieto} L.,  {Santand er-Garc{\'\i}a} M.,  {Quintana-Lacaci}
  G.,   {Cernicharo} J.,  2018, \mn@doi [\aap] {10.1051/0004-6361/201833632},
  \href {https://ui.adsabs.harvard.edu/abs/2018A&A...618A.164S} {618, A164}

\bibitem[\protect\citeauthoryear{{Sch{\"o}nberner}, {Jacob}, {Lehmann},
  {Hildebrand t}, {Steffen}, {Zwanzig}, {Sandin}  \&
  {Corradi}}{{Sch{\"o}nberner} et~al.}{2014}]{Schonberner+2014}
{Sch{\"o}nberner} D.,  {Jacob} R.,  {Lehmann} H.,  {Hildebrand t} G.,
  {Steffen} M.,  {Zwanzig} A.,  {Sandin} C.,   {Corradi} R.~L.~M.,  2014,
  \mn@doi [Astronomische Nachrichten] {10.1002/asna.201412051}, \href
  {https://ui.adsabs.harvard.edu/abs/2014AN....335..378S} {335, 378}

\bibitem[\protect\citeauthoryear{{Schreier}, {Hillel}  \& {Soker}}{{Schreier}
  et~al.}{2019}]{Schreier+2019}
{Schreier} R.,  {Hillel} S.,   {Soker} N.,  2019, \mn@doi [\mnras]
  {10.1093/mnras/stz2914}, \href
  {https://ui.adsabs.harvard.edu/abs/2019MNRAS.490.4748S} {490, 4748}

\bibitem[\protect\citeauthoryear{{Soker}}{{Soker}}{1992}]{Soker1992}
{Soker} N.,  1992, \mn@doi [\apj] {10.1086/171235}, \href
  {https://ui.adsabs.harvard.edu/abs/1992ApJ...389..628S} {389, 628}

\bibitem[\protect\citeauthoryear{{Soker}}{{Soker}}{1997}]{Soker1997}
{Soker} N.,  1997, \mn@doi [\apjs] {10.1086/313040}, \href
  {https://ui.adsabs.harvard.edu/abs/1997ApJS..112..487S} {112, 487}

\bibitem[\protect\citeauthoryear{{Soker} \& {Livio}}{{Soker} \&
  {Livio}}{1989}]{Soker+Livio1989}
{Soker} N.,  {Livio} M.,  1989, \mn@doi [\apj] {10.1086/167294}, \href
  {https://ui.adsabs.harvard.edu/\#abs/1989ApJ...339..268S} {339, 268}

\bibitem[\protect\citeauthoryear{{Staff}, {De Marco}, {Macdonald}, {Galaviz},
  {Passy}, {Iaconi}  \& {Low}}{{Staff} et~al.}{2016}]{Staff+2016}
{Staff} J.~E.,  {De Marco} O.,  {Macdonald} D.,  {Galaviz} P.,  {Passy} J.-C.,
  {Iaconi} R.,   {Low} M.-M.~M.,  2016, \mn@doi [\mnras]
  {10.1093/mnras/stv2548}, \href
  {https://ui.adsabs.harvard.edu/abs/2016MNRAS.455.3511S} {455, 3511}

\bibitem[\protect\citeauthoryear{{Steffen} \& {L{\'o}pez}}{{Steffen} \&
  {L{\'o}pez}}{2004}]{Steffen+Lopez2004}
{Steffen} W.,  {L{\'o}pez} J.~A.,  2004, \mn@doi [\apj] {10.1086/422445}, \href
  {https://ui.adsabs.harvard.edu/\#abs/2004ApJ...612..319S} {612, 319}

\bibitem[\protect\citeauthoryear{{Towns}, {Cockerill}, {Dahan}  \&
  {Foster}}{{Towns} et~al.}{2014}]{xsede}
{Towns} J.,  {Cockerill} T.,  {Dahan} M.,   {Foster} I.,  2014, \mn@doi
  [Computing in Science and Engineering] {10.1109/MCSE.2014.80}, 16, 62

\bibitem[\protect\citeauthoryear{{Vishniac} \& {Ryu}}{{Vishniac} \&
  {Ryu}}{1989}]{Vishniac+1989}
{Vishniac} E.~T.,  {Ryu} D.,  1989, \mn@doi [\apj] {10.1086/167161}, \href
  {https://ui.adsabs.harvard.edu/abs/1989ApJ...337..917V} {337, 917}

\makeatother
\end{thebibliography}



\appendix

\section{Jet injected in equatorial plane}\label{apx:xjet}

\begin{figure*}
    \includegraphics[height = \columnwidth]{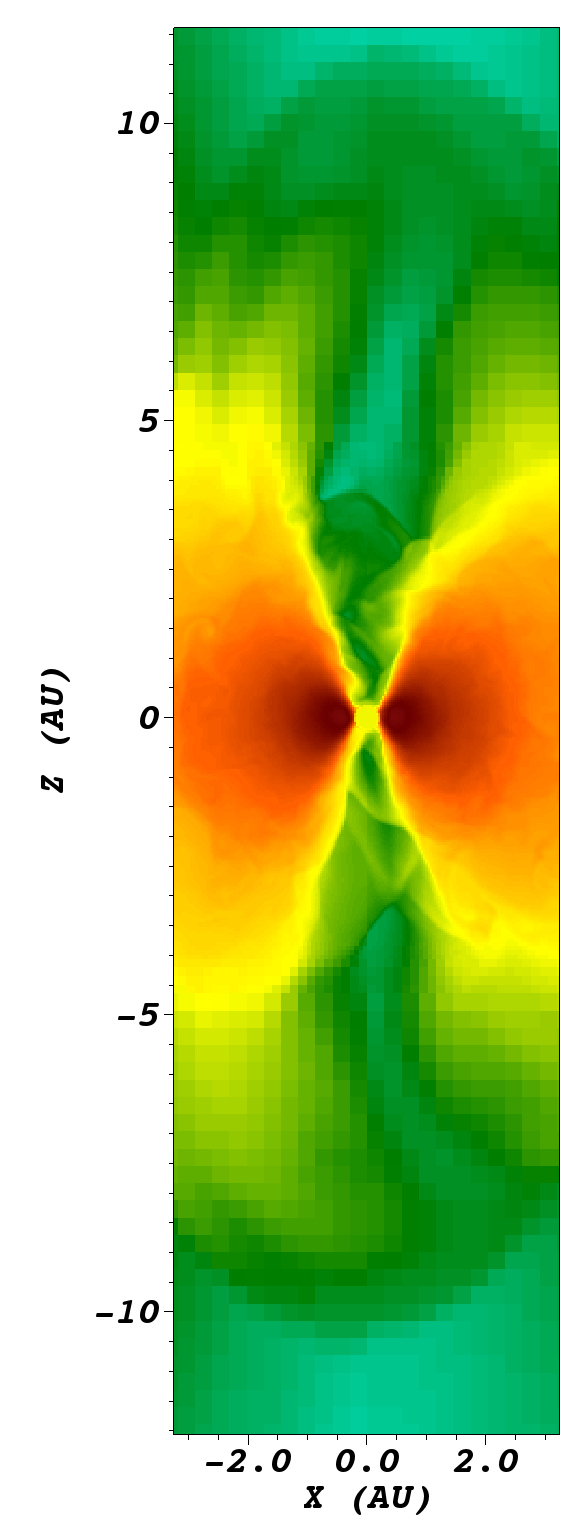}
    \includegraphics[height = \columnwidth]{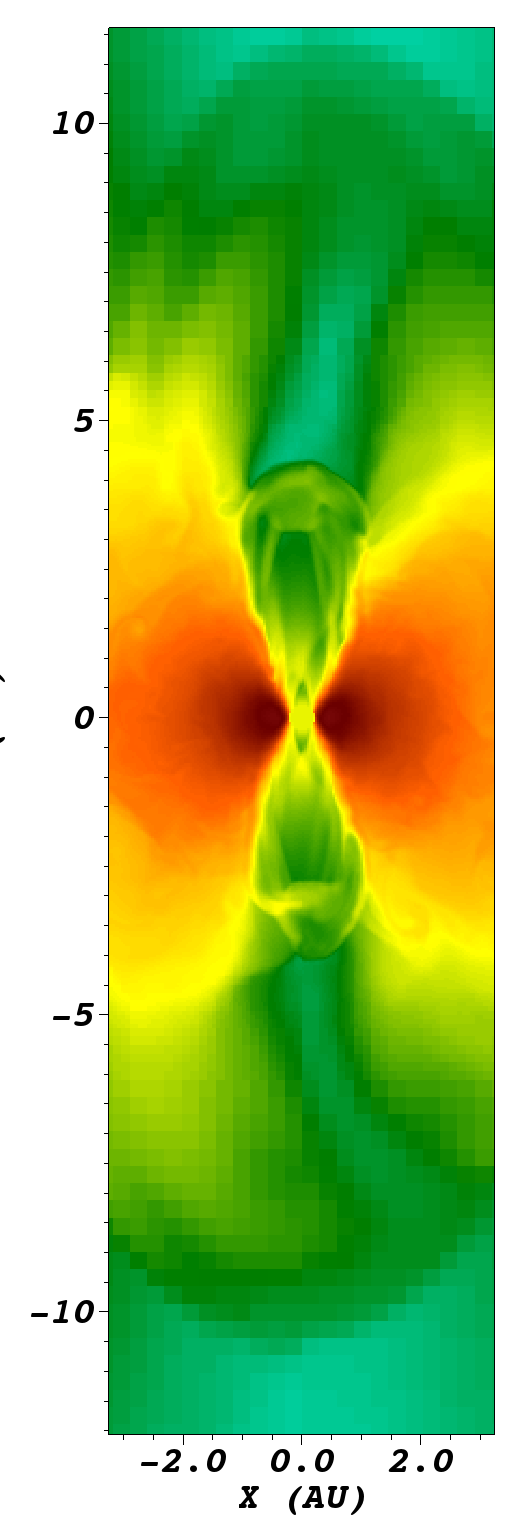}
    \includegraphics[height = \columnwidth]{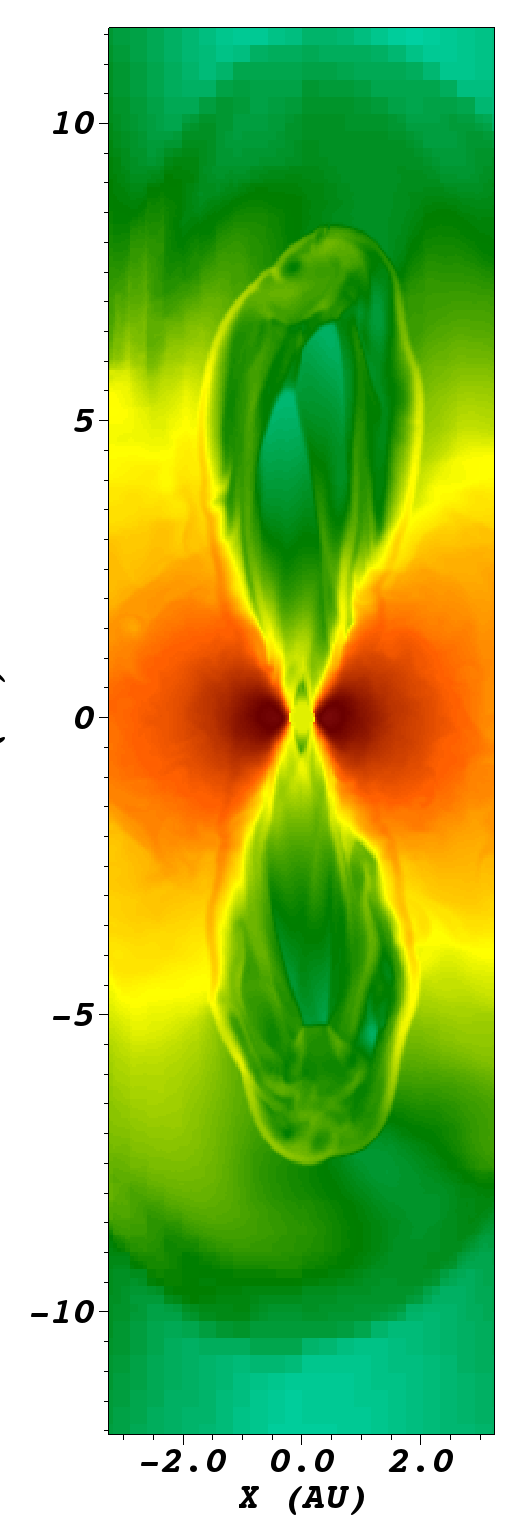}
    \includegraphics[height = \columnwidth]{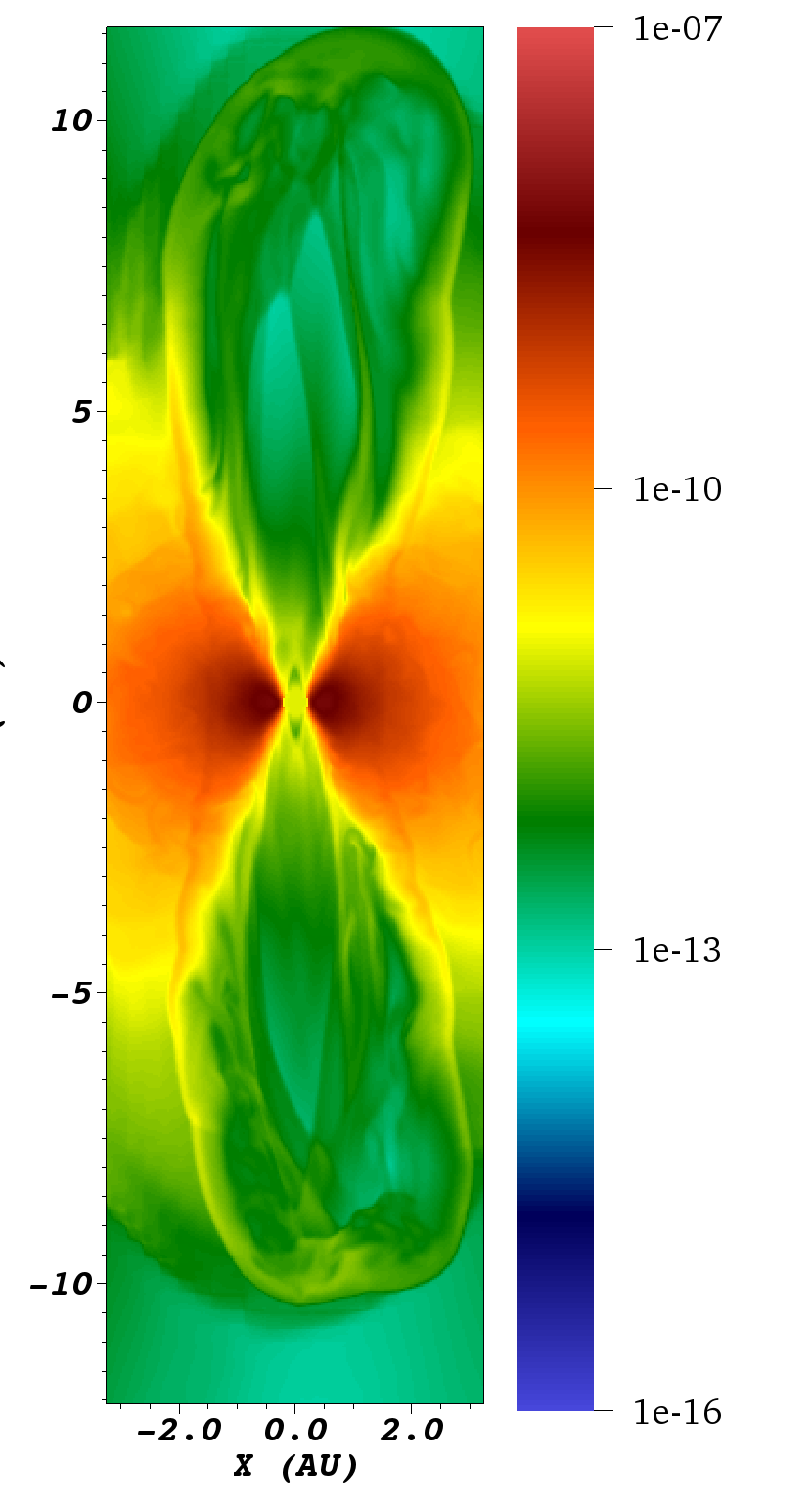}
    \includegraphics[height = \columnwidth]{time-evolutions/rho-new.png}
    \caption{Density ($\mathrm{g\,cm^{-3}}$, in log scale) maps for the test run where the outflow is confined in jets along the x-axis and injected into the equatorial plane. Panels are taken at 20 days apart after starting the fast wind. Note that scale is 10 times smaller compared to figures in the main text. This outflow is again collimated by the CE ejecta and bipolar lobes develop clearly after 40 days.}
    \label{fig:xjet-rho}
\end{figure*}

\begin{figure*}
    \includegraphics[height = \columnwidth]{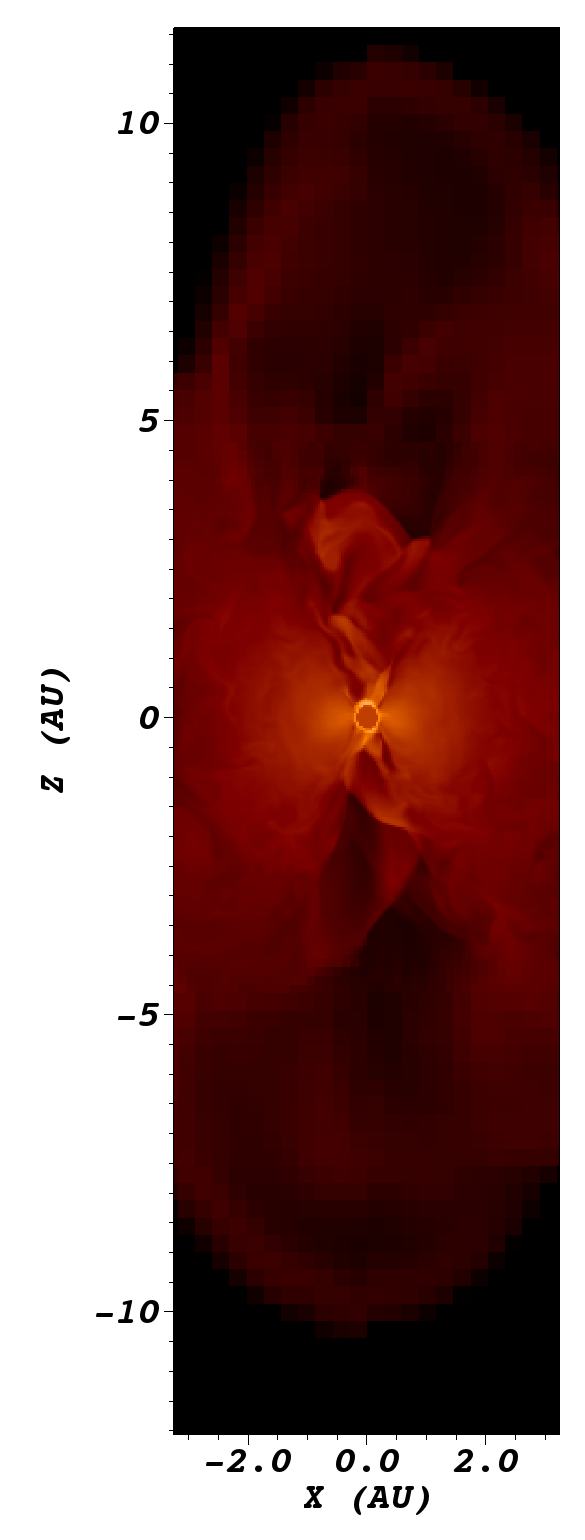}
    \includegraphics[height = \columnwidth]{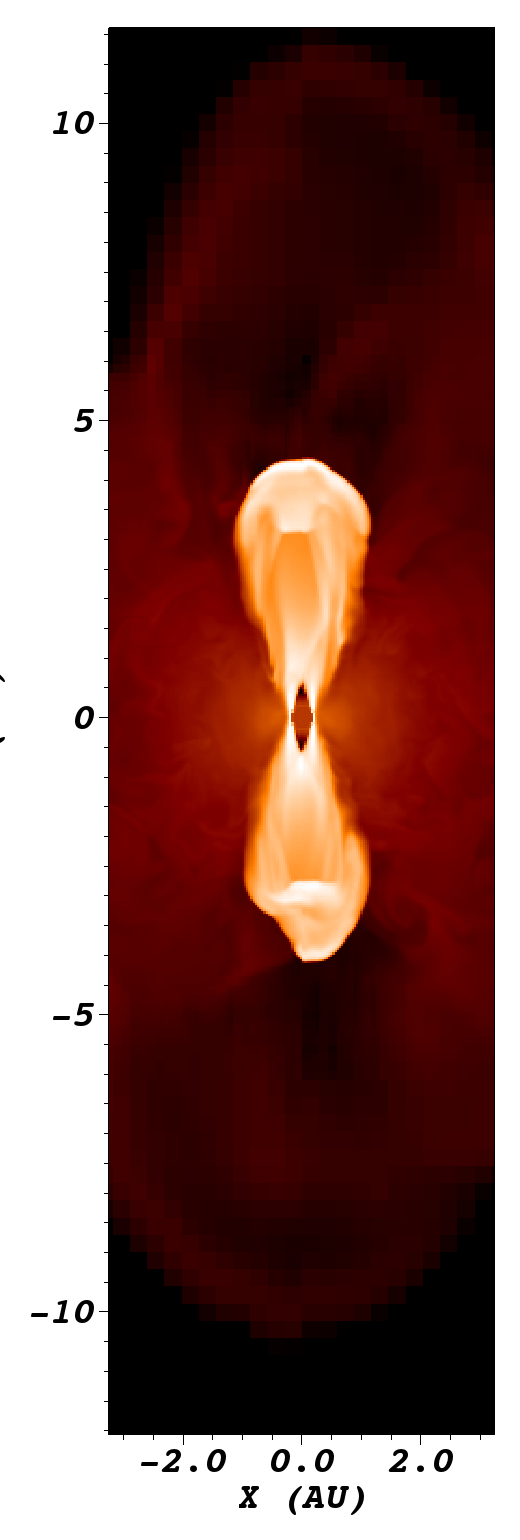}
    \includegraphics[height = \columnwidth]{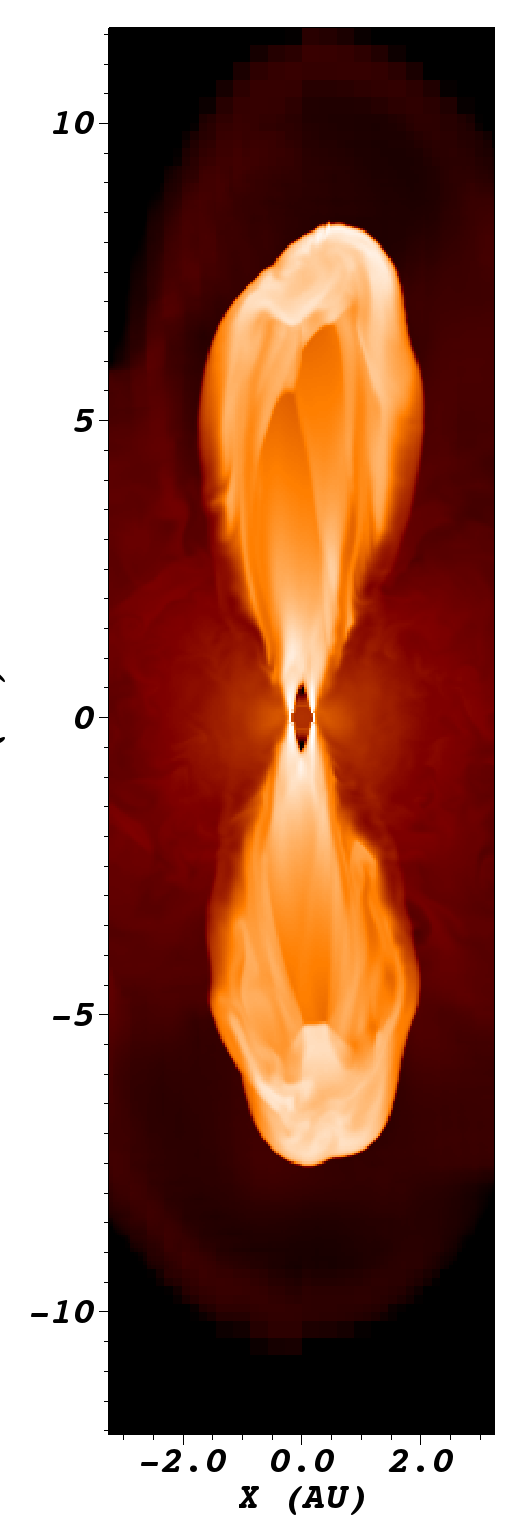}
    \includegraphics[height = \columnwidth]{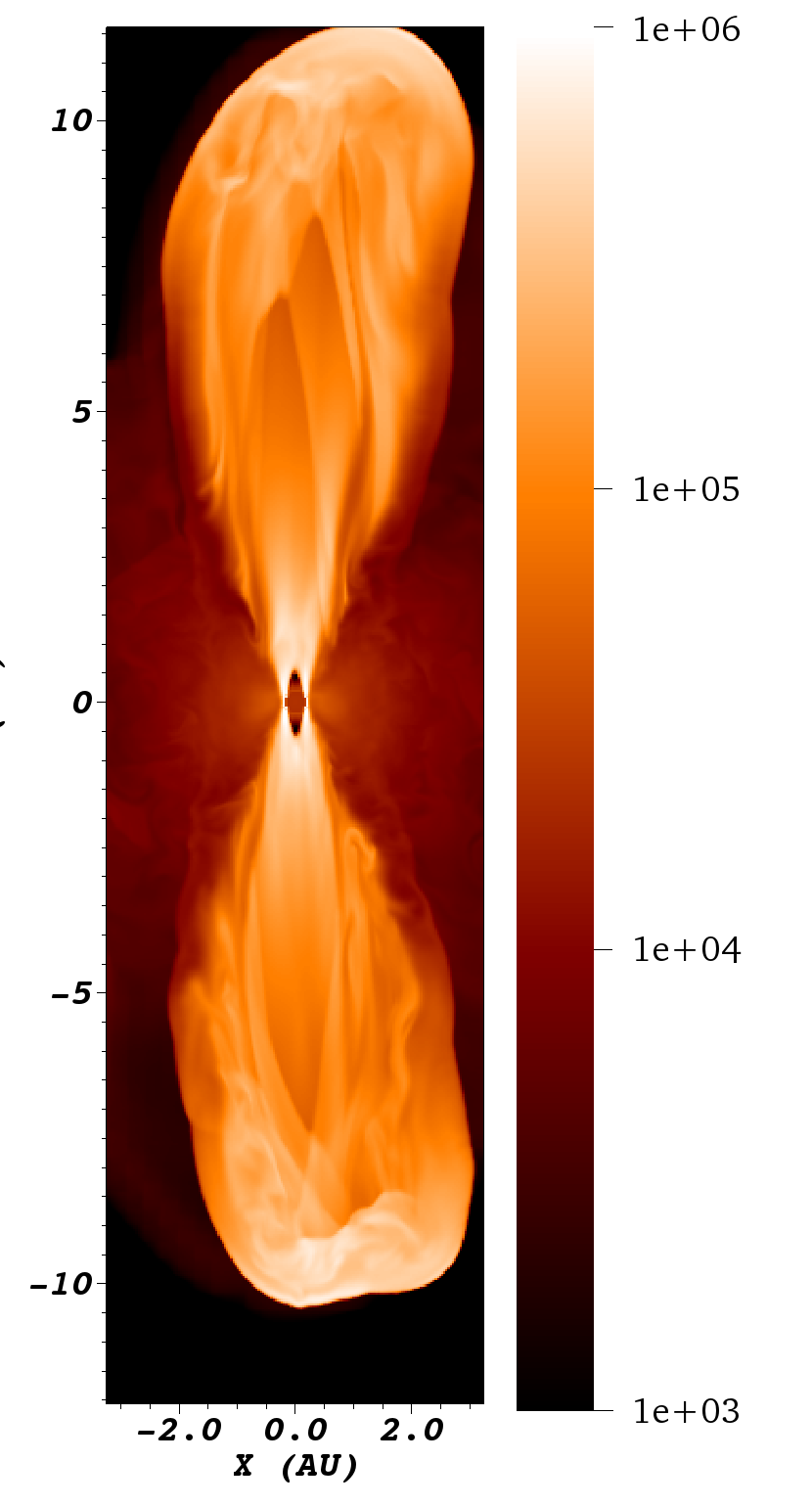}
    \includegraphics[height = \columnwidth]{time-evolutions/T-new.png}
    \caption{Temperature (K, in log scale) maps for the same test run shown in Fig. \ref{fig:xjet-rho}. Panels are taken at 20 days apart after starting the fast wind. Note the elliptical shock, which outlines the central cold region, is visible in all except in the first panel. }
    \label{fig:xjet-Temp}
\end{figure*}

For comparison, we present a test run where the outflow is injected into the equatorial plane. 
The outflow is now confined in two opposite cones oriented along the x-axis. 
Outflow material is directed into $\mathrm{\pm \hat{x}}$-directions, with a half-open angle of $15^{\circ}$,
while the outflow density and speed are set as our high-momentum case. 
Fig.~\ref{fig:xjet-rho} and \ref{fig:xjet-Temp} show evolution of this case in density and 
temperature respectively. 

As seen in other cases, bipolar lobes above and below the equatorial plane quickly developed along z-axis. 
Even though this is another extreme case where outflow is injected orthogonal to the lobes, it has still been redirected to the open funnel.  

\section{Test run without the quiescent period}\label{apx:noIW}
\begin{figure*}
    \includegraphics[height=\columnwidth]{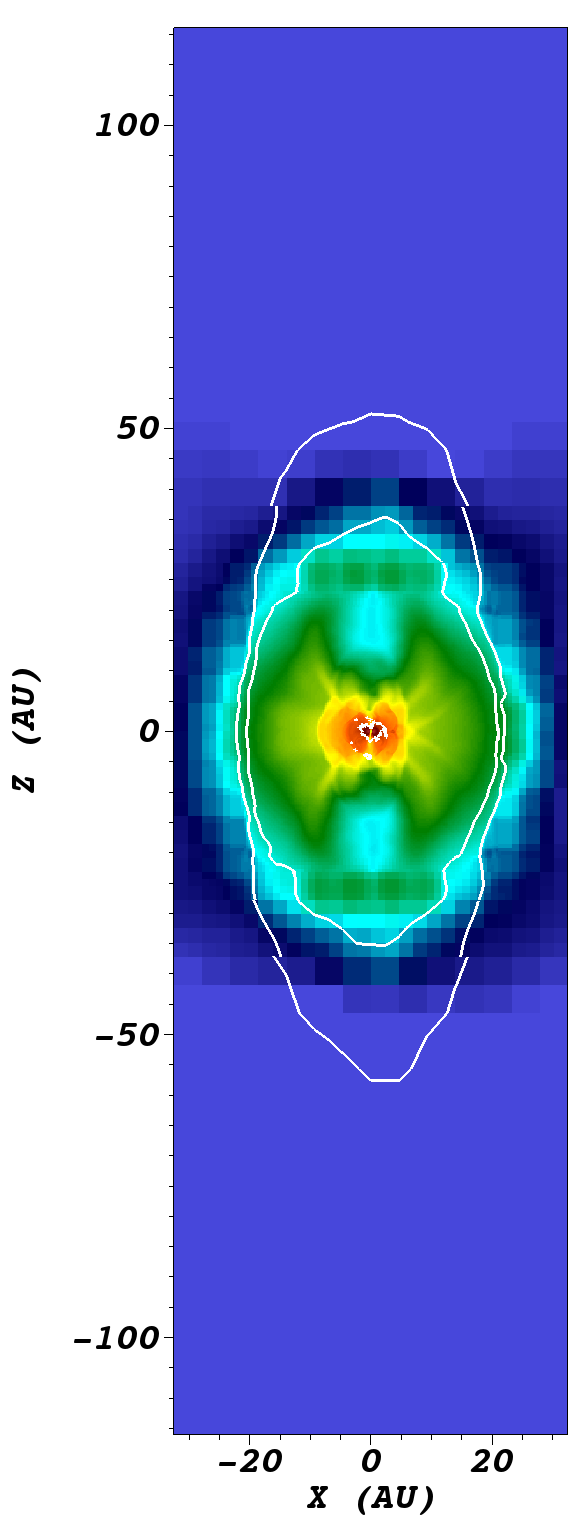}
    \includegraphics[height=\columnwidth]{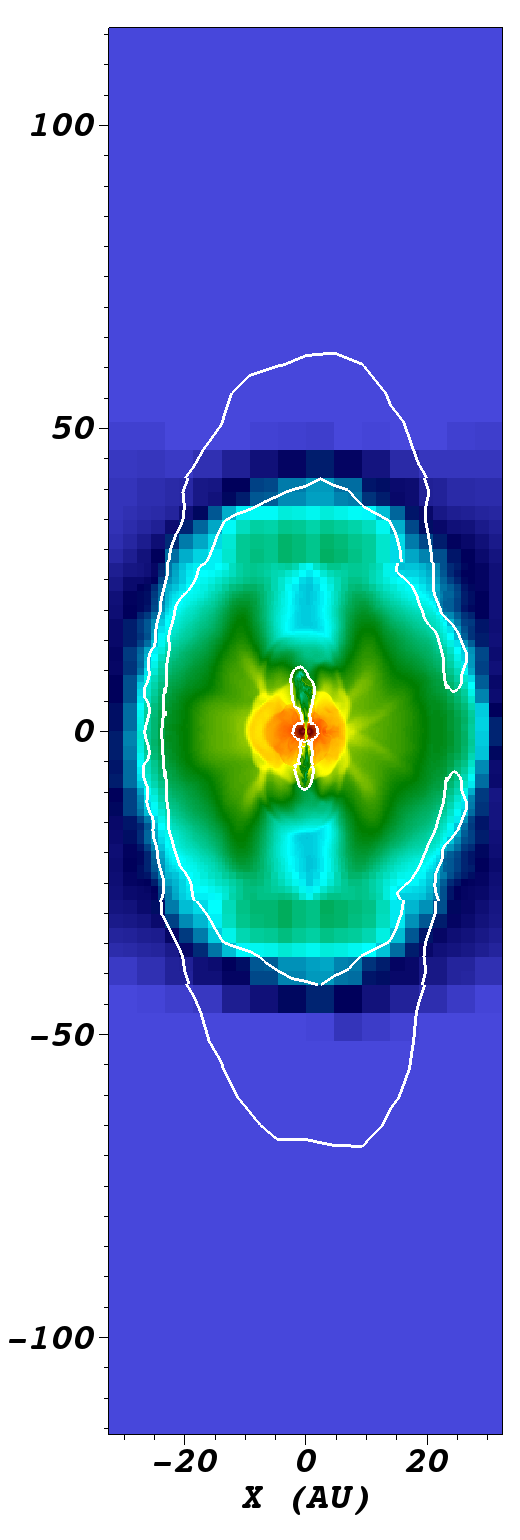}
    \includegraphics[height=\columnwidth]{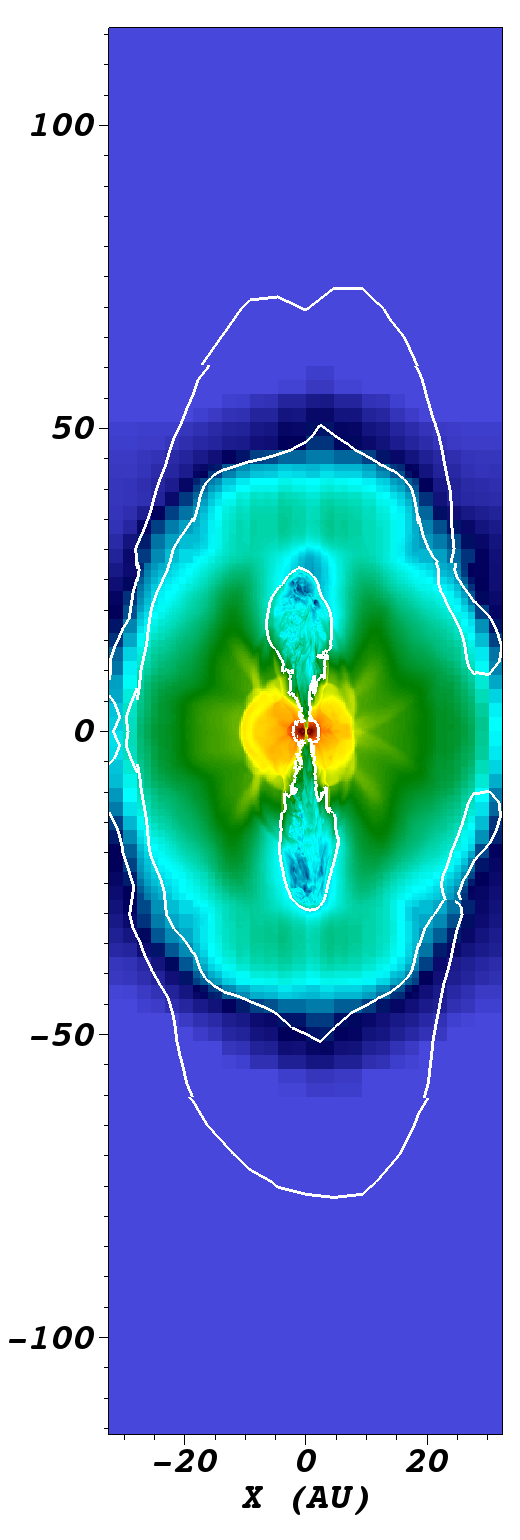}
    \includegraphics[height=\columnwidth]{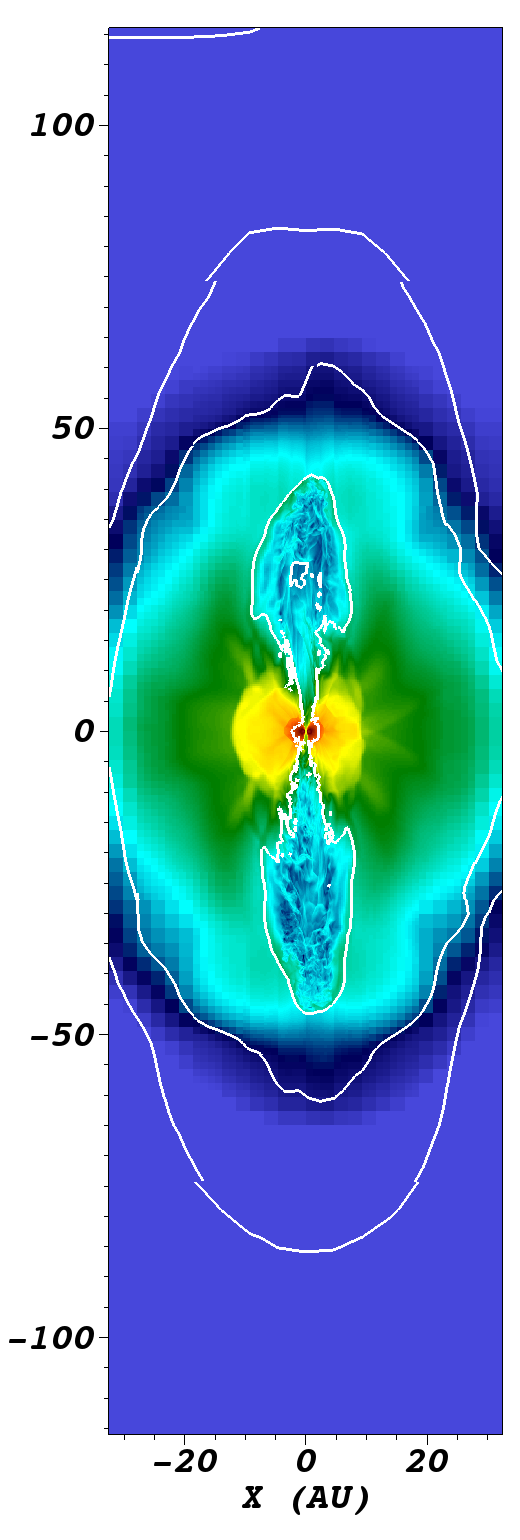}
    \includegraphics[height=\columnwidth]{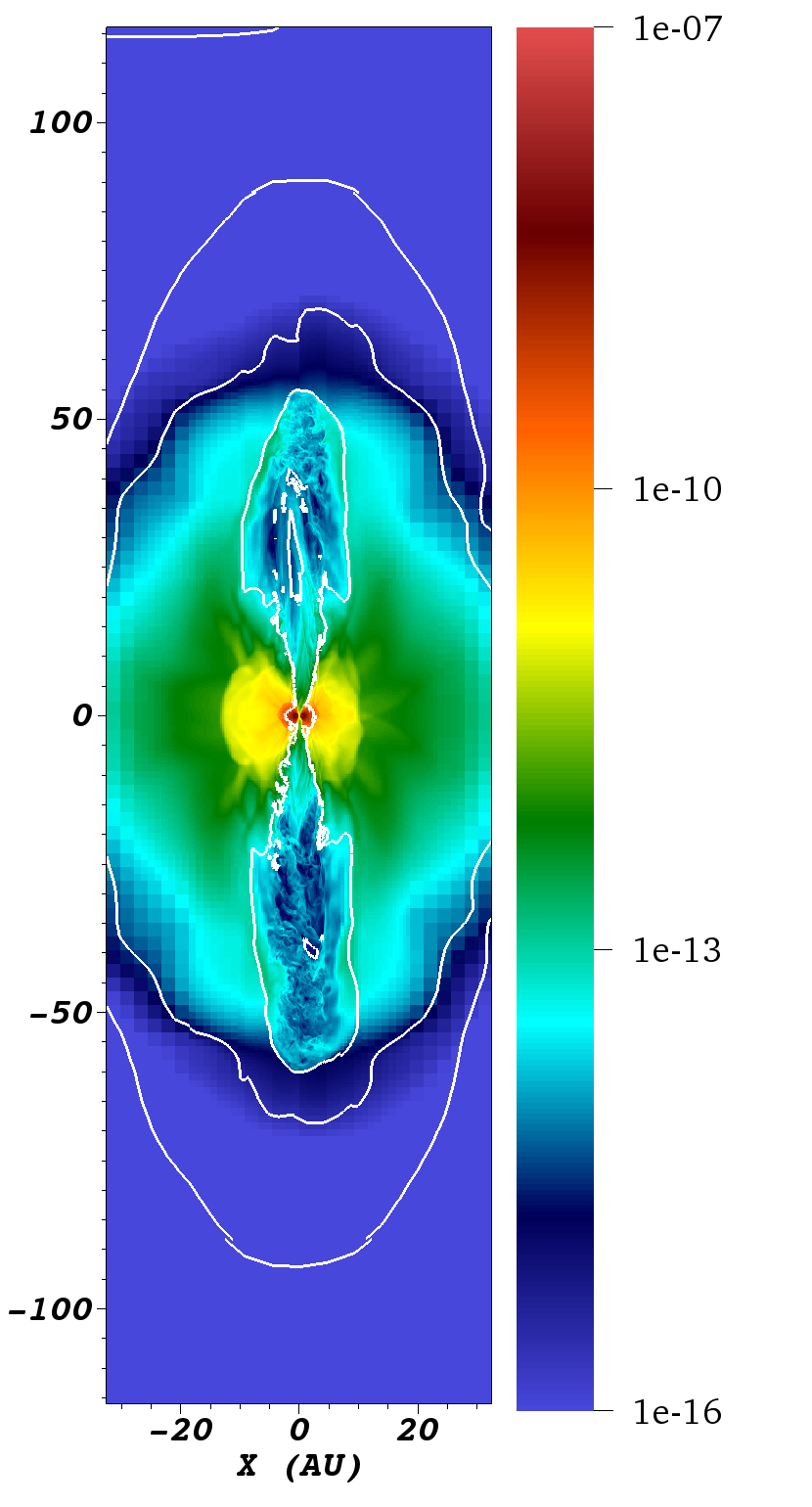}
    \includegraphics[height=\columnwidth]{time-evolutions/rho-new.png}
    \caption{Time evolution of a test run where outflow starts at the beginning of the simulation. Panels are snapshots taken at 260, 480, 700, 920, and 1080 days in the simulation. Pseudocolor shows density ($\mathrm{g\,cm^3}$, in log scale), and the white contours mark constant temperature of 10 000 K. As can be seen the bipolar outflows morphology evolution is qualitatively similar.}
    \label{fig:noIW-rho}
\end{figure*}

As part of our convergence and sensitivity study, we carried out a test run without the initial 3000-day quiescent phase. Here the spherical outflow starts at the beginning of the simulation with the same parameters as in model A. Time evolution of this run is shown in Fig.~\ref{fig:noIW-rho}. It takes longer for the outflow to make it through the central remnant and the bipolar lobes are about half the size compared to those in model A. The funnel is much denser since the CE remnant has no time to relax and expand. Top lobe in earlier time has a left tilt similar to those in models B and C. 

\section{ AMR grids and refinement}
\label{apx:mesh}
\begin{figure*}
    \includegraphics[height=\columnwidth]{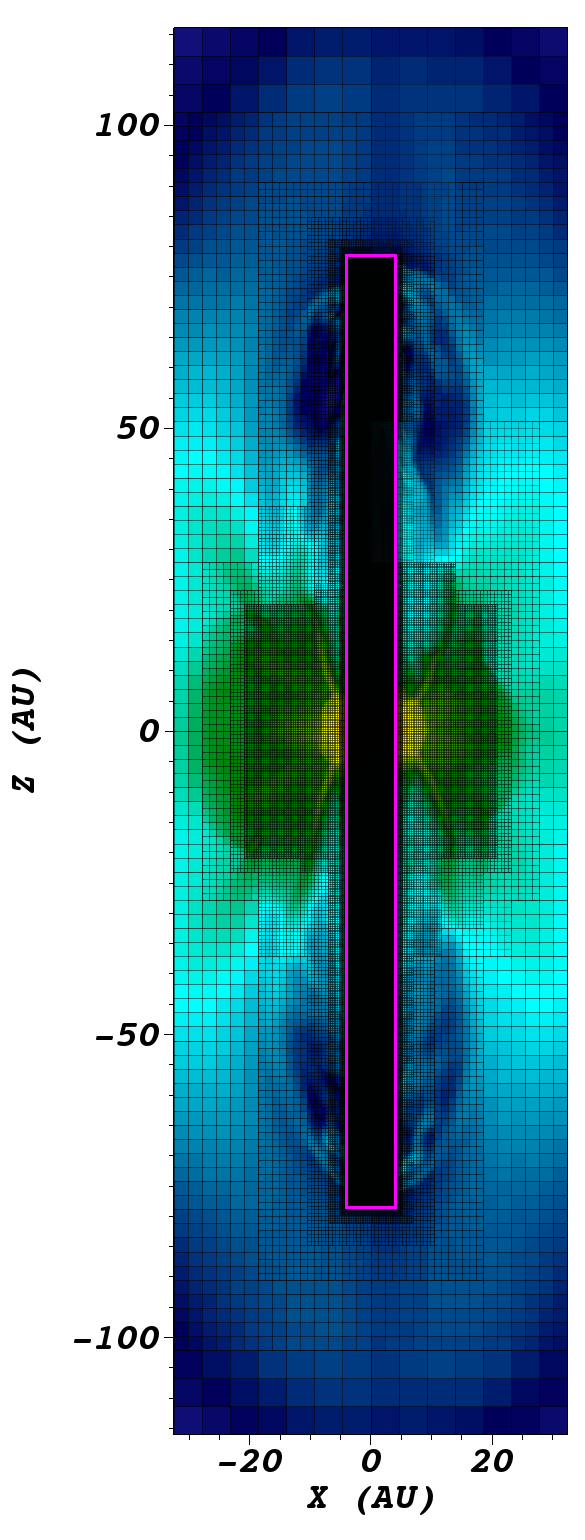}
    \includegraphics[height=\columnwidth]{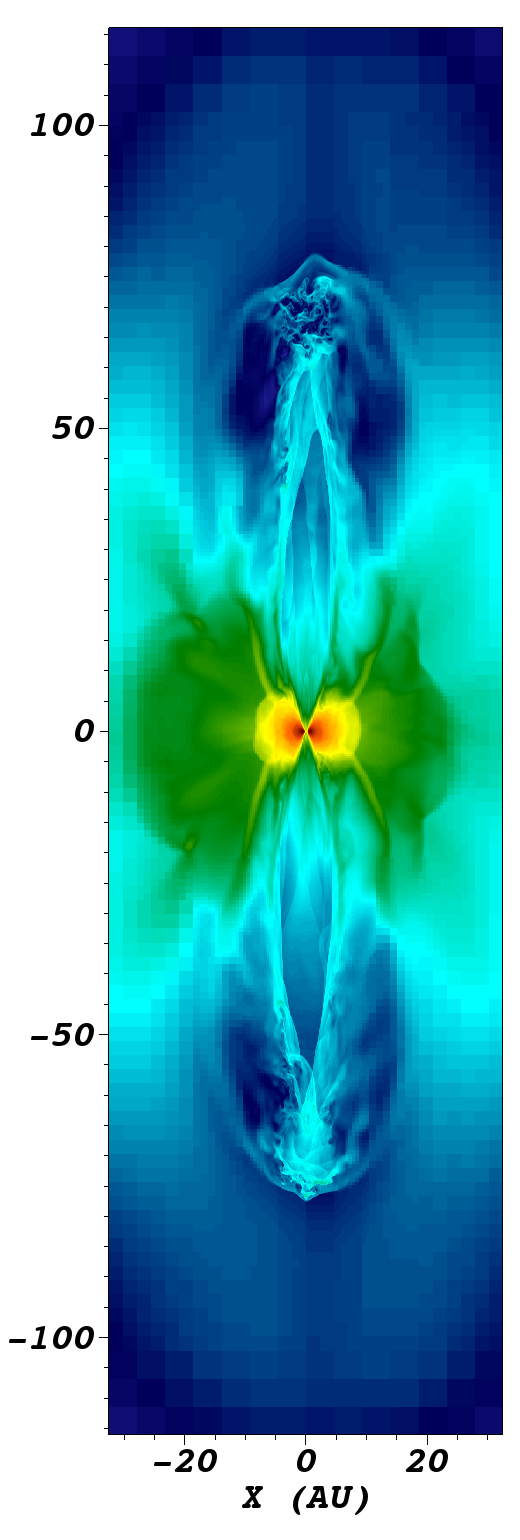}
    \includegraphics[height=\columnwidth]{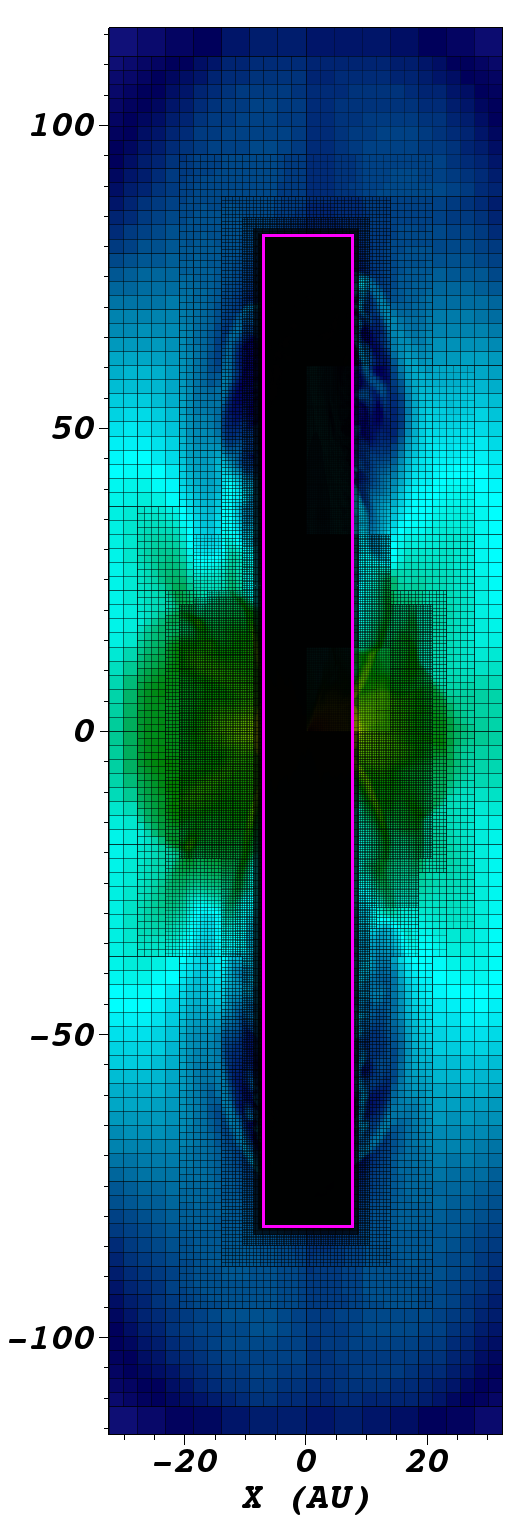}
    \includegraphics[height=\columnwidth]{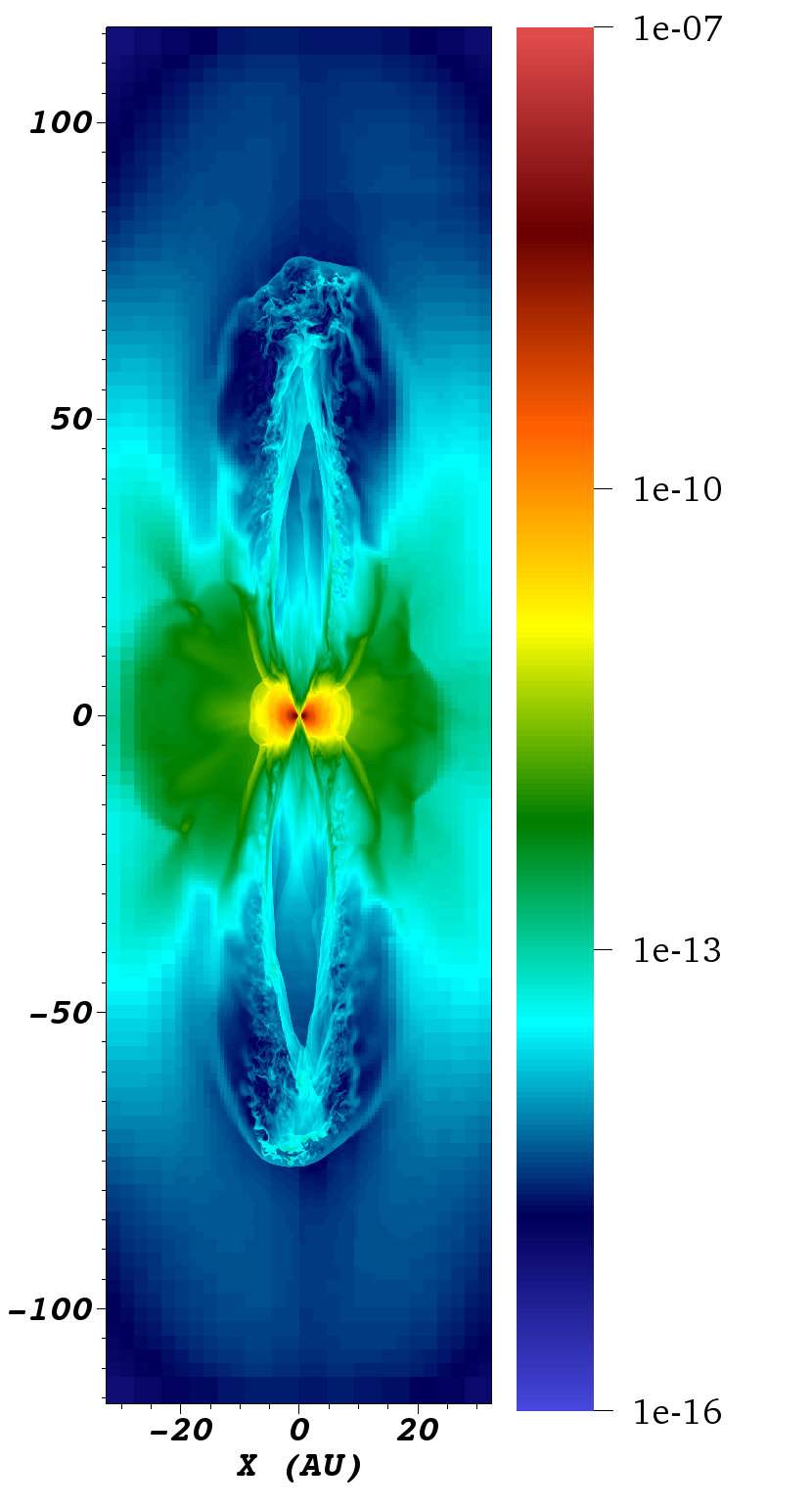}
    \includegraphics[height=\columnwidth]{time-evolutions/rho-new.png}
    \caption{Density maps and AMR grids taken at 3700 days in the simulation (same time as the 3rd panel in Fig.~\ref{fig:A-rho}). Left two panels: model A. Right two panels: test run which doubles the side of the refinement prism. Magenta squares mark the boundary of the rectangular prism, where refinement is forced to be the highest level 7. The global flow pattern remains the same for both runs, while there are more small scale features along the sides of the lobes in the right panels.}
    \label{fig:mesh}
\end{figure*}

To test the effect of the prism region of high refinement we performed a run which doubles the prism's width to $\sim~14 \mathrm{AU}$. Other parameters were held the same as in model A.
Fig.~\ref{fig:mesh} shows the AMR grids plotted over density maps at 3700 days. The left two panels are from the original model A, while the right two panels are from the test run. While we see more small scale features along the sides of the lobes (due to resolving Kelvin-Helmholtz instabilities) in the run with more refinement, the global flow pattern including the shape and extent of the bipolar lobes, the central jet pillar, and the lens-shaped inner shock, remain the same for both runs. 

\bsp	
\label{lastpage}
\end{document}